\documentclass[12pt,runningheads]{article}
\usepackage[utf8]{inputenc}
\usepackage{array,multicol,mathpazo}
\usepackage[normalem]{ulem}
\usepackage{fullpage}
\usepackage{soul}
\usepackage{floatrow}
\usepackage[top=0.9in, bottom=0.9in, left=0.9in, right=0.9in]{geometry}
\usepackage{setspace}
   
\fontsize{12}{0}
\usepackage{amsmath}
\usepackage{amssymb}
\usepackage{bm}
\usepackage{amsthm}
\usepackage{mathtools} 
\usepackage{exscale}
\usepackage[mathscr]{eucal}
\usepackage{bm}
\usepackage{eqlist} % Makes for a nice list of symbols.
\usepackage[final]{graphicx}
\usepackage[dvipsnames]{color}
\usepackage{verbatim}
\usepackage[square,sort,comma]{natbib}
\usepackage{caption}
\usepackage{enumerate}
\usepackage{subcaption}
\usepackage{algpseudocode}
\usepackage{rotating}
\usepackage{etoolbox}
\usepackage{color}
\usepackage[linesnumbered,ruled]{algorithm2e}
\usepackage[thinc]{esdiff}
\usepackage[shortlabels]{enumitem}
\PassOptionsToPackage{hyphens}{url}\usepackage{hyperref}

\makeatletter
\patchcmd{\env@cases}{1.2}{0.6}{}{}
\makeatother

\DeclareGraphicsExtensions{.pdf, .jpg}

\theoremstyle{definition}

\newtheorem{condition}{Condition}

\newtheorem{proposition.a}{Proposition A\ignorespaces}
\newtheorem{proposition.s}{Proposition S\ignorespaces}
\newtheorem{remark}{Remark}
\newtheorem{remark.s}{Remark S\ignorespaces}
\newtheorem{thm}{Theorem}
\newtheorem{thm.s}{Theorem S\ignorespaces}
\newtheorem{cor}{Corollary}
\newtheorem{cor.s}{Corollary S\ignorespaces}
\newtheorem{lem}{Lemma}
\newtheorem{lem.s}{Lemma S\ignorespaces}
\newcommand{\A}{\bm{A}}

\newcommand{\B}{\bm{B}}

\newcommand{\x}{\bm{x}}

\newcommand{\bmd}{\bm{d}}
\newcommand{\y}{\bm{y}}
\newcommand{\X}{\bm{X}}

\newcommand{\h}{{h}}
\newcommand{\Z}{\bm{Z}}
\newcommand{\I}{\bm{I}}
\newcommand{\W}{\bm{W}}

\newcommand{\bmS}{\bm{S}}
\newcommand{\bma}{\bm{a}}

\newcommand{\bmeps}{\bm{\epsilon}}
\newcommand{\bmbeta}{\bm{\beta}}
\newcommand{\bmalpha}{\bm{\alpha}}
\newcommand{\bmeta}{\bm{\eta}}
\newcommand{\1}{\bm{1_n}}
\newcommand{\tE}{\mbox{E}}
\newcommand{\var}{\mbox{Var}}
\newcommand{\diagg}{\mbox{Diag}}
\newcommand{\cov}{\mbox{Cov}}
\newcommand{\tr}{\mbox{tr}}
\newcommand{\bbR}{\mathbb{R}}
\newcommand{\bbC}{\mathbb{C}}

\newcommand{\bmI}{\mbox{I}}
\newcommand{\bmSigma}{\bm{\Sigma}}

\newcommand{\bmPhi}{\bm{\Phi}}
\makeatletter
\newcommand*\rel@kern[1]{\kern#1\dimexpr\macc@kerna}
\newcommand*\widebar[1]{%
  \begingroup
  \def\mathaccent##1##2{%
    \rel@kern{0.8}%
    \overline{\rel@kern{-0.8}\macc@nucleus\rel@kern{0.2}}%
    \rel@kern{-0.2}%
  }%
  \macc@depth\@ne
  \let\math@bgroup\@empty \let\math@egroup\macc@set@skewchar
  \mathsurround\z@ \frozen@everymath{\mathgroup\macc@group\relax}%
  \macc@set@skewchar\relax
  \let\mathaccentV\macc@nested@a
  \macc@nested@a\relax111{#1}%
  \endgroup
}
\makeatother
\usepackage{newfloat}
\DeclareFloatingEnvironment[name={Supplementary Fig.},fileext=lof]{suppfigure}
\DeclareFloatingEnvironment[name={Supplementary Table}]{supptable}
\usepackage[labelfont=bf]{caption}
\usepackage{tabularx}
\usepackage{lipsum}
\title{\LARGE  
Cross-trait prediction accuracy of high-dimensional ridge-type estimators in genome-wide association studies} 
\author{Bingxin Zhao and Hongtu Zhu\\~\\
University of North Carolina at Chapel Hill}
\begin{document}
% Add the title section.
\maketitle
\date{}
% Add an abstract.
\abstract{
Marginal association summary statistics have attracted great attention in statistical genetics, mainly because the primary results of most genome-wide association studies (GWAS) are produced by marginal screening. 
In this paper, we study the prediction accuracy of marginal estimator in dense (or sparsity free) high-dimensional settings with $(n,p,m) \to \infty$, $m/n \to \gamma \in (0,\infty)$, and $p/n \to \omega \in (0,\infty)$. 
We consider a general correlation structure among the $p$ features and allow an unknown subset $m$ of them to be signals. 
As the marginal estimator can be viewed as a ridge estimator with regularization parameter $\lambda \to \infty$, we further investigate a class of ridge-type estimators in a unifying framework, including the popular best linear unbiased prediction (BLUP) in genetics.
We find that the influence of $\lambda$ on out-of-sample prediction accuracy heavily depends on $\omega$.
Though selecting an optimal $\lambda$ can be important when $p$ and $n$ are comparable, it turns out that the out-of-sample $R^2$ of ridge-type estimators becomes near-optimal for any $\lambda \in (0,\infty)$ as $\omega$ increases. 
For example, when features are independent, the out-of-sample $R^2$ is always bounded by $1/\omega$ from above and is largely invariant to $\lambda$ given large $\omega$ (say, $\omega>5$).
We also find that in-sample $R^2$ has completely different patterns and depends much more on $\lambda$ than out-of-sample $R^2$.
In practice, our analysis delivers useful messages for genome-wide polygenic risk prediction and computation-accuracy trade-off in dense high-dimensions.
We numerically illustrate our results in simulation studies and a real data example.

\bigskip

}

%%%%%%%%%%%%%%%%%%%%%%%%%%%%%%%%%%%%%%%%%%%%%%%%%%%%%%%%%%%%%%%%%%%%
%%%%%%%%%%%%%%%%%%%%%%%%%%%%%%%%%%%%%%%%%%%%%%%%%%%%%%%%%%%%%%%%%%%%
%%%%%%%%%%%%%%%%%%%%%%%%%%%%%%%%%%%%%%%%%%%%%%%%%%%%%%%%%%%%%%%%%%%%
\section{Introduction}\label{sec1}
Human complex traits often have a polygenic genetic architecture \citep{o2019extreme,wray2018common,boyle2017expanded}. 
That is, a large number of genetic variants have small but nonzero contributions to phenotypic variation \citep{timpson2018genetic}.
Genome-wide association studies (GWAS) aim to find suspicious genetic risk variants by examining association between complex traits and millions of variants, typically common (minor allele frequency [MAF] $\ge0.05$) single-nucleotide polymorphisms (SNPs) collected across the genome. 
After a decade of GWAS discovery, more than 100 millions of individuals have been genotyped \citep{martin2019clinical} and thousands of unique traits have been studied \citep{visscher201710}. 

In the genetics community, GWAS summary association statistics (e.g., effect size, standard error, $p$-value) of all SNPs for various traits are shared and assembled into large databases.  
Summary statistics from more than $4000$ GWAS are now publicly available \citep{watanabe2019global} and the number rises steeply.
As individual-level SNP data are massive and are often under strict ethical/regulatory protections, it is an active research area to directly use these GWAS summary statistics for various in-sample and out-of-sample analyses \citep{pasaniuc2017dissecting}. 
For example, GWAS summary statistics are used to prioritize causal variants in fine-mapping analysis \citep{schaid2018genome}, 
to quantify genetic overlaps among different traits \citep{bulik2015atlas,speed2019sumher}, 
to perform causal inference among traits via Mendelian randomization \citep{zhao2018statistical},
and to carry out integrative association tests with gene expression data \citep{gamazon2015gene,gusev2016integrative,hu2019statistical}.
Owing primarily to the potential to translate GWAS findings to medical advancements, 
it is of particular interest to predict the personalized genetic risk for new GWAS individuals using results from historical GWAS \citep{torkamani2018personal,sugrue2019polygenic,martin2019clinical}.
One of the state-of-art methods for genetic risk prediction of human complex traits is genome-wide polygenic risk score (PRS) \citep{purcell2009common}, which is a weighted sum of millions of SNPs where each SNP is weighted by their estimated effect size from discovery GWAS. 
As no need to access the personal DNA information of subjects in the training set, PRS is computationally efficient and has widespread applications with more than $3,000$ related publications in 2018 \citep{Zhao447797}. 
Recent efforts have begun to explore the clinical utility of PRS on human diseases, such as heart disease and breast cancer \citep{mavaddat2019polygenic,khera2018genome}.

Though GWAS summary statistics have numerous applications, there is little rigorous theoretical evaluation.  
Without sufficient understanding of their statistical properties, we risk drawing erroneous decisions on summary statistics database design and construction. 
Most, if not all, of publicly shared GWAS summary statistics are marginal effects generated from marginal screening. 
Let $\y$ be an $n \times 1$ vector of continuous trait, a linear polygenic structure between $\y$ and SNP data $\X$ is often assumed in GWAS (e.g., \cite{jiang2016high})
\begin{flalign*}
\y=\X\bmbeta+\bmeps=\sum_{i=1}^{p}\x_i\beta_i+\bmeps=\sum_{i=1}^{m}\x_i\beta_i+\bmeps,
%\label{equ1.0.1}
\end{flalign*}
where $\X=(\x_{1},\cdots,\x_{m},\x_{m+1}, \cdots,\x_{p})$ is an $n \times p$ SNP data matrix with population-level correlation $\bmSigma$ among the $p$ features, $\bmbeta=(\beta_1,\cdots, \beta_{m},$ $ \beta_{m+1},\cdots, \beta_p)^{T}$ is a $p\times 1$ vector of genetic effects such that $(\beta_1,\cdots, \beta_{m})^T$ are $m$ unknown nonzero parameters, and $(\beta_{m+1},\cdots, \beta_p)^T$ are zeros, and the $n\times 1$ vector $\bmeps$ represents independent non-genetic random errors. 
The single SNP analysis in GWAS is given by
\begin{flalign}
\y=\1\mu_i+\x_i\beta_i+\bmeps_i^*
\label{equ1.0.2}
\end{flalign}
for $i=1,\cdots,p$, which is a marginal screening
approach similar to sure independence screening \citep{fan2008sure}. 
Let $\widehat{\bmbeta}_S=(\widehat{\beta}_1{_S},\cdots$ $,  \widehat{\beta}_p{_S})^{T}$ be 
the marginal screening ordinary least squares (OLS) estimator of model~(\ref{equ1.0.2}), the marginal estimators are given by 
\begin{flalign*}
\widehat{\beta_i}_{S}=(\x_i^T\x_i)^{-1}\x_i^T\y, \quad i=1,\cdots,p,
%\label{equ1.0.3}
\end{flalign*}
and thus  $\widehat{\bmbeta}_S=\{\diagg(\X^T\X)\}^{-1}\X^T\y$ is the form of nearly all shared GWAS summary statistics for continuous traits, where $\diagg(\A)$ is the diagonal of matrix $\A$.

For human complex traits, the number of causal SNPs $m$ is trait-specific and population-specific, 
and it can be comparable with $n$, but is not necessarily $p$.
An overwhelming number of empirical evidence supports the polygenicity and pleiotropy of complex traits  (e.g., \cite{watanabe2019global,martin2018predicting,sullivan2019defining}), which can be potentially explained by biological complexity and negative selection \citep{o2019extreme}.   
Statistically, we may have a dense (sparsity free) signal model, while not every feature has a nonzero effect on the outcome. 
This is different from the standard settings in spare regression (e.g.,  \cite{zhao2006model,fan2008sure,feng2017sorted,guo2019optimal}), which often has sparsity restriction on $m$. 

Motivated by GWAS applications, the goal of this paper is to evaluate the out-of-sample and in-sample behaviors of high-dimensional marginal estimator $\widehat{\bmbeta}_S$ in sparsity free settings. 
As marginal estimator can be viewed as a ridge-type estimator with regularization parameter increases to infinity \citep{fan2008sure}, we investigate several popular ridge-type estimators in an unifying framework and allow the regularization parameter goes to zero or infinity.
We first introduce the whole class of ridge-type estimators studied in this paper.
%%%%%%%%%%%%%%%%%%%%%%%%%%%%%%%%%%%%%%%%%%%%%%%%%%%%%%%%%%%%%%%%%%%%
%%%%%%%%%%%%%%%%%%%%%%%%%%%%%%%%%%%%%%%%%%%%%%%%%%%%%%%%%%%%%%%%%%%%
\subsection{The class of ridge-type estimators}\label{sec1.1}
In this section, we summarize the estimators investigated in our analysis and highlight their natural connections. 
\subsubsection{Ridge, marginal, and ridge-less}\label{sec1.1.1}
For simplicity, suppose $\X$ have been column-standardized to have mean zero and variance one, then the marginal estimator can be asymptotically given by 
\begin{flalign}
\widehat{\bmbeta}_{S}=\{\diagg(\X^T\X)\}^{-1}\X^T\y=\{\diagg(\widehat{\bmSigma}_X)\}^{-1}\cdot n^{-1}\X^T\y=n^{-1}\X^T\y,
\label{equ1.1.1}
\end{flalign} 
where $\widehat{\bmSigma}_X=n^{-1}\X^T\X$ is the sample covariance matrix.
The ridge-regularized estimator \citep{hoerl1970ridge,tikhonov1963solution} with regularization parameter $\lambda$ is 
\begin{flalign}
&\widehat{\bmbeta}_{R}(\lambda)
=\big(\X^T\X+\lambda n \I_p \big)^{-1}\X^T\y
%=\big(\widehat{\bmSigma}_X+\lambda \I_p \big)^{-1}\cdot n^{-1}\X^T\y
=\big(\widehat{\bmSigma}_X+\lambda \I_p \big)^{-1}\widehat{\bmbeta}_{S}, \quad \lambda\in (0,\infty).
\label{equ1.1.2}
\end{flalign} 
%where $\bmP_R(\X,\lambda)=\big(\X^T\X+\lambda n \I_p \big)^{-1}$ and $\widehat{\bmSigma}_X=n^{-1}\X^T\X$.
Here $\widehat{\bmSigma}_X+\lambda \I_p$ is a linear combination of $\widehat{\bmSigma}_X$ and diagonal matrix $\lambda \I_p$, and is called linear shrinkage estimator of $\bmSigma$ \citep{ledoit2004well}. In equation~(\ref{equ1.1.2}),
 $\widehat{\bmbeta}_{R}(\lambda)$ can be viewed as the marginal estimator $\widehat{\bmbeta}_{S}$ after ``accounting for $\bmSigma$'' through this linear shrinkage estimator. 
When $\lambda$ is large enough such that $\lambda \I_p$ can dominate $\widehat{\bmSigma}_X$, 
$\widehat{\bmSigma}_X+\lambda \I_p$ becomes asymptotically a diagonal matrix.
%, so is $n\bmP_R(\X,\lambda)$.
Thus, let $\widehat{\bmbeta}_{R}(\infty)=\lim_{\lambda\to \infty}
\widehat{\bmbeta}_{R}(\lambda)$, as $\lambda \to \infty$, we can have
$$\widehat{\bmbeta}_{R}(\infty) \propto \widehat{\bmbeta}_{S}.$$

On the other hand, as $\lambda \to 0^{+}$ (from the right), we have the ridge-less least squares estimator  \citep{hastie2019surprises}
\begin{flalign*}
\widehat{\bmbeta}_{R}(0^{+})
=\lim_{\lambda\to 0^{+}} \widehat{\bmbeta}_{R}(\lambda)
=\big(\X^T\X\big)^{+}\X^T\y
=\widehat{\bmSigma}_X^{ +}\widehat{\bmbeta}_{S},
\end{flalign*} 
where $\A^{+}$ is the Moore-Penrose pseudoinverse of matrix $\A$. When $n>p$ and suppose $\X$ has full column rank, $\widehat{\bmbeta}_{R}(0^{+})$ reduces to the classic OLS estimator $\widehat{\bmbeta}_{O}$ given by 
\begin{flalign*} 
\widehat{\bmbeta}_{O}=
\widehat{\bmbeta}_{R}(0)
=\big(\X^T\X\big)^{-1}\X^T\y
=\widehat{\bmSigma}_X^{-1}\widehat{\bmbeta}_{S}.
\end{flalign*} 

\subsubsection{BLUP and BLUP-less}\label{sec1.1.2}
In addition, ridge estimators have natural connection with the following best linear unbiased prediction (BLUP)
\begin{flalign*}
&\widehat{\bmbeta}_{B}(\tau)=\X^T\big(\X\X^T+\tau p \I_n \big)^{-1}\y, \quad \tau \in (0,\infty).
\end{flalign*} 
BLUP is originally from linear mixed effects model (LMM) \citep{henderson1975best,henderson1950best} and has been widely applied in genetics to tackle dense genetic effects (e.g., \cite{yang2010common}). 
Similar to $\widehat{\bmbeta}_{R}(0^{+})$, we can define the BLUP-less
estimator $\widehat{\bmbeta}_{B}(0^{+})$ by letting $\tau \to 0$
\begin{flalign*}
&\widehat{\bmbeta}_{B}(0^{+})=\lim_{\tau\to 0^{+}}
\widehat{\bmbeta}_{B}(\tau)=\X^T\big(\X\X^T \big)^{+}\y.
\end{flalign*} 
When $n<p$ and suppose $\X$ has full row rank, $\X^T\big(\X\X^T \big)^{+}$ reduces to $\X^T\big(\X\X^T \big)^{-1}$,
which has been used for variable selection \citep{wang2016high} and also has many applications in genetics, such as OmicKriging  \citep{wheeler2014poly}.
It can be shown that \citep{wang2016high}
\begin{flalign*}
\X^T\big(\X\X^T+\tau p \I_n \big)^{-1}\y=
\big(\X^T\X+\tau p \I_p \big)^{-1}\X^T\y.
\end{flalign*} 
It follows that  $\widehat{\bmbeta}_{B}(\tau)=\widehat{\bmbeta}_{R}(\tau\omega )$ and there is one-to-one correspondence between ridge estimator and BLUP.

Similar to $\widehat{\bmbeta}_{S}$, all of the ridge-type \textit{conditional} estimators  
$\widehat{\bmbeta}_{R}(\lambda)$, 
$\widehat{\bmbeta}_{B}(\tau)$, 
$\widehat{\bmbeta}_{R}(0^{+})$,
$\widehat{\bmbeta}_{B}(0^{+})$, and $\widehat{\bmbeta}_{O}$ can be shared as GWAS summary-level data for following-up in-sample and out-of-sample applications.
However, their computational complexity can be totally different in large training dataset with both $n$ and $p \to \infty$. 
Particularly, marginal estimator 
$\widehat{\bmbeta}_{S}$ is usually much less computationally expensive than conditional estimators.
Thus, understanding their connections and differences are important to determine the ``best" GWAS summary-level data to share while considering the computation-accuracy trade-off.  
In the rest of this paper, we analyze and compare these estimators in an unifying framework. We name them the class of ridge-type estimators.

%%%%%%%%%%%%%%%%%%%%%%%%%%%%%%%%%%%%%%%%%%%%%%%%%%%%%%%%%%%%%%%%%%%%
%%%%%%%%%%%%%%%%%%%%%%%%%%%%%%%%%%%%%%%%%%%%%%%%%%%%%%%%%%%%%%%%%%%%
\subsection{Overview of main results}\label{sec1.2}
Below we briefly summarize some phenomena we observed for high-dimensional dense signal prediction and then provide some practical guidelines on GWAS applications.
\begin{itemize}
	\item (Cross-validation free)
	The optimal regularizer $\lambda^*$ (or $\tau^*$) for out-of-sample $R^2$ is independent of $\bmSigma$ and is
	solely determined by $\omega$ and the global signal strength $\h^2$ (i.e., the genetic heritability).
	Given consistent estimator of $\h^2$ \citep{jiang2016high,ma2019mahalanobis}, optimal ridge-type estimator can be obtained without the need for cross-validation in ridge estimator or BLUP.  
	\item (Sparsity free)
	Given $\omega$ and $\h^2$, out-of-sample $R^2$ depends on the eigenvalue distribution of $\bmSigma$, but is independent of the unknown $m$. 
	For example, $\bmSigma$ affects the prediction accuracy of marginal estimator $\widehat{\bmbeta}_S$ through the first three moments of its eigenvalue distribution. 
	Thus, the prediction accuracy of feature datasets can be assessed and compared by their eigenvalue distributions, without the knowledge of the sparsity of signals. 
	\item (Is it important to select optimal $\lambda$?)
	The sensitivity of out-of-sample $R^2$ to $\lambda$ highly depends on $\omega$. 
    $\lambda$ plays an important role when $n$ and $p$ are similar (for example, $\omega \approx 1$), in which 
	$\widehat{\bmbeta}_{R}(\lambda)$ with small $\lambda$ may be over-fitted and have very poor out-of-sample performance. 
	Particularly, the out-of-sample $R^2$ of $\widehat{\bmbeta}_{R}(0^{+})$ can equal to zero while its in-sample $R^2$ can be one regardless of $\h^2$.
	In addition, model under-fitting with large $\lambda$ can be substantially suboptimal.
	However, as $\omega$ increases, $\widehat{\bmbeta}_R(\lambda)$ becomes near-optimal for any $\lambda \in (0,\infty)$. Therefore, selecting optimal $\lambda$ for out-of-sample prediction becomes less essential as feature dimensionality increases. 
	Instead, computational efficiency might be more important and thus marginal estimator becomes more favorable. 
	\item (Goodness-of-fit)
	Compared to out-of-sample $R^2$, in-sample $R^2$ is much more sensitive to $\lambda$. 
    For example, the in-sample $R^2$ of $\widehat{\bmbeta}_{R}(\lambda)$  with small $\lambda$ is usually larger than $\h^2$ and can easily become one when $\omega >1$ regardless of $\h^2$. 
	On the other hand, the in-sample $R^2$ of $\widehat{\bmbeta}_{S} \propto \widehat{\bmbeta}_{R}(\infty)$ may be lower than $\h^2$ and 
	can only equal to one as $\omega \to \infty$. 
	Thus, $\widehat{\bmbeta}_R(\lambda)$ with different $\lambda$ may differ a lot for their in-sample goodness-of-fit, though their out-of-sample $R^2$ can be very similar. 
	\item (Meta-analysis/Distributed computation) Different from ridge-type conditional estimators \citep{dobriban2019one}, marginal estimator $\widehat{\bmbeta}_{S}$ can have zero prediction accuracy loss in distributed computation given optimal weights. 
	This mainly because $\widehat{\bmbeta}_{S}$ has no need of calculating the inverse of sample covariance matrix. 
\end{itemize}
Now we deliver some useful messages for GWAS applications.    
\begin{itemize}
\item (Diverse populations) 
We provide analytic results on how the prediction accuracy of GWAS summary statistics is related to the linkage 
disequilibrium (LD) structure $\bmSigma_G$ of the SNP data.
For example, we show that the out-of-sample $R^2$ of $\widehat{\bmbeta}_{S}$ is determined by the first three moments of eigenvalues of $\bmSigma_G$. 
More generally the out-of-sample $R^2$ of $\widehat{\bmbeta}_{R}(\lambda^*)$ is related to the Stieltjes transform.
The LD structure is known to be consistent within each population, but substantially different among different populations. 
Following our results, we can calculate and compare the population-specific prediction accuracy using some simple functions of eigenvalue distribution.

\item (Marginal summary statistics) Since $\omega$ remains large in most of current GWAS, we provide statistical guarantees that marginal estimator is computation-accuracy efficient in the sense that it can have near-optimal out-of-sample $R^2$ and is much faster than ridge-type conditional estimators.  
Moreover, the zero efficiency loss of $\widehat{\bmbeta}_{S}$ in meta-analysis is desirable for summary-level data sharing and assembling. 
\item (Comparable $n$ and $p$)
However, one should be aware that as $n$ and $p$ become more comparable, or even $n>p$, $\widehat{\bmbeta}_{R}(\lambda^*)$ (or $\widehat{\bmbeta}_{B}(\tau^*)$) may have substantially better performance than $\widehat{\bmbeta}_{S}$.
As GWAS sample size increases dramatically in recent few years, efficient estimation of $\widehat{\bmbeta}_{R}(\lambda^*)$ (or $\widehat{\bmbeta}_{B}(\tau^*)$) can become more important for out-of-sample prediction in the near future.
When $n<p$, one good alternative of marginal estimator is the optimal BLUP estimator $\widehat{\bmbeta}_{B}(\tau^*)$ from LMM; and 
when $n>p$, the optimal ridge estimator $\widehat{\bmbeta}_{R}(\lambda^*)$ is a good choice. Again, $\tau^*$ and 
$\lambda^*$ have one-to-one correspondence and can both be directly estimated from the data without the need for cross-validation. 

\end{itemize}

%%%%%%%%%%%%%%%%%%%%%%%%%%%%%%%%%%%%%%%%%%%%%%%%%%%%%%%%%%%%%%%%%%%%
%%%%%%%%%%%%%%%%%%%%%%%%%%%%%%%%%%%%%%%%%%%%%%%%%%%%%%%%%%%%%%%%%%%%
\subsection{Related work and novelty}\label{sec1.3}

The present work is related to literature on the studies of high-dimensional linear model without sparsity assumption, most of which are on the asymptotic behavior of high-dimensional ridge estimator, including  \cite{dicker2013optimal,dicker2016ridge,dobriban2018high,karoui2013asymptotic,el2018impact,hastie2019surprises,hsu2011random} and \cite{pluta2017adaptive}.
For example, 
\cite{dicker2013optimal,dicker2016ridge} studies dense signal ridge problems with Gaussian assumption of data and allows general correlation structure $\bmSigma$ among predictors. 
\cite{dobriban2018high} study ridge estimator without Gaussian assumption and recently extend their results to distributed computing problem \citep{dobriban2019one}.
\cite{karoui2013asymptotic,el2018impact} studies ridge estimator in robust regression. 
Motivated by interpolation in machine learning, 
\cite{hastie2019surprises} analyze the ridge-less estimator by taking a limit on regularization parameter $\lambda$.
In addition, our results for $\omega \in (0,1)$ are related to studies of OLS estimator in moderate-dimensions \citep{guo2018moderate,yang2018quadratic}.

Our analysis is also related to previous studies on high-dimensional
LMM \citep{jiang2016high,steinsaltz2018statistical,dicker2017flexible,ma2019mahalanobis}, in which the authors mainly focus on the in-sample inference of LMM model parameters (such as $h^2$) and 
do not pay attention to BLUP and out-of-sample predictions. 
On the other hand, BLUP has been a popular method in genetics and agriculture  for a long time \citep{robinson1991blup}. 
Thus, there are studies of BLUP in genetics community, sometimes named genomic BLUP (gBLUP), such as \cite{goddard2009genomic,daetwyler2010impact,de2013prediction,speed2014multiblup}, and some Bayesian or ridge alternatives, such as \cite{zhou2013polygenic} and \cite{li2014improving}.

This study is motivated by the increasing applications of high-dimensional GWAS marginal estimator, especially in out-of-sample polygenic risk prediction.
A few studies (such as  \cite{daetwyler2008accuracy,dudbridge2013power,chatterjee2013projecting}, and \cite{Zhao447797}) have explored the prediction accuracy of GWAS summary statistics in the special case $\bmSigma=\I_p$.
To the best of our knowledge, 
there is no study on the behavior of marginal estimator in dense high-dimensional settings with general $\bmSigma$.
In the present work, we build our analysis on random matrix theory (RMT) and allow a arbitrary correlation structure $\bmSigma$ among SNPs. 
Moreover, we link marginal estimator to ridge estimator and BLUP, and study them in an unifying framework.
Driven by read data applications, we also generalize our results to cover cross-trait prediction and meta-analysis.

Different from most of previous studies on ridge-type estimators, we focus on $R^2$ instead of mean squared prediction error (MSE). 
$R^2$ is between zero and one and can be viewed as a normalized version of MSE. Different from MSE, $R^2$ is invariant to linear transformations of predictors, such as scaling and adding constants. This enables us to quantify and compare the performance of different estimators in a unified manner.
More importantly, using $R^2$ allows us to generalize our analysis to study cross-trait prediction where the traits in training and testing data can be different. 
From a practical perspective, $R^2$ and pseudo $R^2$ are  standard measures used in GWAS (and many other areas) to evaluate the out-of-sample prediction performance and in-sample goodness-of-fit. 
As shown in later sections, $R^2$ can yield very intuitive asymptotic results when comparing these estimators.

%%%%%%%%%%%%%%%%%%%%%%%%%%%%%%%%%%%%%%%%%%%%%%%%%%%%%%%%%%%%%%%%%%%%
%%%%%%%%%%%%%%%%%%%%%%%%%%%%%%%%%%%%%%%%%%%%%%%%%%%%%%%%%%%%%%%%%%%%
%%%%%%%%%%%%%%%%%%%%%%%%%%%%%%%%%%%%%%%%%%%%%%%%%%%%%%%%%%%%%%%%%%%%
\subsection{Outline and notation}\label{sec1.4}
This paper proceeds as follows.
In Section~\ref{sec2}, we introduce the detailed model setups and  assumptions for cross-trait prediction. 
In Section~\ref{sec3}, we provide the results for marginal estimator.  
Section~\ref{sec4} investigates the whole class of ridge-type estimators. 
Section~\ref{sec7} performs simulation studies and real data analysis to numerically verify our asymptotic results in finite samples. 
We discuss a few future topics in Section~\ref{sec8}. 
Most of the special case results, MSE analysis, and technical lemmas and details are provided in supplementary file.

We make use of the following notations frequently. $\tr(\A)$ is the trace of matrix $\A$, $\diagg(\A)$ is the diagonal of matrix $\A$, $\A^{-}$ is the inverse of matrix $\A$,  $\A^{T}$ is the transpose of matrix $\A$, and $\A^{+}$ is the Moore-Penrose pseudoinverse of matrix $\A$.
$\to$ donates the convergence of a series of real numbers, $\to_p$ represents the in probability convergence of a series of random variables, and $\to_{a.s.}$ is the  almost surely convergence of a series of random variables.
$\lambda_i(\A)$ is the $i$th eigenvalue of matrix $\A$, $\bmI(\cdot)$ is the indicator function, and
$\Vert\x\Vert^2=\x^T\x={\sum_{i=1}^p x^2_i}$ is the squared $l_2$ norm of $p\times 1$ vector $\x$, and $\Vert\x\Vert^2_{\bmSigma}=\x^T\bmSigma\x$ is the norm induced by $\bmSigma$. In addition, 
$o(1)$ and $O(1)$ define the small $o$ and big $O$, $o_p(1)$ and $O_p(1)$ define the small $o$ and big $O$ in probability, and $c,C$ are some generic constant numbers.  
%%%%%%%%%%%%%%%%%%%%%%%%%%%%%%%%%%%%%%%%%%%%%%%%%%%%%%%%%%%%%%%%%%%%
%%%%%%%%%%%%%%%%%%%%%%%%%%%%%%%%%%%%%%%%%%%%%%%%%%%%%%%%%%%%%%%%%%%%
%%%%%%%%%%%%%%%%%%%%%%%%%%%%%%%%%%%%%%%%%%%%%%%%%%%%%%%%%%%%%%%%%%%%
%%%%%%%%%%%%%%%%%%%%%%%%%%%%%%%%%%%%%%%%%%%%%%%%%%%%%%%%%%%%%%%%%%%%
%%%%%%%%%%%%%%%%%%%%%%%%%%%%%%%%%%%%%%%%%%%%%%%%%%%%%%%%%%%%%%%%%%%%
%%%%%%%%%%%%%%%%%%%%%%%%%%%%%%%%%%%%%%%%%%%%%%%%%%%%%%%%%%%%%%%%%%%%
\section{Modeling framework}\label{sec2}
In this section, we introduce the modeling framework, including the genetic architecture, assumptions on SNP data and genetic effects. 
\subsection{Cross-trait prediction}\label{sec2.1}
Consider two independent GWAS that are conducted for two traits with the same $p$ SNPs (features):
\begin{enumerate}[(a).]
\item Training GWAS: $(\X,\y)$, with $\X=[\X_{(1)},\X_{(2)}] \in \bbR^{n\times p}$, $\X_{(1)} \in \bbR^{n \times m_{\beta}}$, and $\y \in \bbR^{n \times 1}$. 
\item Testing GWAS: $(\Z,\y_{z})$, with $\Z=[\Z_{(1)},\Z_{(2)}] \in \bbR^{n_{z}\times p}$, $\Z_{(1)} \in \bbR^{n_{z} \times m_{\eta}}$, and $\y_{z} \in \bbR^{n_{z} \times 1}$. 
\end{enumerate}
Here  $\y$ and $\y_{z}$ are two continuous phenotypes measured in two independent groups of individuals with sample sizes $n$ and $n_z$, respectively. 
The $\X_{(1)}$ is an $n \times m_{\beta}$ matrix of the SNP data with nonzero effects, and
$\X_{(2)}$ is an $n \times (p-m_{\beta})$ matrix of the null SNPs, resulting in an $n\times p$ matrix of all SNPs, donated by $\X=[\X_{(1)},\X_{(2)}]=(\x_{1},\cdots,\x_{m_{\beta}},\x_{m_{\beta}+1}, \cdots,\x_{p})$, where $\x_{i}$ is an $n\times 1$ vector of the SNP $i$, $i=1,\cdots,p$.   
Similarly, $\Z_{(1)}$ denotes the causal SNPs of $\y_{z}$ and $\Z_{(2)}$ donate the null SNPs. 
We allow $\y$ and $\y_{z}$ to be two different traits. 
That is, we consider a general cross-trait prediction problem, such as predicting cognitive ability by educational attainment \citep{lee2018gene} or neuroimaging traits \citep{zhao2019genome}, and treat same-trait prediction as a special case. 
Thus, $m_{\beta}$ and $m_{\eta}$ can be  different numbers  and $\X_{(1)}$ and $\Z_{(1)}$ correspond to two different sets of causal SNPs in general.  
The linear polygenic models assume
\begin{flalign}
\y= \X_{(1)}\bmbeta_{(1)}+\bmeps  \quad \text{and}  \quad
\y_{z}= \Z_{(1)}\bmeta_{(1)}+\bmeps_{z},  
\label{equ1.2.1}
\end{flalign}
where $\bmbeta_{(1)}^T=(\beta_1,\cdots, \beta_{m_\beta})^{T}$ and $\bmeta_{(1)}^T=(\eta_1,\cdots, \eta_{m_\eta})^{T}$
are vectors of nonzero causal SNP effects, and $\bmeps$ and $\bmeps_{z}$ represent independent random error vectors. We let 
$\bmbeta^T=\big(\bmbeta_{(1)}^T,\bmbeta_{(2)}^T\big)$ and $\bmeta^T=\big(\bmeta_{(1)}^T,\bmeta_{(2)}^T\big)$, 
in which elements in  $\bmbeta_{(2)}^T=(\beta_{{m_\beta}+1},\cdots, \beta_p)^T$ and $\bmeta_{(2)}^T=(\eta_{{m_\eta}+1},\cdots, \eta_p)^T$ are all  zeros. 
We model $\bmbeta_{(1)}$ and $\bmeta_{(1)}$ as random variables \citep{dobriban2018high} and will introduce the detailed distribution assumptions in the following section. 
The overall genetic heritability of $\y$ and $\y_{z}$ are given by  
\begin{flalign}
\h^2_{\beta}=\frac{\var(\X\bmbeta)}{\var(\y)}=\frac{\bmbeta^T\X^T\X\bmbeta}{\bmbeta^T\X^T\X\bmbeta+\bmeps^T\bmeps} \quad \mbox{and}\quad
\h^2_{\eta}=\frac{\var(\Z\bmeta)}{\var(\y_{z})}=\frac{\bmeta^T\Z^T\Z\bmeta}{\bmeta^T\Z^T\Z\bmeta+\bmeps_z^T\bmeps_z},
\label{equ1.2.3}
\end{flalign}
which measure the proportion of variation in phenotype that can be explained by additive genetic factors. 
We assume $h^2_{\beta}$ and $h^2_{\eta} \in (0, 1]$. 

\subsection{Model assumptions and definitions}\label{sec2.2}
Since $m_{\beta}$ and $m_{\eta}$ can be different and the causal SNPs of different traits may partially overlap, we 
let $m_{\beta\eta} \le \mbox{min}(m_{\beta},m_{\eta})$ be the number of overlapping causal SNPs of $\y$ and $\y_{z}$. 
\paragraph{SNP data} The assumptions on SNP data $\X$ and $\Z$ are summarized in Condition~\ref{con1}.
\begin{condition}
\label{con1}
\begin{enumerate}[(a).]
\item SNP data satisfy $\X={\X_0}\bmSigma^{1/2}$, $\Z={\Z_0}\bmSigma^{1/2}$, and 
entries of $\X_0$ and $\Z_0$ are real-value i.i.d. random variables with mean zero, variance one and a finite 12th order moment.
$\bmSigma$ is a $p\times p$ population level deterministic positive definite matrix with $0<c\le \lambda_{min}(\bmSigma) \le \lambda_{max}(\bmSigma)\le C$ for all $n,p$ and some constants $c,C$, where 
$\lambda_{min}(\bmSigma)$ and $\lambda_{max}(\bmSigma)$ are the smallest and largest eigenvalues of $\bmSigma$, respectively. 
$\bmSigma^{1/2}$ is any nonnegative square root of $\bmSigma$. For simplicity, we assume $\bmSigma_{ii}=1$, $i=1,\cdots,p$, or equivalently, $\X$ and $\Z$ have been column-standardized. 
In summary, $\bmSigma$ is assumed to be a correlation matrix with uniformly bounded eigenvalues. 
\item 
As $\mbox{min}(n,n_z,m_{\beta\eta}) \to \infty$, we assume
\begin{gather*}
m_{\beta}/n \to \gamma, 
\quad
m_{\eta}/n_z \to \gamma_{z},
\quad
\text{for} \quad \gamma, \gamma_{z} \in(0, \infty);  \\
p/n \to \omega, 
\quad
p/n_z \to \omega_{z}, 
\quad
\text{for} \quad \omega, \omega_{z} \in (0,\infty); \quad \text{and}\\
m_{\beta\eta}/\sqrt{m_{\beta}m_{\eta}} \to \kappa_{\beta\eta},
\quad
\text{for} \quad \kappa_{\beta\eta} \in (0,1].
\end{gather*}
\item 
For the $p\times p$ population level correlation matrix $\bmSigma$, we define its empirical spectral distribution (ESD) as $F^{\bmSigma}_p(x)=p^{-1}\cdot\sum^{p}_{i=1}\bmI(\lambda_i(\bmSigma)\le x )$, $x \in \bbR$. As $p \to \infty$, let $\{\bmSigma_p\}_{p>1}$ be a sequence of matrices, 
we assume the sequence of corresponding ESDs $\{F^{\bmSigma}_p(x)\}_{p>1}$ converges weakly to a limit probability distribution $H(x)$, $x \in \bbR$, named the limiting spectral distribution (LSD) of $\bmSigma$. 
\end{enumerate}
\end{condition}
Conditions~\ref{con1}~(a).~and~\ref{con1}~(c). are frequently used in high-dimensional data analysis \citep{ledoit2011eigenvectors,dobriban2018high,hastie2019surprises}.
In modern GWAS, the number of the SNPs $p$ in Condition~\ref{con1}~(b). is usually in millions \citep{tam2019benefits}, and the current training GWAS simple size $n$ is often smaller or substantially smaller than $p$ \citep{watanabe2019global}.
The genetics community has striven to increase the GWAS sample size in recent
years, and for a few traits the sample size has been larger than one million  (e.g., \citet{lee2018gene}).
With these great efforts, the $\omega$ is decreasing from a large number towards one for many traits.

\paragraph{Genetic effects and random errors}
Let $F(0,V)$ represent a generic distribution with mean zero, (co)variance $V$, and finite fourth order moments.
%Without loss of generality, 
We introduce the following condition on genetic effects $\bmbeta_{(1)}$ and $\bmeta_{(1)}$ and random error vectors $\bmeps$ and $\bmeps_{z}$.
\begin{condition}\label{con2}
We assume the distributions of $\bmbeta$ and $\bmeta$ are independent of $\bmSigma$. Moreover, nonzero elements 
$\beta_i$ and $\eta_j$ are independent random variables satisfying
\begin{gather*}
%\begin{align}
\beta_i \sim F(0,\sigma^2_{\beta}/p), \quad i=1,...,m_{\beta}; \qquad
 \eta_j \sim F(0,\sigma^2_{\eta}/p), \quad 
j=1,...,m_{\eta} \mathbf{.}
%\end{align}
\end{gather*}
The $m_{\beta\eta}$ overlapping nonzero effects $(\beta_k,\eta_k)$s of  
$(\y$,$\y_{z})$ 
satisfy
\begin{flalign*}
\begin{pmatrix} 
\beta_k\\
\eta_k 
\end{pmatrix}
\sim F 
\left \lbrack
\begin{pmatrix} 
0\\
0 
\end{pmatrix},
p^{-1}\cdot
\begin{pmatrix} 
\sigma^2_{\beta} & \sigma_{\beta\eta} \\
\sigma_{\beta\eta} & \sigma^2_{\eta}
\end{pmatrix}
\right \rbrack,
\end{flalign*} 
where 
$\sigma_{\beta\eta}=\rho_{\beta\eta} \cdot \sigma_{\beta}\sigma_{\eta}$.
And $\epsilon_i$ and $\epsilon_{z_j}$ are independent random variables satisfying
\begin{gather*}
\epsilon_{i} \sim F(0,\sigma^2_{\epsilon}), \quad i=1,...,n ; \qquad
\epsilon_{z_j} \sim F(0,\sigma^2_{\epsilon_{z}}), \quad j=1,...,n_{z}.
\end{gather*}
\end{condition}

\paragraph{Genetic correlation and heritability}
Given the above assumptions, we define the genetic correlation between $\y$ and $\y_{z}$ as 
\begin{flalign*}
\varphi_{\beta\eta}=\frac{\bm\beta^T\bmSigma\bmeta}{\Vert\bm\beta\Vert_{\bmSigma}\cdot\Vert\bmeta\Vert_{\bmSigma}}\cdot \bmI(\Vert\bm\beta\Vert_{\bmSigma}\cdot\Vert\bmeta\Vert_{\bmSigma}>0),  
\end{flalign*}
and we assume $\varphi_{\beta\eta} \in [-1,1]$. Following Conditions~\ref{con1} and~\ref{con2}, as $\mbox{min}(n,n_z,m_{\beta\eta}) \to \infty$, the genetic correlation between $\y$ and $\y_{z}$ is asymptotically given by 
\begin{flalign*}
\varphi_{\beta\eta}=\frac{m_{\beta\eta} \sigma_{\beta\eta} \tr(\bmSigma)/p^2 }{\{m_{\beta} \sigma_{\beta}^2 \tr(\bmSigma)/p^2\}^{1/2}\{m_{\eta} \sigma_{\eta}^2 \tr(\bmSigma)/p^2\}^{1/2}}
=\kappa_{\beta\eta}\cdot\rho_{\beta\eta}+o_p(1).
%\label{equ1.2.4}
\end{flalign*}
Similarly, the heritability $h^2_{\beta}$ and $h^2_{\eta}$ defined in equation~(\ref{equ1.2.3}) can be asymptotically represented as
\begin{flalign*}
\h^2_{\beta}=\frac{\Vert\bmbeta\Vert_{\bmSigma}}{\Vert\bmbeta\Vert_{\bmSigma}+\sigma^2_{\epsilon}}=\frac{m_{\beta}\sigma^2_{\beta}\tr(\bmSigma)/p^2}{m_{\beta}\sigma^2_{\beta}\tr(\bmSigma)/p^2+\sigma^2_{\epsilon}}
\quad \text{and}  \quad
\h^2_{\eta}=\frac{\Vert\bmeta\Vert_{\bmSigma}}{\Vert\bmeta\Vert_{\bmSigma}+\sigma^2_{\epsilon_z}}=\frac{m_{\eta}\sigma^2_{\eta}\tr(\bmSigma)/p^2}{m_{\eta}\sigma^2_{\eta}\tr(\bmSigma)/p^2+\sigma^2_{\epsilon_{z}}}.
%\label{equ2.14}
\end{flalign*}
With $\bmSigma_{ii}=1$, $i=1,\cdots,p$, we have $\tr(\bmSigma)/p=1$, and thus we have the same definitions of $h^2_{\beta}$ and $h^2_{\eta}$ as those in \cite{jiang2016high} and \cite{guo2019optimal} for the special case $\bmSigma=\I_p$.

Our theoretical analysis of large-scale GWAS below 
is based on classic random matrix theory (e.g., \cite{bai2010spectral,paul2014random,yao2015sample}),  and some recent advances of trace functionals (e.g., \cite{ledoit2011eigenvectors,wang2015shrinkage,dobriban2018high,hastie2019surprises}). We introduce some useful RMT results and our lemmas in  Section~\ref{sec9.1} of supplementary file.

%%%%%%%%%%%%%%%%%%%%%%%%%%%%%%%%%%%%%%%%%%%%%%%%%%%%%%%%%%%%%%%%%%%%
%%%%%%%%%%%%%%%%%%%%%%%%%%%%%%%%%%%%%%%%%%%%%%%%%%%%%%%%%%%%%%%%%%%%
%%%%%%%%%%%%%%%%%%%%%%%%%%%%%%%%%%%%%%%%%%%%%%%%%%%%%%%%%%%%%%%%%%%%
%%%%%%%%%%%%%%%%%%%%%%%%%%%%%%%%%%%%%%%%%%%%%%%%%%%%%%%%%%%%%%%%%%%%
\section{Marginal estimator}\label{sec3}
Let $\widehat{\bmbeta}$ be a generic $p\times 1$ estimator of $\bmbeta$, the out-of-sample predictor and in-sample estimation are given by $\widehat{\bmS}_{\Z}=\Z\widehat{\bmbeta}$ and $\widehat{\bmS}_{\X}=\X\widehat{\bmbeta}$, respectively.
The out-of-sample and in-sample $R^2$ are respectively defined as $A^2$ and $E^2$, where 
\begin{flalign}
A=\frac{\y_{z}^T\widehat{\bmS}_{\Z}}{\big\Vert\y_{z}\big\Vert\cdot\big\Vert\widehat{\bmS}_{\Z}\big\Vert}
\quad \mbox{and} \quad
E=\frac{\y^T\widehat{\bmS}_{\X}}{\big\Vert\y\big\Vert\cdot\big\Vert\widehat{\bmS}_{\X}\big\Vert}.
\label{equ3.1}
\end{flalign}
In this section, we present the results of $A^2$ and $E^2$ for marginal estimator $\widehat{\bmbeta}_{S}$, donated as $A^2_S$ and $E^2_S$, respectively. 
%%%%%%%%%%%%%%%%%%%%%%%%%%%%%%%%%%%%%%%%%%%%%%%%%%
%%%%%%%%%%%%%%%%%%%%%%%%%%%%%%%%%%%%%%%%%%%%%%%%%%
\subsection{Asymptotic limits}\label{sec3.1}
The asymptotic limits of $A^2_S$ and $E^2_S$ are given in the following theorem. 
%%%%%%%%%%%%%%%%%%%%%%%%%%%%%%%%%%%%%%%%%%%%%%%%%%
\begin{thm}\label{thm1}
Under polygenic model~(\ref{equ1.2.1}) and Conditions~\ref{con1} and~\ref{con2}, as $\mbox{min}(n$, $n_z$, $m_{\beta\eta}$, $p)\rightarrow\infty$,
for any $\omega \in (0,\infty)$, $\h_{\beta}^2, \h_{\eta}^2\in (0,1]$, $\varphi_{\beta\eta} \in [-1,1]$, and $\bmSigma$, we have 
\begin{flalign*}
A^2_{S}&=
\h_{\eta}^2\varphi_{\beta\eta}^2\cdot\frac{n\{\tr(\widehat{\bmSigma}_X\widehat{\bmSigma}_Z)\}^2\cdot \h_{\beta}^2}{n\tr(\widehat{\bmSigma}_Z)\tr(\widehat{\bmSigma}_X\widehat{\bmSigma}_Z\widehat{\bmSigma}_X)\cdot\h_{\beta}^2+\tr(\widehat{\bmSigma}_Z)\tr(\widehat{\bmSigma}_X)\tr(\widehat{\bmSigma}_Z\widehat{\bmSigma}_X)\cdot(1-\h_{\beta}^2)}+o_p(1)\\
%&=\h_{\eta}^2 \varphi_{\beta\eta}^2\cdot
%\frac{n\{\tr(\bmSigma^2)\}^2\h_{\beta}^2}{n\tr(\bmSigma)\tr(\widehat{\bmSigma}_X^2\bmSigma)\h_{\beta}^2+\{\tr(\bmSigma)\}^2\tr(\bmSigma^2)(1-\h_{\beta}^2)} \\
&=\h_{\eta}^2 \varphi_{\beta\eta}^2\cdot \Big\{\frac{b_3(\bmSigma)}{b_2(\bmSigma)^2} +
\frac{\omega}{b_2(\bmSigma)} \cdot\frac{1}{\h_{\beta}^2}
\Big\}^{-1}+o_p(1),
%\label{equ3.1.0}
\end{flalign*}
and
\begin{flalign*}
E^2_{S}&=\frac{\{n\tr(\widehat{\bmSigma}_X^2)\cdot\h_{\beta}^2+\tr(\widehat{\bmSigma}_X)^2\cdot(1-\h_{\beta}^2)\}^2}
{n^2\tr(\widehat{\bmSigma}_X)\tr(\widehat{\bmSigma}_X^3)\cdot\h_{\beta}^2+n\tr(\widehat{\bmSigma}_X)^2\tr(\widehat{\bmSigma}_X^2)\cdot(1-\h_{\beta}^2)}+o_p(1)\\
&=
\frac{\{b_2(\bmSigma)\cdot\h_{\beta}^2+\omega\}^2}{\{b_2(\bmSigma)\cdot\h_{\beta}^2+\omega\}^2+b_2(\bmSigma)\omega+\{b_3(\bmSigma)-b_2(\bmSigma)^2\cdot\h_{\beta}^2\}\cdot\h_{\beta}^2}+o_p(1).
\end{flalign*}
For $\bmSigma=\I_p$, we have $b_k(\bmSigma)=1$ for any positive integral $k$, it follows that 
\begin{flalign}
A^2_{S}=\h_{\eta}^2\varphi_{\beta\eta}^2\cdot \frac{ \h_{\beta}^2}{\h_{\beta}^2+\omega}+o_p(1)
\quad \mbox{and} \quad
E^2_{S}=
\frac{(\h_{\beta}^2+\omega)^2}{(\h_{\beta}^2+\omega)^2+\omega+\h_{\beta}^2(1-\h_{\beta}^2)}+o_p(1).
\label{equ3.1.1}
\end{flalign}
\end{thm}
Here $\h_{\eta}^2\varphi_{\beta\eta}^2$ represents the signal strength, which is the ultimate upper bound for out-of-sample prediction.
Equation~(\ref{equ3.1.1}) indicates that $A^2_{S}$ is linearly decayed by the nonzero $\omega$ when $\bmSigma=\I_p$. It is also easy to see that 
$A^2_{S}$ and $E^2_{S}$ are the same if $\omega=0$ and $\varphi_{\beta\eta}=1$, but are quite different given nonzero $\omega$.
This represents the difference between low and high-dimensions. Next remark further illustrates that $\bmSigma$ has bidirectional effects on $A^2_{S}$ when $\omega>0$. 
%%%%%%%%%%%%%%%%%%%%%%%%%%%%%%%%%%%%%%%%%%%%%%%%%%
\begin{remark}\label{rmk1}
(Bidirectional influences of feature-wise correlation)
For general correlation $\bmSigma\neq \I_p$, $A^2_{S}$ depends on the first three moments of the LSD $H(t)$ of $\bmSigma$ through the two terms $b_2(\bmSigma)^2/b_3(\bmSigma)$ and $\omega/b_2(\bmSigma)$.
Note that for general $\bmSigma\neq \I_p$, $b_3(\bmSigma)> b_2(\bmSigma)^2$ and $b_2(\bmSigma)> 1$ by Cauchy–Schwarz inequality. 
In classic linear model theory where $p$ is fixed, or $\omega=0$, we have $A^2_{S}=\h_{\eta}^2\varphi_{\beta\eta}^2\cdot b_2(\bmSigma)^2/b_3(\bmSigma)$.
Thus, $A^2_{S}$ is reduced by a factor $b_2(\bmSigma)^2/b_3(\bmSigma)$ due to the unadjusted feature-wise correlation. 
On the other hand, when $\omega>0$, further decay of $A^2_{S}$ is introduced by the nonzero term $\omega/b_2(\bmSigma)$. Thanks to the fact that $b_2(\bmSigma)> 1$, correlation among features can delay such decay. 
This makes sense, because correlation among $p$ features can be viewed as a reduction of signal dispersion in high-dimensions.
Together, there is a transition point for whether or not feature-wise correlation can help achieve higher prediction accuracy in high-dimensions. 
Formally, we can define the \textit{prediction relative efficiency} (PRE) for $\bmSigma\neq \I_p$ to quantify the bidirectional effects of $\bmSigma$ on $A^2_{S}$ and identify the transition point. Let $\delta_S(\bmSigma)=A^2_{S}(\bmSigma)/A^2_{S}(\I_p)$, we have 
\begin{flalign*}
\delta_S(\bmSigma)=\frac{\h^2_{\beta}+\omega}
{\h^2_{\beta}\cdot\frac{b_3(\bmSigma)}{b_2(\bmSigma)^2}+\omega\cdot\frac{1}{b_2(\bmSigma)}}+o_p(1),
\end{flalign*}
%which is determined by the first three moments of $H(t)$,
and it follows that 
\begin{flalign*}
\delta_S(\bmSigma) 
\begin{array}{lll}
> \\ 
= \\
< \\
\end{array}
1+o_p(1)
\quad \mbox{if} \quad
\omega
\begin{array}{lll}
> \\ 
= \\
< \\
\end{array}
\h_{\beta}^2\cdot \frac{b_3(\bmSigma)-b_2(\bmSigma)^2}{b_2(\bmSigma)^2-b_2(\bmSigma)}.
\end{flalign*}
\end{remark}
%%%%%%%%%%%%%%%%%%%%%%%%%
\begin{remark}\label{rmk2}(In-sample v.s. out-of-sample)
For the optimal case where $\h_{\beta}^2=\h_{\eta}^2=\varphi_{\beta\eta}^2=1$, i.e., predicting fully heritable traits with absolute genetic correlation one, we have 
\begin{flalign*}
A^2_{S}&=\frac{\big\{\tr(\widehat{\bmSigma}_X\widehat{\bmSigma}_Z)\big\}^2}{\tr(\widehat{\bmSigma}_Z)\tr(\widehat{\bmSigma}_Z\widehat{\bmSigma}_X^2)}+o_p(1) \quad \mbox{and} \quad
E^2_{S}=\frac{\big\{\tr(\widehat{\bmSigma}_X^2)\big\}^2}
{\tr(\widehat{\bmSigma}_X)\tr(\widehat{\bmSigma}_X^3)}+o_p(1).
\end{flalign*}
This optimal case reveals more insights into the difference between in-sample and out-of-sample $R^2$, and the difference between low- and high-dimensions.
We note that $\tr(\widehat{\bmSigma}_X)=\tr(\widehat{\bmSigma}_Z)=\tr(\bmSigma)$, and $\tr(\widehat{\bmSigma}_X\widehat{\bmSigma}_Z)=\tr(\bmSigma^2)$. That is, trace of sample covariance is always the same as 
the trace of population covariance, and similar result holds for the product of two independent sample covariances. 
However, by Lemma~\ref{lemma3} in Section~\ref{sec9.1} of supplementary file, such concordance no longer holds for the trace of higher order products in high-dimensions with nonzero $\omega$. 
Specifically, we have $\tr(\widehat{\bmSigma}_X^2)=\tr(\bmSigma^2)+\omega\tr(\bmSigma)^2$, $\tr(\widehat{\bmSigma}_X^3)=\tr(\bmSigma^3)+3\omega\tr(\bmSigma)\tr(\bmSigma^2)+\omega^2\tr(\bmSigma)^3$, and $\tr(\widehat{\bmSigma}_Z\widehat{\bmSigma}_X^2)=\tr(\bmSigma\widehat{\bmSigma}_X^2)$ 
$=\tr(\bmSigma^3)+n^{-1}\tr(\bmSigma)\tr(\bmSigma^2)$. Therefore, 
we have 
 \begin{flalign*}
A^2_{S}&=\frac{\big\{\tr(\bmSigma^2)\big\}^2}{\tr(\bmSigma)
\cdot\big\{\tr(\bmSigma^3)+n^{-1}\tr(\bmSigma)\tr(\bmSigma^2)+o_p(1)\big\}}+o_p(1) 
\end{flalign*}
and
\begin{flalign*}
E^2_{S}=\frac{\big\{\tr(\bmSigma^2)+\omega\tr(\bmSigma)^2\big\}^2}
{\tr(\bmSigma)\cdot\big\{\tr(\bmSigma^3)+3\omega\tr(\bmSigma)\tr(\bmSigma^2)+\omega^2\tr(\bmSigma)^3  \big\}}+o_p(1).
\end{flalign*}
It is clear that in-sample and out-of-sample $R^2$ are completely different and both can be much less than one given nonzero $\omega$.  
\end{remark}
%%%%%%%%%%%%%%%%%%%%%%%%%
In summary, the asymptotic performance of marginal estimator is solely determined by heritability, genetic correlation, $\omega$,
and the first three moments of $H(t)$. 
These parameters are independent from the unknown number $m$.
Such properties enable us to easily evaluate the prediction accuracy of given SNP dataset. 
In addition, the PRE measures the influence of $\bmSigma$ on $A^2_S$, which can also be used to compare the prediction accuracy among different structures of feature-wise correlation. 
In next section, we illustrate how to apply the Theorem~\ref{thm1} to estimate $A^2_S$ in GWAS applications.
%%%%%%%%%%%%%%%%%%%%%%%%%%%%%%%%%%%%%%%%%%%%%%%%%%
%%%%%%%%%%%%%%%%%%%%%%%%%%%%%%%%%%%%%%%%%%%%%%%%%%
\subsection{Prediction accuracy estimation and comparison}\label{sec3.2}
In GWAS, different global populations (e.g., African, Latino, East Asian) have different SNP correlation structure  $\bmSigma$, and $\bmSigma$ is known to be largely consistent within each population \citep{gurdasani2019genomics}. Thus, given the same $\omega$ and $\h^2_{\eta},\h^2_{\beta},\varphi^2_{\beta\eta}$, the prediction accuracy of GWAS data varies across different populations.
To evaluate and compare the prediction accuracy of GWAS data in diverse populations, we need to study the LSD $H(x)$ of $\bmSigma$.
Here we discuss two approaches to evaluate the prediction accuracy $A^2_S$ for each global population.
%\subsubsection{Prediction accuracy }
\paragraph{Asymptotic estimator}(External reference panel)
The asymptotic estimator is based on the asymptotic limits. It is clear that we only need to estimate the first three moments
$b_1(\bmSigma)$, $b_2(\bmSigma)$, $b_3(\bmSigma)$ of $H(t)$, which have known relationships with $b_1(\widehat{\bmSigma})$, $b_2(\widehat{\bmSigma})$, $b_3(\widehat{\bmSigma})$ according to Lemma~\ref{lemma3}.
Therefore, we can estimate $b_k(\widehat{\bmSigma})$ from SNP data then obatin $b_k(\bmSigma)$, for $k=1,2,3$. 
In practice, this can be done using external data in publicly available LD reference panels \citep{tam2019benefits}, such as the 1000 Genomes Project \citep{10002015global}.
Let the reference data be $\W \in R^{n_w \times p}$, and  
let $\widehat{\bmSigma}_{W}=n^{-1}_{w} \W^T\W$, then 
$b_k(\widehat{\bmSigma}_{W})=p^{-1}\tr(\widehat{\bmSigma}_{W}^k)=p^{-1} \sum_{i=1}^p \lambda_i(\widehat{\bmSigma}_{W})^k$, $k=1,2,3$. 
Thus, all we need are the eigenvalues of $\widehat{\bmSigma}_{W}$, $\lambda_i(\widehat{\bmSigma}_{W})$, $i=1,\cdots,p$. 
When $n_w<p$, we may instead focus on the $n\times n$ companion matrix $\widehat{\bmPhi}_{W}=n^{-1}_{w} \W\W^T$ to obtain these moments.
%since the nonzero eigenvalues of $\widehat{\bmPhi}_{W}$ and $\widehat{\bmSigma}_{W}$ are the same. 
\paragraph{Empirical estimator}(Individual-level data)
When SNP data $\X$ and $\Z$ are available, one can also directly estimate the prediction accuracy by evaluating the four traces $\tr(\widehat{\bmSigma}_X)$, $\tr(\widehat{\bmSigma}_Z)$, $\tr(\widehat{\bmSigma}_X\widehat{\bmSigma}_Z)$, and $\tr(\widehat{\bmSigma}_X^2\widehat{\bmSigma}_Z)$.
Since $\tr(\widehat{\bmSigma}_X)=\tr(\widehat{\bmSigma}_Z)=p$, we only need to estimate $\tr(\widehat{\bmSigma}_X\widehat{\bmSigma}_Z)$ and $\tr(\widehat{\bmSigma}_X^2\widehat{\bmSigma}_Z)$. Estimating $\widehat{\bmSigma}_X$ and $\widehat{\bmSigma}_Z$ can be computationally expensive when both $n$ and $p$ are large. However, some tools have been developed to tackle this challenge \citep{quick2018emerald,das2016next}. 
Moreover, we may need to additionally account for the population stratification when population substructures exist
\citep{sun2017set}. One common solution is to remove the top few ``outlier'' eigenvalues, which often represent population substructures if any, since the population substructures are usually much stronger than the local SNP correlations.

To estimate the prediction accuracy for a specific pair of traits 
$\y$ and $\y_z$, we also need to estimate $\h_{\beta}^2$, $\h_{\eta}^2$ and $\varphi_{\beta\eta}$. Various estimators of these parameters have been proposed in GWAS context, we provide a brief review and discussion of these estimators in Section~\ref{sec9.2} of supplementary file. Next, we discuss two more potential usages of the prediction accuracy results. 

\paragraph{Diverse populations}
Based on these estimators, one can compare the prediction accuracy among diverse populations using their PREs. 
For example, suppose population~1 has $\bmSigma_1$ and population~2 has $\bmSigma_2$, then their relative prediction accuracy can be written by the ratio of their PREs 
\begin{flalign*}
\frac{\delta_S(\bmSigma_1)}{\delta_S(\bmSigma_2)}
=\frac{\h^2_{\beta}\cdot\frac{b_3(\bmSigma_2)}{b_2(\bmSigma_2)^2}+\omega\cdot\frac{1}{b_2(\bmSigma_2)}}
{\h^2_{\beta}\cdot\frac{b_3(\bmSigma_1)}{b_2(\bmSigma_1)^2}+\omega\cdot\frac{1}{b_2(\bmSigma_1)}}+o_p(1).
\end{flalign*}
It is clear that when $\omega$ is much larger than $\h^2_{\beta}$ and $b_3(\bmSigma)$ is comparable to $b_2(\bmSigma)^2$, $b_2(\bmSigma)$ plays an important role in the relative prediction accuracy. 

\paragraph{LD-based pruning}
In practice, it is quite common to first perform LD-based pruning with predefined threshold to remove highly related SNPs (e.g., remove one of a pair of SNPs that have correlation larger than the threshold) before out-of-sample prediction. The choice of the predefined threshold is often arbitrary. 
Using Theorem~\ref{thm1}, it is possible to input a series of thresholds, estimate the corresponding prediction accuracies, and then make a decision about the ``optimal'' threshold for SNP pruning.

%%%%%%%%%%%%%%%%%%
%%%%%%%%%%%%%%%%%%
\begin{figure}
\includegraphics[page=1,width=0.8\linewidth]{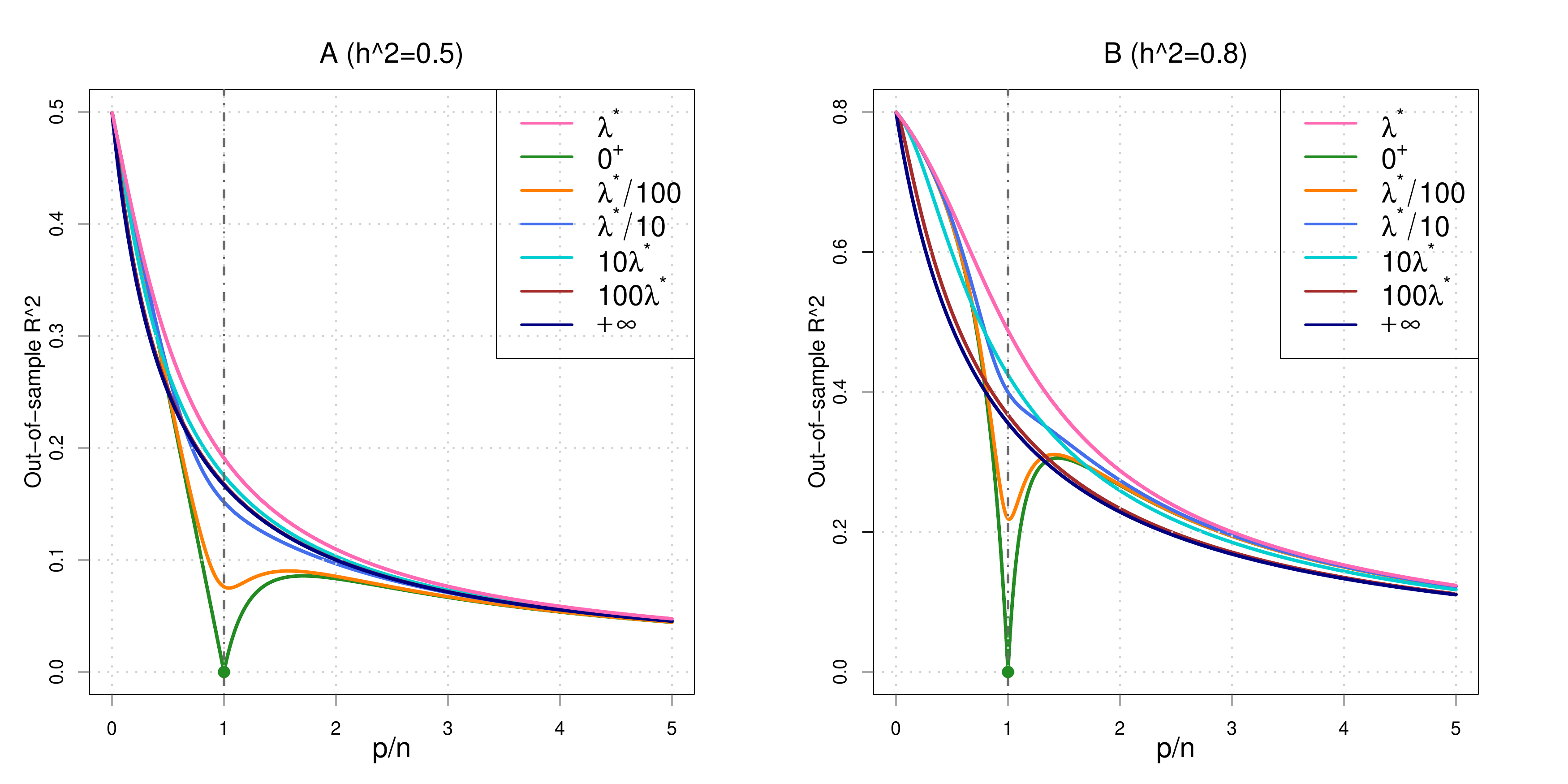}
  \caption{Out-of-sample $R$-squared $A^2_{R}(\lambda)$ of ridge-type estimators given different $\lambda$ and heritability  when $\bmSigma=\I_p$. $\lambda^*$ is the optimal $\lambda$ value, $0^{+}$ corresponds to the case $\lambda \to 0^+$, and $+\infty$ represents $\lambda \to +\infty$. 
  We set $\varphi_{\beta\eta}=1$, $\h_{\beta}^2=\h_{\eta}^2=(0.5,0.8)$ in A and B, respectively. 
}
\label{fig1}
\end{figure}
%%%%%%%%%%%%%%%%%%%
%%%%%%%%%%%%%%%%%%
%%%%%%%%%%%%%%%%%%%%%%%%%%%%%%%%%%%%%%%%%%%%%%%%%%%%%%%%%%%%%%%%%%%%
%%%%%%%%%%%%%%%%%%%%%%%%%%%%%%%%%%%%%%%%%%%%%%%%%%%%%%%%%%%%%%%%%%%%
%%%%%%%%%%%%%%%%%%%%%%%%%%%%%%%%%%%%%%%%%%%%%%%%%%%%%%%%%%%%%%%%%%%%
%%%%%%%%%%%%%%%%%%%%%%%%%%%%%%%%%%%%%%%%%%%%%%%%%%%%%%%%%%%%%%%%%%%%
\subsection{Meta-analysis of marginal estimator}
%\label{sec6.1}
Suppose there are $k$ independent GWAS $(\X_i,\y_i)$, $i=1,\cdots,k$, on the same trait with genetic effects $\bmbeta$, we have the following results when aggregating the summary statistics $\widehat{\bmbeta}_1,\cdots,\widehat{\bmbeta}_k$ from the $k$ studies.
%%%%%%%%%%%%%%%%%%%%%%%%%%%%%%%%%%%%%%%%%%%%%%%%%%%%%%%%%%%%%%%%%%%%
\begin{thm}\label{thm5}
Under polygenic model~(\ref{equ1.2.1}) and Conditions~\ref{con1} and~\ref{con2}, 
suppose we have independent GWAS $(\X_i,\y_i)$, with sample sizes $(n_i,\cdots,n_k)$ and $p$ SNPs, $i=1,\cdots,k$, $k \in (0,\infty)$, let  $\widehat{\B}=(\widehat{\bmbeta}_1^T,\cdots,\widehat{\bmbeta}_k^T)$ be the $p\times k$ matrix of marginal estimators from the $k$ GWAS. Let $\bmd=(d_i,\cdots,d_k)^T$ be an $k\times 1$ vector of weights, and let $\widehat{\B}(\bmd)=\widehat{\B}\bmd$
be the aggregated summary statistics.
As $\mbox{min}(n_1$,$\cdots,n_k$, $n_z$, $m_{\beta\eta}$, $p)\rightarrow\infty$,
for any $\omega \in (0,\infty)$, $\h_{\beta}^2, \h_{\eta}^2\in (0,1]$, $\varphi_{\beta\eta} \in [-1,1]$, $\bmd$, and $\bmSigma$, we have 
\begin{flalign*}
A^2_{S}(\bmd)= \h_{\eta}^2\varphi_{\beta\eta}^2\cdot \Big\{ 
\frac{b_3(\bmSigma)}{b_2(\bmSigma)^2}+
\frac{\sum_{i=1}^k d_i^2/n_i}{(\sum_{i=1}^k d_i)^2}\cdot \frac{p}{b_2(\bmSigma)\h_{\beta}^2}
\Big\}^{-1}+o_p(1).
\end{flalign*}
For $\bmd \equiv \bmd^*=(n_i,\cdots,n_k)^T$, we have 
\begin{flalign*}
A^2_{S}(\bmd^*)=\h_{\eta}^2\varphi_{\beta\eta}^2\cdot
\Big\{\frac{b_3(\bmSigma)}{b_2(\bmSigma)^2} +
\frac{\widetilde{\omega}_k}{\h_{\beta}^2} \cdot\frac{1}{b_2(\bmSigma)}
\Big\}^{-1}+o_p(1),
\end{flalign*}
where $\widetilde{\omega}_k=p/\sum_{i=1}^k n_i$.
$A^2_{S}(\bmd^*)$ is the same as the out-of-sample $R^2$ for one single GWAS with sample size $\sum_{i=1}^k n_i$. 
Particularly, when $\bmSigma=\I_p$, we have 
\begin{flalign*}
A^2_{S}(\bmd^*)=\h_{\eta}^2\varphi_{\beta\eta}^2\cdot \frac{\h_{\beta}^2}{\h_{\beta}^2
+\widetilde{\omega}_k }
+o_p(1).
\end{flalign*}
\end{thm}
%%%%%%%%%%%%%%%%%%%%%%%%%%%%%%%%%%%%%%%%%%%%%%%%%%%%%%%%%%%%%%%%%%%%
Theorem~\ref{thm5} shows that marginal screening has no prediction accuracy loss in distributed computation followed by meta-analysis with weights $\bmd^*$. 
Thus, aggregating summary statistics from independent training GWAS has the same asymptotic prediction accuracy as one big GWAS that trains all the individual-level data together.  
This is a favorable property of high-dimensional marginal estimator. It is known that both OLS  \citep{dobriban2018distributed} and ridge \citep{dobriban2019one} estimators may have prediction accuracy loss in high-dimensional distributed computation. 
Similar results also hold for in-sample $R^2$. For example, when $\bmd =\bmd^*$, we have 
\begin{flalign*}
&E^2_{S}(\bmd^*)=
\frac{\{b_2(\bmSigma)\cdot\h_{\beta}^2+\widetilde{\omega}_k\}^2}{\{b_2(\bmSigma)\cdot\h_{\beta}^2+\widetilde{\omega}_k\}^2+b_2(\bmSigma)\widetilde{\omega}_k+\{b_3(\bmSigma)-b_2(\bmSigma)^2\cdot\h_{\beta}^2\}\cdot\h_{\beta}^2}+o_p(1).
\end{flalign*}
And for $\bmSigma=\I_p$, we further have 
\begin{flalign*}
&E^2_{S}(\bmd^*)=\frac{(\h_{\beta}^2+\widetilde{\omega}_k)^2}{(\h_{\beta}^2+\widetilde{\omega}_k)^2+\widetilde{\omega}_k+\h_{\beta}^2(1-\h_{\beta}^2)}
\cdot\{1+o_p(1)\}.
\end{flalign*}

%%%%%%%%%%%%%%%%%%%%%%%%%%%%%%%%%%%%%%%%%%%%%%%%%%%%%%%%%%%%%%%%%%%%%
%%%%%%%%%%%%%%%%%%%%%%%%%%%%%%%%%%%%%%%%%%%%%%%%%%%%%%%%%%%%%%%%%%%%%
%%%%%%%%%%%%%%%%%%%%%%%%%%%%%%%%%%%%%%%%%%%%%%%%%%%%%%%%%%%%%%%%%%%%%
\section{The class of ridge-type estimators}\label{sec4}
In this section, we present the results for the following ridge-type conditional estimators:  
$\big\{\widehat{\bmbeta}_{R}(\lambda)$, 
$\widehat{\bmbeta}_{B}(\tau)$,
$\widehat{\bmbeta}_{R}(0^{+})$,
$\widehat{\bmbeta}_{B}(0^{+})$, 
$\widehat{\bmbeta}_{O}\big\}$. 
We define their out-of-sample $R^2$ as 
$\big\{A^2_R(\lambda)$, $A^2_B(\tau)$, $A^2_R(0^{+})$, $A^2_B(0^{+})$, $A^2_O\big\}$,
and in-sample $R^2$ as $\big\{E^2_R(\lambda)$, $E^2_B(\tau)$, $E^2_R(0^{+})$, $E^2_B(0^{+})$, $E^2_O\big\}$. 
%%%%%%%%%%%%%%%%%%%%%%%%%%%%%%%%%%%%%%%%%%%%%%%%%%%%%%%%%%%%%%%%%%%
%%%%%%%%%%%%%%%%%%%%%%%%%%%%%%%%%%%%%%%%%%%%%%%%%%%%%%%%%%%%%%%%%%%
%%%%%%%%%%%%%%%%%%%%%%%%%%%%%%%%%%%%%%%%%%%%%%%%%%%%%%%%%%%%%%%%%%%
\subsection{Out-of-sample \textit{R}-squared}\label{sec4.1}
We have the following results on $\big\{A^2_R(\lambda)$, $A^2_B(\tau)$, $A^2_R(0^{+})$, $A^2_B(0^{+})$, $A^2_O\big\}$.
\begin{thm}\label{thm2}
Under polygenic model~(\ref{equ1.2.1}) and Conditions~\ref{con1} and~\ref{con2}, as $\mbox{min}(n$, $n_z$, $m_{\beta\eta}$, $p)\rightarrow\infty$,
for any $\omega \in (0,\infty)$, $\h_{\beta}^2, \h_{\eta}^2\in (0,1]$, $\varphi_{\beta\eta} \in [-1,1]$,  and $\bmSigma$, we have 
\begin{flalign*}
&A^2_{R}(\lambda)=A^2_{B}(\lambda/\omega)\\
&=\h_{\eta}^2\varphi_{\beta\eta}^2\cdot
\frac{\big[1+\frac{\lambda}{\omega}\{ 1-\frac{1}{\lambda v(-\lambda)}\} \big]^2\cdot \h_{\beta}^2}
{\Big[1+\frac{\lambda}{\omega}\big\{2-\frac{1}{\lambda v(-\lambda)}- \frac{ v^{'}(-\lambda)}{v(-\lambda)^2}\big\}\Big]\cdot\h_{\beta}^2
+ \big\{\frac{v^{'}(-\lambda)}{v(-\lambda)^2}-1 \big\} \cdot(1-\h_{\beta}^2)}
+o_p(1),
\end{flalign*}
and
\begin{flalign*}
A^2_{R}(0^{+})=A^2_{B}(0^{+})=\h_{\eta}^2\varphi_{\beta\eta}^2\cdot
\frac{\big\{1-\frac{1}{v(0^{+})\omega} \big\}^2\cdot \h_{\beta}^2}
{ \big\{1-\frac{1}{v(0^{+})\omega} \big\}\cdot \h_{\beta}^2
+\big\{\frac{v^{'}(0^{+})}{v(0^{+})^2}-1 \big\}\cdot(1-\h_{\beta}^2) }
+o_p(1)
,
\end{flalign*}
where $v(0^{+})=\lim_{\lambda \to 0^{+}}v(-\lambda)$ and  $v^{'}(0^{+})=\lim_{\lambda \to 0^{+}}v^{'}(-\lambda)$. 
$A^2_{R}(0^{+})$ reduces to $A^2_{O}$ if $\omega <1$, which is given by
\begin{flalign*}
&A^2_{O}=\h_{\eta}^2\varphi_{\beta\eta}^2\cdot \Big\{1+\frac{1-\h_{\beta}^2}{\h_{\beta}^2}\cdot \frac{\omega}{1-\omega}\Big\}^{-1}
+o_p(1). 
\end{flalign*}
If $\h_{\beta}^2\in (0,1)$, $A^2_{R}(\lambda)$ is maximized at $\lambda=\lambda^{*}\equiv \omega\cdot(1-\h_{\beta}^2)/\h_{\beta}^2$, and the optimal out-of-sample $R^2$ is given by 
\begin{flalign*}
A^2_{R}(\lambda^{*})=A^2_{B}(\lambda^{*}/\omega)=
\h_{\eta}^2\varphi_{\beta\eta}^2\cdot\Big\{ \frac{1}{\h_{\beta}^2}-\frac{1}{v(-\lambda^{*})\omega}\Big\}
+o_p(1).
\end{flalign*}
If $\h_{\beta}^2=1$, i.e., $\y$ is a fully heritable trait, the optimal out-of-sample $R^2$ is obtained as $\lambda \to 0^{+}$, and we have 
\begin{flalign*}
A^2_{R}(0^{+})=A^2_{B}(0^{+})=  \left\{ 
\begin{array}{lll}
\h_{\eta}^2\varphi_{\beta\eta}^2+o_p(1), & \mbox{if \quad $\omega<1$;}\\ 
\h_{\eta}^2\varphi_{\beta\eta}^2\cdot\big\{ 1-\frac{1}{v(0^{+})\omega}\big\}+o_p(1), & \mbox{if \quad $\omega>1$.} \\
\end{array} \right .
\end{flalign*}
\end{thm}

Theorem~\ref{thm2} shows that $A^2_O$ is invariant to $\bmSigma$ and always has closed-form expression.
For other estimators, due to the linear shrinkage induced by nonzero $\lambda$, $\bmSigma$ still has influence on $A^2$ through the limits of Stieltjes transform $v(-\lambda)$ and its first order derivative $v^{'}(-\lambda)$. When $\lambda=\lambda^{*}$, $v^{'}(-\lambda)$ cancels out and thus the optimal out-of-sample $R^2$ depends on $\bmSigma$ only through $v(-\lambda)$. 
Let $\mbox{STN}(\h_{\beta}^2)=\h_{\beta}^2/(1-\h_{\beta}^2)$ be the signal to noise ratio, $\lambda^{*}$ can be rewritten as 
$$\lambda^{*}=\omega/\mbox{STN}(\h_{\beta}^2).$$
Thus, in linear shrinkage estimator $\widehat{\bmSigma}_X+\lambda \I_p$, the optimal weight for $\I_p$ is proportional to $\omega$ and inversely proportional to the signal to noise ratio, matching previous results on MSE \citep{dobriban2018high}.

When $\bmSigma=\I_p$, the closed-form expressions for $v(-\lambda)$ and $v^{'}(-\lambda)$ are available, and thus we have closed-form expressions for $\big\{A^2_R(\lambda)$, $A^2_B(\tau)$, $A^2_R(0^{+})$, $A^2_B(0^{+})$, $A^2_O\big\}$, which are given in the following corollary.

\begin{cor}\label{cor1}
Under the same conditions as in Theorem~\ref{thm2}, when $\bmSigma=\I_p$, 
we have 
\begin{flalign*}
A^2_{R}(\lambda)&=A^2_{B}(\lambda/\omega)=
\h_{\eta}^2\varphi_{\beta\eta}^2\cdot\\
&
\frac{\big\{1-\lambda g(-\lambda) \big\}^2\cdot \h_{\beta}^2}
{\big\{1-2\lambda g(-\lambda)+ 
\lambda^2 g^{'}(-\lambda) \big\}\cdot\h_{\beta}^2
+\big\{\omega g(-\lambda)-\omega\lambda g^{'}(-\lambda) \big\}
\cdot(1-\h_{\beta}^2)}
+o_p(1)
\mathbf{,}
\end{flalign*}
where closed-form expressions of $g(\cdot)$ and $ g^{'}(\cdot)$ can be found in equations (\ref{equ2.3.1})~and~(\ref{equ2.3.2}).
In addition,
\begin{flalign*}
A^2_{R}(0^{+})=A^2_{B}(0^{+})&=\h_{\eta}^2\varphi_{\beta\eta}^2\cdot
\frac{ \big\{ \frac{1+\omega-|\omega-1|}{2\omega} \big\}^2 \cdot \h_{\beta}^2}{ \frac{1+\omega-|\omega-1|}{2\omega} \cdot\h_{\beta}^2 + \frac{\omega+1-|\omega-1|}{2|\omega-1|}\cdot (1-\h_{\beta}^2) } +o_p(1)\\
& =\left\{ 
\begin{array}{lll}
\h_{\eta}^2\varphi_{\beta\eta}^2\cdot\frac{\h_{\beta}^2}{\h_{\beta}^2+\frac{\omega}{1-\omega}\cdot(1-\h_{\beta}^2)}+o_p(1), & \mbox{if \quad $\omega<1$;}\\ 
\h_{\eta}^2\varphi_{\beta\eta}^2\cdot\frac{\h_{\beta}^2}{\h_{\beta}^2\cdot\omega+\frac{\omega^2}{\omega-1}\cdot(1-\h_{\beta}^2)}+o_p(1), & \mbox{if \quad $\omega>1$.} \\
\end{array} \right .
\end{flalign*}
If $\h_{\beta}^2\in (0,1)$, $A^2_{R}(\lambda)$ is maximized at $\lambda^{*}$, and the optimal out-of-sample $R^2$  is given by 
\begin{flalign*}
A^2_{R}(\lambda^{*})=A^2_{B}(\lambda^{*}/\omega)&=h_{\eta}^2\varphi_{\beta\eta}^2\cdot\big\{1-\lambda^{*}g(-\lambda^{*})\big\}+o_p(1)\\
&=\h_{\eta}^2\varphi_{\beta\eta}^2\cdot
\frac{ \omega+h_{\beta}^2-\sqrt{(\omega-h_{\beta}^2)^2+4\omega h_{\beta}^2(1-h_{\beta}^2)}}{2\omega \h_{\beta}^2}
+o_p(1).
\end{flalign*}
If $\h_{\beta}^2=1$, the optimal $A^2_{R}(\lambda)$ is obtained as $\lambda \to 0^{+}$, and we have 
\begin{flalign*}
A^2_{R}(0^{+})=A^2_{B}(0^{+})&= \h_{\eta}^2\varphi_{\beta\eta}^2\cdot\frac{\omega+1-|\omega-1|}{2\omega}+o_p(1)= 
\left\{ 
\begin{array}{lll}
\h_{\eta}^2\varphi_{\beta\eta}^2+o_p(1), & \mbox{if \quad $\omega<1$;}\\ 
\h_{\eta}^2\varphi_{\beta\eta}^2\cdot \frac{1}{\omega} +o_p(1), & \mbox{if \quad $\omega>1$.} \\
\end{array} \right .
\end{flalign*}
\end{cor}

%%%%%%%%%%%%%%%%%%
\begin{figure}
\includegraphics[page=1,width=0.8\linewidth]{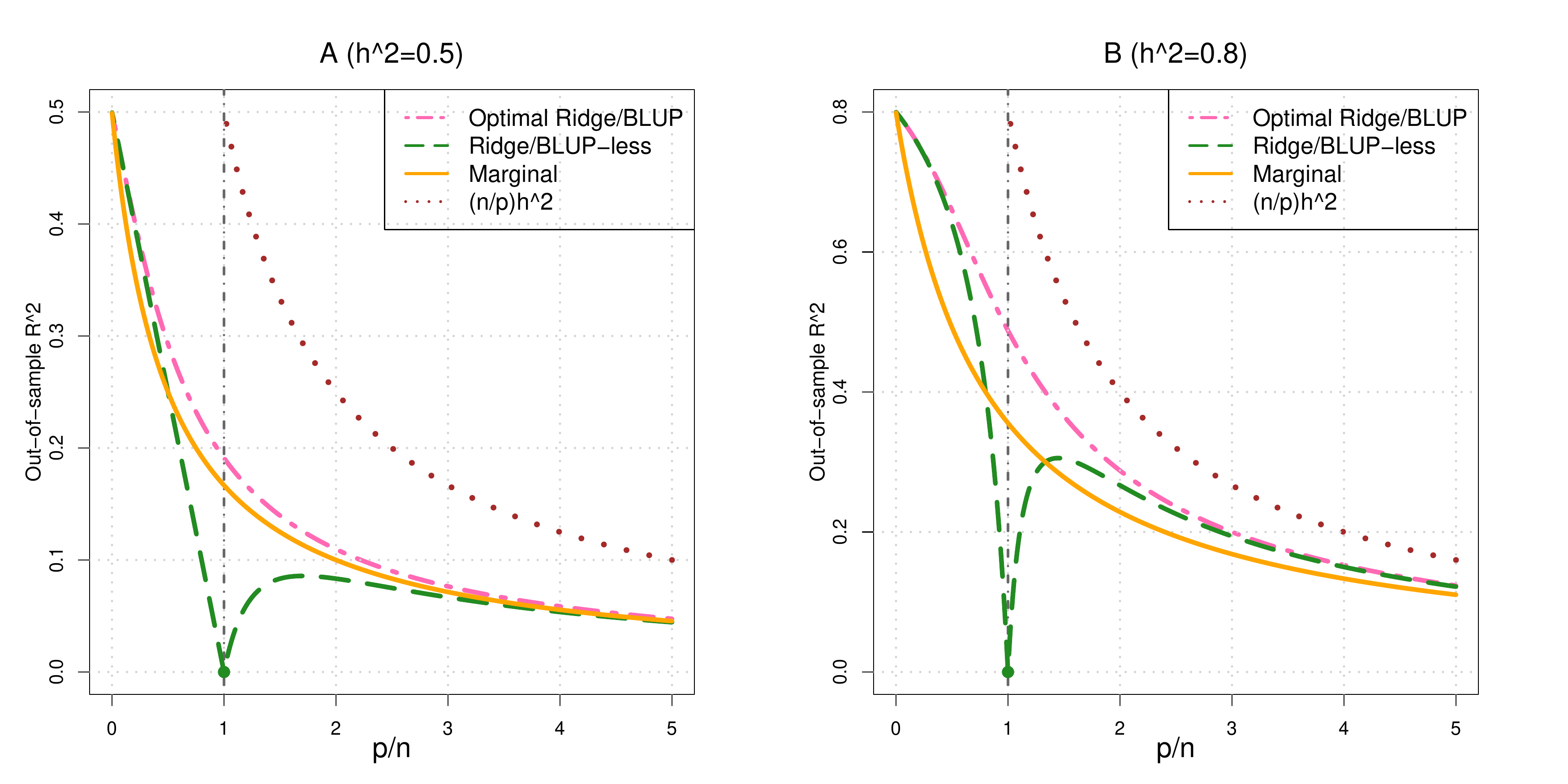}
  \caption{Out-of-sample  $R$-squared of optimal ridge/BLUP estimators ($A^2_{R}(\lambda^*)=A^2_{B}(\lambda^*/\omega)$), ridge/BLUP-less estimators ($A^2_{R}(0^{+})=A^2_{B}(0^{+})$), and marginal estimator ($A^2_{S}$) when $\bmSigma=\I_p$.  We set $\varphi_{\beta\eta}=1$, $\h_{\beta}^2=\h_{\eta}^2=(0.5,0.8)$ in A and B, respectively. The dash line represents the upper limit $(n/p)\cdot \h_{\beta}^2$.
}
\label{fig2}
\end{figure}
%%%%%%%%%%%%%%%%%%%%%%%%%%%%%%%%%%%%%%%%%%%%%%%%%%%%%%%%%%%%%%%%%%%%%%%%
%%%%%%%%%%%%%%%%%%%%%%%%%%%%%%%%%%%%%%%%%%%%%%%%%%%%%%%%%%%%%%%%%%%%%%%%
In Corollaries~S\ref{cor2}~and~S\ref{cor3} of supplementary file, we quantify the relative prediction accuracy between marginal estimator and the optimal ridge estimator in closed-form expressions for the special case $\bmSigma=\I_p$.
The following remarks provide more insights into high-dimensional dense signal prediction. 
\begin{remark}\label{rmk4} (Optimal regularizer and prediction accuracy)
The optimal values $\lambda^{*}$ and $\tau^{*}$ for out-of-sample $R^2$ (when $\h_{\beta}^2\in (0,1)$) are functions of  $\h_{\beta}^2$ and $\omega$, and are independnet of all other parameters including $\bmSigma$.
Since $\omega$ is known and consistent estimator of $\h_{\beta}^2$ is available, in practice cross-validation techniques are not required to obtain optimal regularizers for ridge estimator or BLUP. 
Moreover, $A^2_{R}(\lambda^{*})$
can be estimated by additionally calculating $s_{F_n}(\lambda^*)$, which is a consistent estimator of $v(-\lambda^*,\bmSigma)$. 
To quantify the influence of $\bmSigma$ on $A^2_{R}(\lambda^*)$, we can define the PRE for $\bmSigma\neq \I_p$ as  $\delta_R(\lambda^*,\bmSigma)=A^2_{R}(\lambda^*,\bmSigma)/A^2_{R}(\lambda^*,\I_p)$, and then
\begin{flalign*}
\delta_R(\lambda^*,\bmSigma)=\frac{\omega+h_{\beta}^2-\sqrt{(\omega-h_{\beta}^2)^2+4\omega h_{\beta}^2(1-h_{\beta}^2)}}
{2\omega-2\h_{\beta}^2/v(-\lambda^{*},\bmSigma)}+o_p(1).
\end{flalign*}
\end{remark}
%%%%%%%%%%%%%%%%%%
\begin{remark}\label{rmk5} (Over-and-under fitting)
Figure~\ref{fig1} and Supplementary Figure~\ref{sfig1} display $A^2_{R}(\lambda)$ across different $\lambda$, $\omega$, and $\h^2$. 
It is clear that $A^2_{R}(\lambda)$ is near-optimal for any $\lambda$ when $\omega$ is big (e.g., $\omega>5$), especially when $\h^2$ is not high.
In contrast, when $\omega \approx 1$, 
model over-fitting with small $\lambda$ should be avoided and model under-fitting with large $\lambda$ can be substantially suboptimal.
Notably, $A^2_{R}(0^+)$ can become surprisingly small.
For $\omega>1$, $A^2_{R}(0^+)$ is not a monotone function of $\omega$, and the optimal value is achieved at 
$\omega=1+\sqrt{1-\h_{\beta}^2}$. 
When $\omega$ decreases from $1+\sqrt{1-\h_{\beta}^2}$ towards $1$, $A^2_{R}(0^+)$ reduces dramatically. 
\end{remark}
%%%%%%%%%%%%%%%%%%%%%%%%%%%%%%%%%%%%
\begin{remark}\label{rmk6} (Blessing of dimensionality)
When $\omega$ is large, $A^2_{R}(\lambda)$ has almost identical performance for all $\lambda$.
Particularly, the out-of-sample $R^2$ of $\widehat{\bmbeta}_{S}$ is similar to that of $\widehat{\bmbeta}_{R}(\lambda^*)$, then $\widehat{\bmbeta}_{S}$ can be a good choice for out-of-sample applications because it is much more computationally efficient than $\widehat{\bmbeta}_{R}(\lambda^*)$. 
High dimensionality indeed reduces the required computational burden to obtain good prediction performance, which is quite counterintuitive. 
\end{remark}
%%%%%%%%%%%%%%%%%%%%%%%%%%%%%%%%%%%%
\begin{remark}\label{rmk7} (Curse of dimensionality)
On the other hand, the upper limit of the prediction accuracy of all ridge-type estimators might be not satisfactory when $\omega$ is large. 
For example, when $\bmSigma=\I_p$, consider the optimal case where $\h_{\eta}^2=\h_{\beta}^2=\varphi_{\beta\eta}^2=1$, the asymptotic optimal out-of-sample $R^2$ of ridge-type estimators is one for $n>p$.
However, when $n<p$, the out-of-sample $R^2$ has a upper bound of $n/p=1/\omega$, which can be viewed as the ratio of sample size and model complexity.
These results reveal the fundamental challenge in  high-dimensional dense signal prediction. 
In addition, $1/\omega$ can hardly be achieved in practical situations if any. 
To see this, consider $\h_{\eta}^2=\h_{\beta}^2 \in (0,1)$ and $\varphi_{\beta\eta}^2=1$, 
we can rewrite the optimal out-of-sample $R^2$ as
\begin{flalign*}
A^2_{R}(\lambda^{*})=A^2_{B}(\lambda^{*}/\omega)=
\frac{p+n\h_{\beta}^2-|p-n\h_{\beta}^2|\cdot\sqrt{1+\Delta}}{2p}
+o_p(1),
\end{flalign*}
where $\Delta=4\h_{\beta}^2(1-\h_{\beta}^2)/(\omega^{1/2}-\omega^{-1/2}\h_{\beta}^2)^2>0$. Note that $\Delta\approx 0$ only for large $\omega$ and $\h_{\beta}^2 \approx 1$. 
Then $A^2_{R}(\lambda^{*})$ is close to the upper bound $n/p\cdot \h_{\beta}^2$ only when $\omega$ is large for highly heritable traits prediction (Figure~\ref{fig2} and Supplementary Figure~\ref{sfig2}). 
For general $\bmSigma$, 
similar to the case of marginal estimator, 
feature-wise correlation can delay the negative influences of growing dimensionality, but the general pattern remains the same.
\end{remark}
%%%%%%%%%%%%%%%%%%%%%%%%%%%%%%%%%%%%
\begin{remark}\label{rmk8} [Unboundedness of $\tr\{(\X^T\X)^{-1}\}$]
$\widehat{\bmbeta}_{R}(\lambda^*)$ has better out-of-sample $R^2$ than 
$\widehat{\bmbeta}_{O}$ for $\omega \in (0,1)$. 
As shown in Figure~\ref{fig2}, when $\omega$ is close to zero, $\widehat{\bmbeta}_{R}(\lambda^*)$ and $\widehat{\bmbeta}_{O}$ are close to each other, which matches the classic results in linear models. However, as $\omega$ moves towards one, the performance of $\widehat{\bmbeta}_{O}$ is much worse than $\widehat{\bmbeta}_{R}(\lambda^*)$. 
One way to explain this surprising behavior of $\widehat{\bmbeta}_{O}$ is that $\tr\{(\X^T\X)^{-1}\}$ can become very large when $\omega \to 1^{-}$.
To see this, when $\omega<1$, note that
\begin{flalign*}
A^2_{O}=A^2_{R}(0)=\h_{\eta}^2\varphi_{\beta\eta}^2\cdot \frac{\h_{\beta}^2}{\h_{\beta}^2+ \tr\{(\X^T\X)^{-1}\}\cdot(1-\h_{\beta}^2)}+o_p(1).
\end{flalign*}
In Gaussian case, $(\X^T\X)^{-1}$ follows the 
inverse Wishart distribution and the mean of 
$\tr\{(\X^T\X)^{-1}\}$ is $\omega/(1-\omega-1/n)$, 
which can be large as $\omega \to 1^{-}$ \citep{guo2018moderate}.
Without the need for Gaussianity, \cite{hastie2019surprises} show that  $\tr\{(\X^T\X)^{-1}\}=\lim_{\lambda\to 0^{+}} \omega g(-\lambda)=\omega/(1-\omega)$. 
Then, a tiny small nonzero error term $1-\h_{\beta}^2$ can ruin the out-of-sample performance of OLS estimator (and also ridge-less/BLUP-less estimators) when $\omega$ is close to one.
Ridge estimator
$\widehat{\bmbeta}_{R}(\lambda)$ avoids the unboundedness of $\tr\{(\X^T\X)^{-1}\}$ by introducing a nonzero shrinkage term $\lambda$.  
In marginal estimator $\widehat{\bmbeta}_{S}$, the estimator of $(\X^T\X)^{-1}$ is simply  $\{\mbox{Diag}(\X^T\X)\}^{-1}$, which can be viewed as an extreme case of banded covariance estimator \citep{bickel2008regularized} with zero bandwidth. Thus, $\widehat{\bmbeta}_{S}$ can avoid the issue of $\tr\{(\X^T\X)^{-1}\}$. However, the price is that $\widehat{\bmbeta}_{S}$ may have larger squared bias. See Section~\ref{sec9.4} of supplementary file for more details, in which we illustrate the bias-variance decomposition using mean squared prediction errors of these estimators. 
\end{remark}
 
%%%%%%%%%%%%%%%%%%%%%%%%%%%%%%%%%%%%%%%%%%%%%%%%%%%%%%%%%%%%%%%%%%%%%%%%
%%%%%%%%%%%%%%%%%%%%%%%%%%%%%%%%%%%%%%%%%%%%%%%%%%%%%%%%%%%%%%%%%%%%%%%%
\subsection{In-sample \textit{R}-squared}\label{sec4.3}
In this section, we present the results for in-sample $R^2$, which measures the goodness-of-fit and is related to the performance of many in-sample applications of GWAS summary statistics (e.g., \cite{barbeira2018exploring}). In-sample $R^2$ has completely different pattern compared to out-of-sample $R^2$. 
The asymptotic results are summarized in the following theorem.
%%%%%%%%%%%%%%%%%%
%%%%%%%%%%%%%%%%%%
\begin{figure}
\includegraphics[page=1,width=0.8\linewidth]{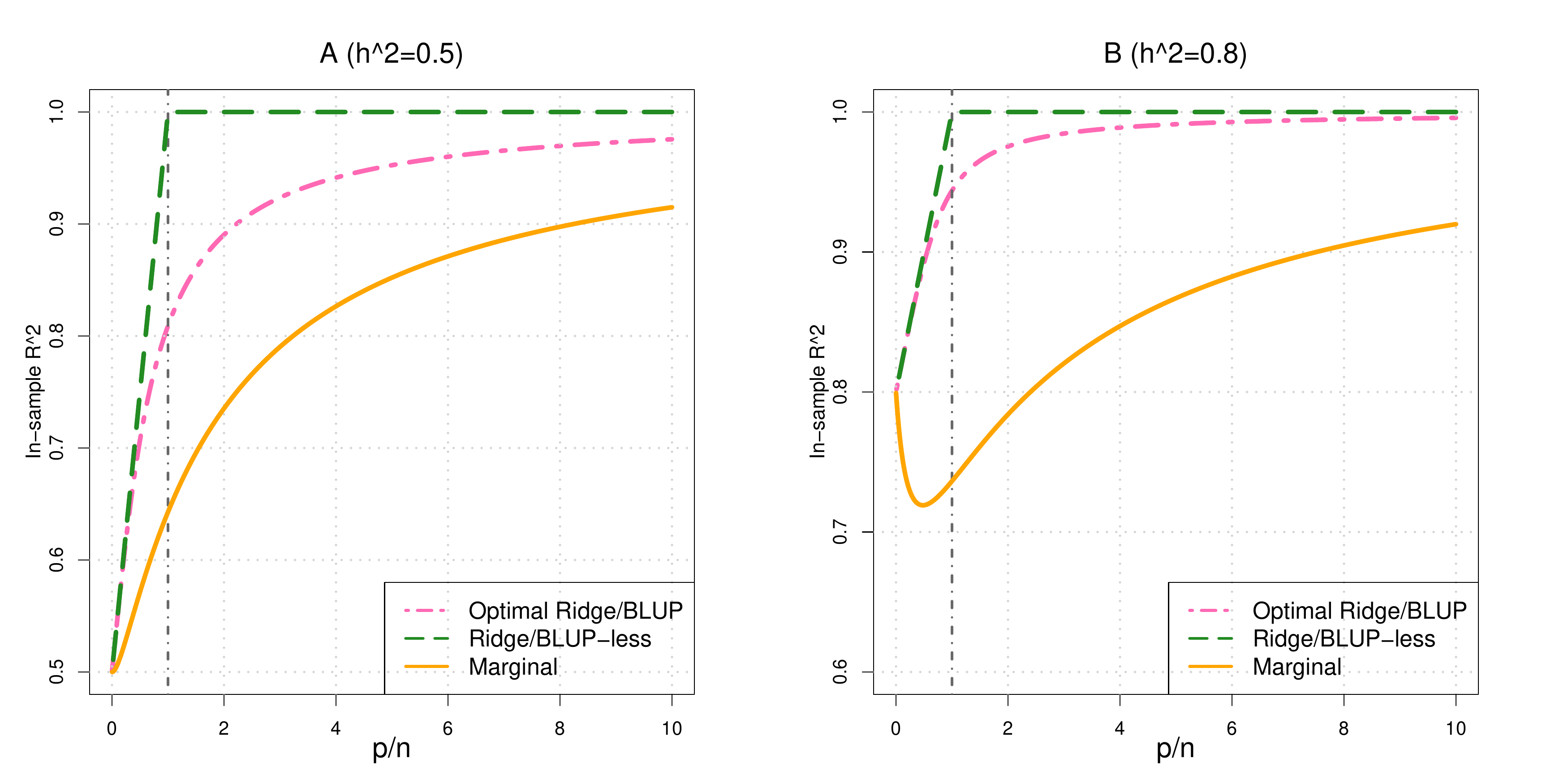}
  \caption{In-sample $R$-squared of ridge/BLUP-less estimators ($E^2_{R}(0^{+})=E^2_{B}(0^{+})$), optimal (out-of-sample) ridge/BLUP estimators ($E^2_{R}(\lambda^*)=E^2_{B}(\lambda^*/\omega)$), and marginal estimator ($E^2_{S}$) when $\bmSigma=\I_p$. We set $\h_{\beta}^2=(0.5,0.8)$ in A and B, respectively.}
\label{fig3}
\end{figure}
%%%%%%%%%%%%%%%%%%
%%%%%%%%%%%%%%%%%%
\begin{thm}\label{thm3}
Under polygenic model~(\ref{equ1.2.1}) and Conditions~\ref{con1} and~\ref{con2}, as $\mbox{min}(n$, $m_{\beta}$, $p)\rightarrow\infty$,
for any $\omega \in (0,\infty)$, $\h_{\beta}^2 \in (0,1]$ and $\bmSigma$, we have 
\begin{flalign*}
&E^2_{R}(\lambda)=E^2_{B}(\lambda/\omega)\\
&=
\frac{\Big[\h_{\beta}^2 \cdot \big\{1-\lambda+\lambda^2g(-\lambda)\big\} +(1-\h_{\beta}^2 )\cdot \omega  \big\{ 1-\lambda g(-\lambda)\big\} \Big]^2}
{ \h_{\beta}^2 \cdot\big\{1-2\lambda+3\lambda^2g(-\lambda)-\lambda^3 g^{'}(-\lambda) \big\}
+(1-\h_{\beta}^2 )\cdot \omega \big\{1-2\lambda+\lambda^2g^{'}(-\lambda)\big\}}
+o_p(1).
\end{flalign*}
$E^2_{R}(\lambda)$ is maximized as $\lambda\to 0^{+}$, and we have
\begin{flalign*}
E^2_{R}(0^{+})=E^2_{B}(0^{+})&=\big\{\h_{\beta}^2  +(1-\h_{\beta}^2 )\cdot \frac{\omega+1-|\omega-1|}{2} \big\}
+o_p(1)
=  \left\{ 
\begin{array}{lll}
E^2_{O}, & \mbox{if \quad $\omega<1$;}\\ 
1+o_p(1), & \mbox{if \quad $\omega>1$,} \\
\end{array} \right .
\end{flalign*}
where $E^2_{O}=\big\{\h_{\beta}^2 +(1-\h_{\beta}^2 )\cdot \omega\big\}+o_p(1)$.
In addition, we have 
\begin{flalign*}
E^2_{R}(\lambda^{*})=
\frac{\h_{\beta}^2 }{1-\lambda^*+\lambda^{*2}g(-\lambda^{*})}
+o_p(1).
\end{flalign*}
When $\bmSigma=\I_p$, the closed-form expression of $E^2_{R}(\lambda^{*})$ is given by
\begin{flalign*}
&E^2_{R}(\lambda^{*})=\frac{2\h_{\beta}^6}{(1-\h_{\beta}^2)\cdot \big\{\sqrt{(\omega-\h_{\beta}^2)^2+4\omega\h_{\beta}^2(1-\h_{\beta}^2)}-\omega\big\}+\h_{\beta}^2\cdot (3\h_{\beta}^2-1)}+o_p(1).
\end{flalign*}
\end{thm}
The optimal in-sample $R$-squared
$E^2_{R}(0^{+}) \ge \h^2_{\beta}$ for any $\h^2_{\beta} \in(0,1]$ and is a linear function of $\omega \in(0,1)$ (Figure~\ref{fig3} and Supplementary Figure~\ref{sfig3}).  The term 
$(1-\h^2_{\beta})\cdot \omega$ in $E^2_{O}$ represents the degree of model over-fitting due to spurious correlations \citep{fan2012variance,fan2018discoveries}. 
For $\omega >1$, the limit of $E^2_{R}(0^{+})$ is one, which indicates that the ridge-less estimator can have zero training error for $\y$ given any $\h^2_{\beta}\in(0,1]$. 
On the other hand, $E^2_{S}$ may not be a monotone fuction of $\omega$. 
When $\h^2_{\beta}\in (0,0.5]$, $E^2_{S}$ increases with $\omega$.
When $\h^2_{\beta}\in (0.5,1]$, interestingly,
$E^2_{S}$ decreases first as $\omega$ increases and can become much smaller than $\h^2_{\beta}$. 
More results of in-sample $R^2$ can be found in Corollary~S\ref{cor4} of supplementary file.

%%%%%%%%%%%%%%%%%%%%%%%%%%%%%%%%%%%%%%%%%%%%%%%%%%%%%%%%%%%%%%%%%%%%%
%%%%%%%%%%%%%%%%%%%%%%%%%%%%%%%%%%%%%%%%%%%%%%%%%%%%%%%%%%%%%%%%%%%%%
\begin{figure}
\includegraphics[page=1,width=0.7\linewidth]{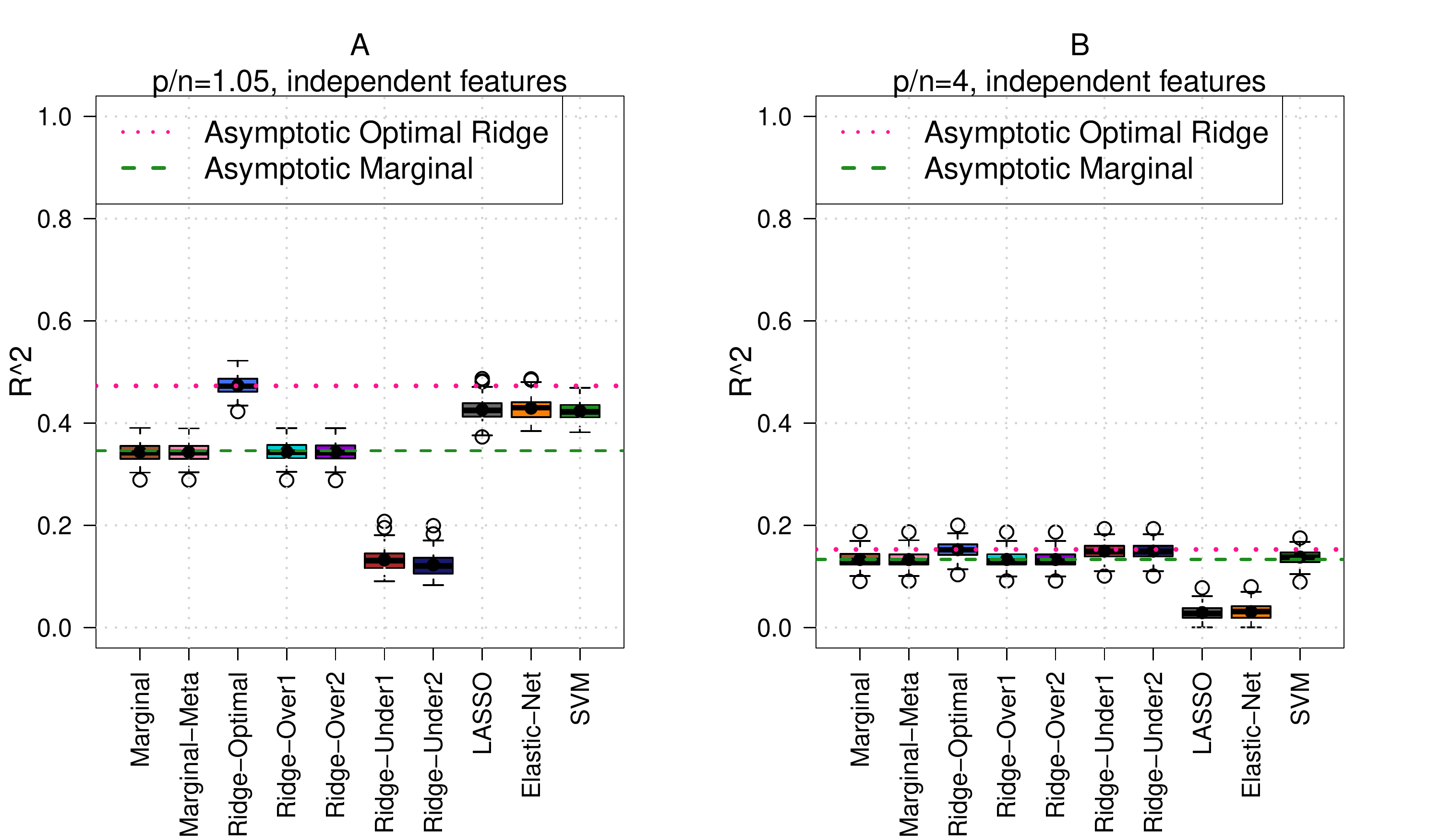}
  \caption{Out-of-sample $R$-squared of different estimators for independent features. Marginal: $\widehat{\bmbeta}_S$; Marginal-meta: meta-analyzed $\widehat{\bmbeta}_S$; Ridge-Optimal: $\widehat{\bmbeta}_R(\lambda^*)$; 
  Ridge-Over1: $\widehat{\bmbeta}_R(n\lambda^*)$; 
  Ridge-Over2: $\widehat{\bmbeta}_R(n^2\lambda^*)$; 
  Ridge-Under1: $\widehat{\bmbeta}_R(\lambda^*/n)$; 
  Ridge-Under2: $\widehat{\bmbeta}_R(\lambda^*/n^2)$;  
  %More details can be found in Section~\ref{sec7.1}.
  We set $n=2000$, and $\omega=$ $1.05$ and $4$ in A and B, respectively. The dash lines represent the asymptotic limits of ridge (red) and  marginal (green) estimators.
  }
\label{fig5}
\end{figure}
%%%%%%%%%%%%%%%%%%%
%%%%%%%%%%%%%%%%%%
%%%%%%%%%%%%%%%%%%%%%%%%%%%%%%%%%%%%%%%%%%%%%%%%%%%%%%%%%%%%%%%%%%%%
%%%%%%%%%%%%%%%%%%%%%%%%%%%%%%%%%%%%%%%%%%%%%%%%%%%%%%%%%%%%%%%%%%%%
%%%%%%%%%%%%%%%%%%%%%%%%%%%%%%%%%%%%%%%%%%%%%%%%%%%%%%%%%%%%%%%%%%%%
\section{Numerical results}\label{sec7}
\subsection{Simulation}\label{sec7.1}
We first numerically evaluate our theoretical results with $n=2000$, and $\omega= 1.05$, $2$, $4$ and $8$.  
Each entry of $\X$ and $\Z$ is independently generated from $N(0,1/p)$, and the ratio  $m/p$ is set to be $0.8$. 
We simulate a trait with heritability $0.8$ from model~(\ref{equ1.2.1}), and predict the same trait in the testing data (i.e., $\h_{\beta}^2=\h_{\eta}^2=0.8$, $\varphi_{\beta\eta}=1$, $\bmbeta_{(1)}=\bmeta_{(1)}$).
The nonzero genetic effects $\bmbeta_{(1)}$ and entries of $\bmeps$ and $\bmeps_z$ are generated from Normal distribution according to Condition~\ref{con2}. 
We evaluate the following estimators: 
1) marginal estimator defined in model~(\ref{equ1.1.1}) (Marginal); 
2) a meta-analyzed version of marginal estimator with weights equal to sample sizes ($400$ and $1600$, respectively; Marginal-meta); 
3) ridge estimator defined in model~(\ref{equ1.1.2}) with optimal regularizer $\lambda^*$ (Ridge-Optimal);
3) ridge estimator with $n\lambda^*$ (Ridge-Over1);
4) ridge estimator with $n^2\lambda^*$ (Ridge-Over2);
5) ridge estimator with $\lambda^*/n$ (Ridge-Under1); and
6) ridge estimator with $\lambda^*/n^2$ (Ridge-Under2).
In addition, we examine three other methods in our settings, including Lasso \citep{tibshirani1996regression}, Elastic-Net \citep{zou2005regularization}, and support vector machines (SVM) \citep{cortes1995support}.
A total of $100$ replicates is conducted, and we calculate the in-sample and out-of-sample $R$-squared ($A^2$ and $E^2$) defined in equation~(\ref{equ3.1}). 
The results are summarized in Figure~\ref{fig5} and Supplementary Figures~\ref{sfig5}~-~\ref{sfig6}.
As expected, the finite sample performance of marginal and ridge estimators supports our asymptotic results. 
For example, when $\omega=1.05$, the optimal ridge estimator clearly outperforms marginal estimator, and marginal estimator has similar $R^2$ to ridge estimator with large $\lambda$ (Figure~\ref{fig5}). 
Ridge estimator with small $\lambda$ performs poorly for $\omega=1.05$.
However, when $\omega$ becomes $4$, marginal estimator and all ridge estimators have similar $R^2$. 
In addition, meta-analyzed marginal estimator shows no decay of prediction accuracy. Lasso and Elastic-Net have poor performances for both $A^2$ and $E^2$ as $\omega$ becomes large, and SVM shows similar pattern to ridge estimators (Supplementary Figures~\ref{sfig5}-~\ref{sfig6}). We also observe that the in-sample $R^2$ has smaller variance compared to out-of-sample $R^2$.

%%%%%%%%%%%%%%%%%%%
%%%%%%%%%%%%%%%%%%
\begin{figure}
\includegraphics[page=1,width=0.7\linewidth]{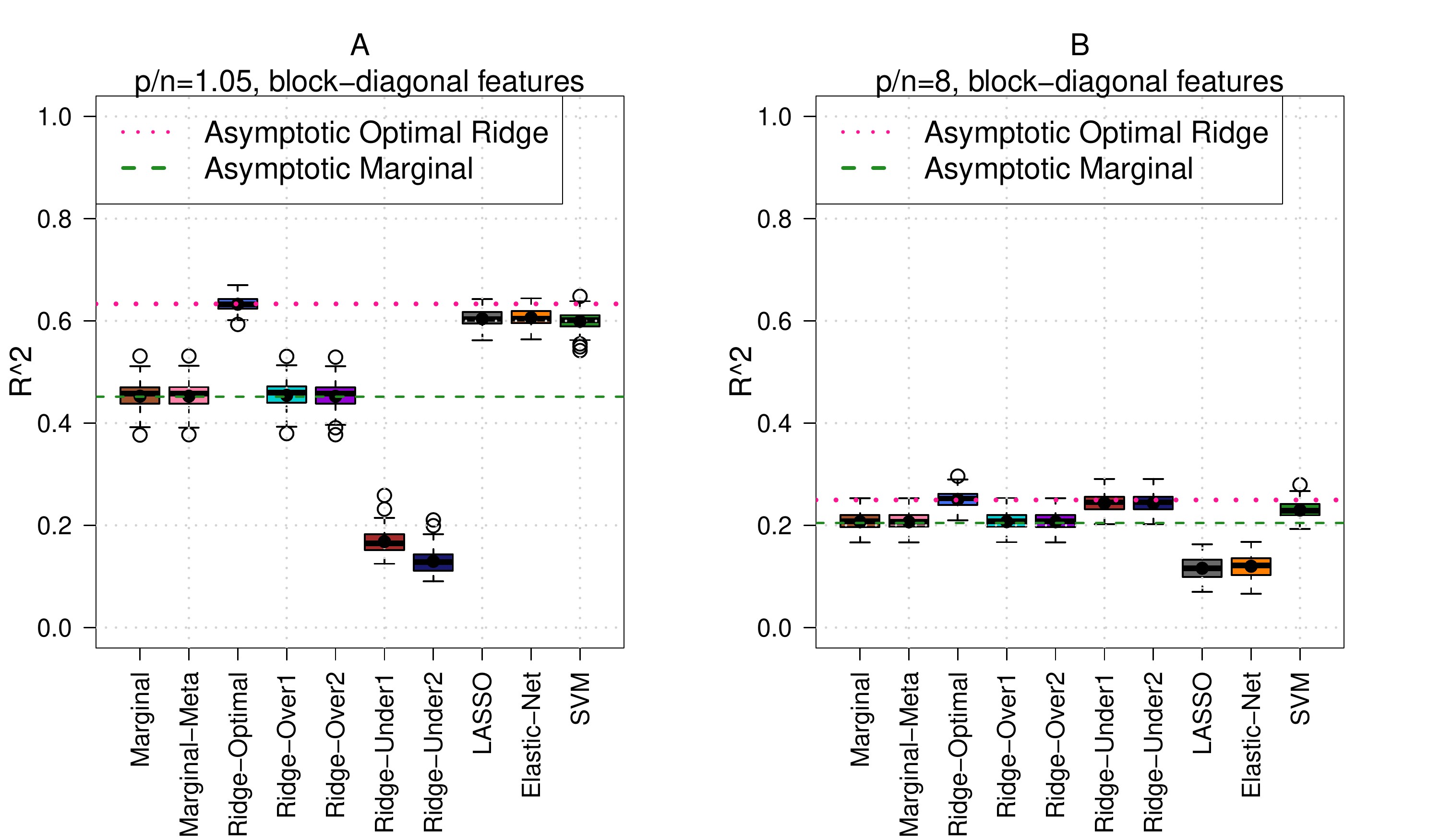}
  \caption{Out-of-sample $R$-squared of different methods for features with block-diagonal correlation structure. See Figure~\ref{fig5} for figure notations.
  We set $n=2000$, and $\omega=$ $1.05$ and $8$ in A and B, respectively. The dash lines represent the asymptotic limits of ridge (red) and  marginal (green) estimators.
  }
\label{fig6}
\end{figure}
%%%%%%%%%%%%%%%%%%%
%%%%%%%%%%%%%%%%%%

To mimic the LD structure of SNP data, we also construct $\bmSigma$ with a block-diagonal structure (block size $=20$). Features within the block have pair-wise correlation $\rho_b=0.8$, and features belong to different blocks are independent. Other settings are exactly the same as in $\bmSigma=\I_p$. 
The results are shown in Figure~\ref{fig6} and Supplementary Figures~\ref{sfig7}~-~\ref{sfig8}. 
Again, the performance of marginal and ridge estimators matches our theoretical limits, and the general pattern remains the same as in $\bmSigma=\I_p$. In addition,  the prediction accuracy is improved due to the feature-wise correlation, verifying that the decay of prediction accuracy due to dimensionality can be delayed by correlation among features. 
In this situation, it is easier for the optimal ridge estimator to outperform marginal estimator.

%%%%%%%%%%%%%%%%%%%
%%%%%%%%%%%%%%%%%%
\begin{figure}
\includegraphics[page=1,width=1\linewidth]{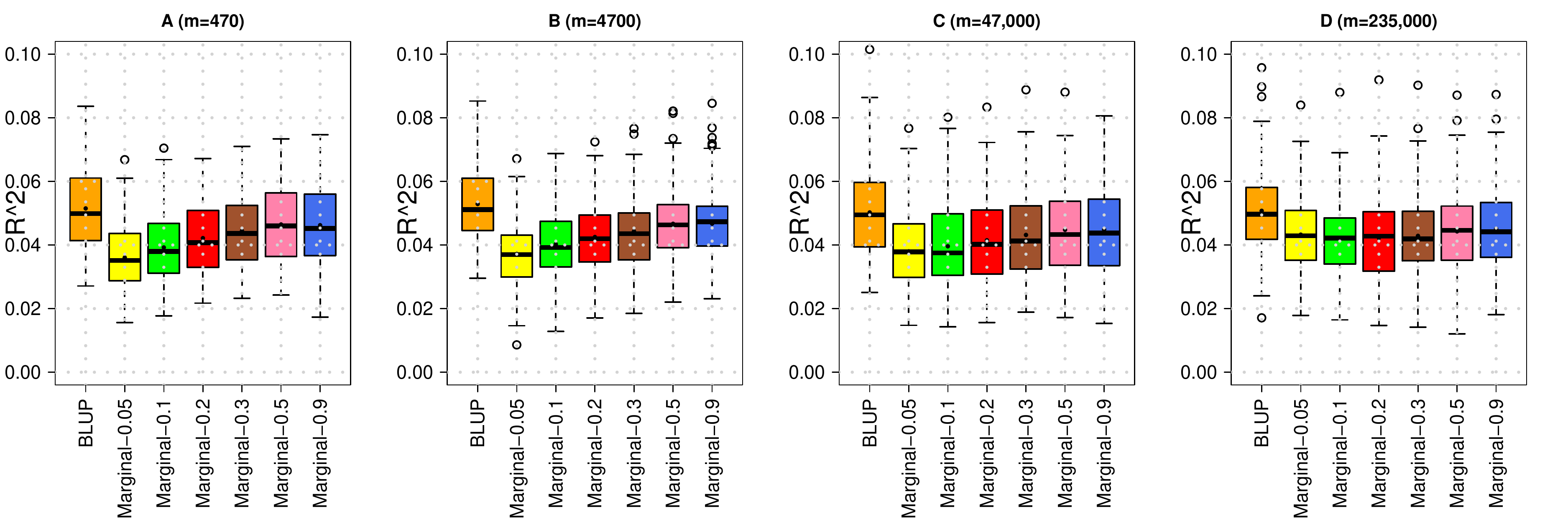}
  \caption{Out-of-sample $R$-squared $A^2$ of different BLUP and margimal estimator across different sparsity $m/p$.  }
\label{fig7}
\end{figure}
%%%%%%%%%%%%%%%%%%%
%%%%%%%%%%%%%%%%%%
%%%%%%%%%%%%%%%%%%%%%%%%%%%%%%%%%%%%%%%%%%%%%%%%%%%%%%
%%%%%%%%%%%%%%%%%%%%%%%%%%%%%%%%%%%%%%%%%%%%%%%%%%%%%%
\subsection{UKB data simulation}\label{sec7.2}
Next, we perform simulation based on real GWAS data from the UK Biobank (UKB) resources \citep{sudlow2015uk}. 
There are $461,488$ common genotyped genetic variants (most of which are unimputed SNPs) after standard quality control (QC) procedures detailed in the supplementary file. 
We randomly select $10,000$ individuals of British ancestry as training samples, and test the prediction accuracy of these genetic variants on another $1,000$ randomly selected individuals. 
Causal variants are randomly selected, and the number $m$ is set to $470$, $4700$, $47,000$, and $235,000$, respectively. The nonzero genetic effects are independently generated from $N(0,1/p)$, and the heritability $\h^2$ is set to $80\%$. 
Marginal estimtaor is generated using PLINK \citep{purcell2007plink}.
Following practical guidelines \citep{Choi416545}, we perform LD-based clumping for the marginal estimtaor via PLINK to obtain a list of relatively independent genetic variants for prediction. 
With the default window size ($250$ kb), we vary the clumping parameter $C_r^2$ and set it to $0.05$, $0.1$, $0.2$, $0.3$, $0.5$, and $0.9$. Smaller $C_r^2$ results in more stringent selection and more filtered variants by clumping. When $C_r^2=0.9$, most of the variants remain. 
The BLUP is obtained from GCTA \citep{yang2011gcta} using all genetic variants.

The results are displayed in Figure~\ref{fig7}, which shows that BLUP has slightly better prediction accuracy than marginal estimator across signal sparsity $m$ and clumping parameter $C_r^2$. The slightly higher prediction accuracy  matches our theoretical results. 
We also find that strict clumping with small $C_r^2$ may reduce the prediction accuracy when genetic signals are sparse. 
As $m/p$ increases, marginal estimator has more consistent performance across clumping parameters. 

%%%%%%%%%%%%%%%%%%%%%%%%%%%%%%%%%%%%%%%%%%%%%%%%%%%%%%%%%%%%%%%
%%%%%%%%%%%%%%%%%%%%%%%%%%%%%%%%%%%%%%%%%%%%%%%%%%%%%%%%%%%%%%%
\subsection{Real data analysis}
In this real data example, we aim to use the GWAS results of neuroimaging traits to predict cognitive test scores.  
We focus on $14$ volumetric traits of seven left/right pairs of brain regions of interest (refer to as ROI volumes),  including
left/right thalamus proper, left/right caudate, 
left/right putamen, left/right pallidum, 
left/right hippocampus, left/right amygdala, and left/right accumbens area.
These subcortical ROI volumes are quantified by magnetic resonance imaging (MRI) and are known to be associated with cognitive functions
\citep{miller2016multimodal}.
We use the UKB samples ($n$ = $19,629$, $p=8,944,375$) as training data for these $14$ ROI volumes, and using the UKB results to predict two cognitive test scores (IBAM and list sort total scores) on subjects in the Pediatric Imaging, Neurocognition, and Genetics (PING, $n$ = $905$) study \citep{jernigan2016pediatric}. 
More details about data processing, quality control procedures, and cohort information can be found in supplementary file.
We perform prediction using both marginal estimator (with $C_r^2=0.2$) and BLUP with all overlapping genetic variants between training and testing data. 
The UKB marginal estimators are from  \url{https://github.com/BIG-S2/GWAS/}, and the BLUP is generated from the GCTA tool. The association between the predicted and observed phenotype is estimated and tested in linear regression, adjusting for the effects of age and sex. The associated partial $R^2$ is used to measure the prediction accuracy. 
The partial $R^2$ of BLUP and marginal estimator on the two cognitive tests are displayed in Supplementary Table~1. 
We find that BLUP has slightly higher prediction accuracy than marginal estimator, but their partial $R^2$ are within similar range (Supplementary Figure~10).
For example, the mean partial $R^2$ of IBAM score are $0.176\%$ and $0.168\%$ for BLUP and marginal estimator, respectively. Such small partial $R^2$  are widely reported in GWAS prediction of cognitive and mental health traits \citep{bogdan2018polygenic}, which may due to the small genetic effects and the extremely polygenic architecture of brain-related complex traits \citep{o2019extreme}. 
%%%%%%%%%%%%%%%%%%%
%%%%%%%%%%%%%%%%%%
%%%%%%%%%%%%%%%%%%%
%%%%%%%%%%%%%%%%%%
\begin{figure}
\centering
\includegraphics[page=1,width=0.7\linewidth]{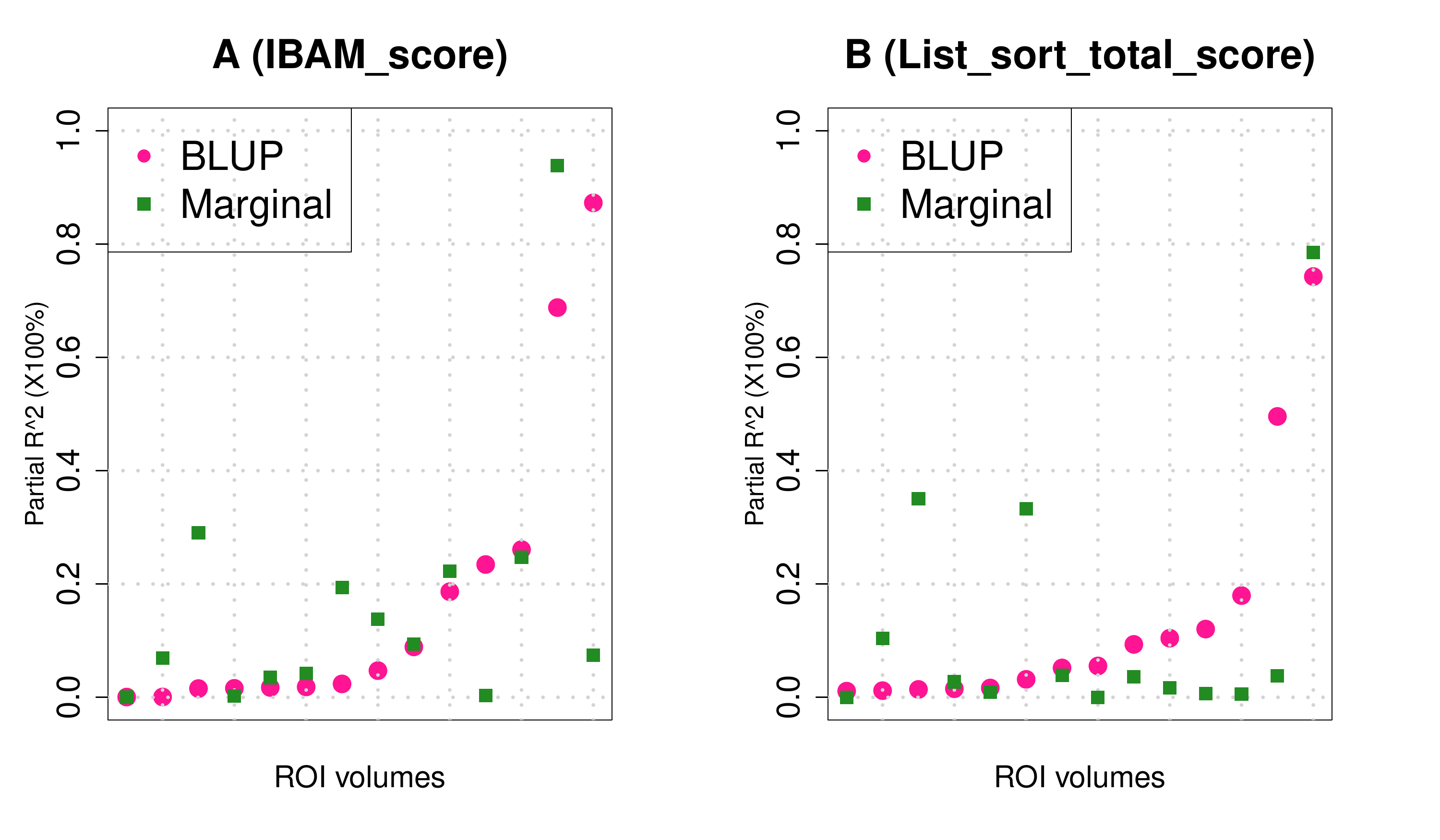}
  \caption{Partical $R$-squared ($\times 100\%$) of  $14$ subcortical ROI volumes to predict IBAM and LST cognitive test scores in the PING cohort. 
  Each point represents one ROI volume phenotype.
  The partial $R$-squared is estimated from linear regression while adjusting for the effects of age and gender.
 BLUP: best linear unbiased prediction; Marginal: marginal estimator. 
  IBAM: IBAM score; LST: list sort total score.}
\label{fig8}
\end{figure}

%%%%%%%%%%%%%%%%%%%%%%%%%%%%%%%%%%%%%%%%%%%%%%%%%%%%%%%%%%%%%%%
%%%%%%%%%%%%%%%%%%%%%%%%%%%%%%%%%%%%%%%%%%%%%%%%%%%%%%%%%%%%%%%%%%%%
%%%%%%%%%%%%%%%%%%%%%%%%%%%%%%%%%%%%%%%%%%%%%%%%%%%%%%%%%%%%%%%%%%%%
%%%%%%%%%%%%%%%%%%%%%%%%%%%%%%%%%%%%%%%%%%%%%%%%%%%%%%%%%%%%%%%%%%%%

\section{Discussion}\label{sec8}
In this paper, we study out-of-sample predictions on large-scale GWAS data with general $\bmSigma$ using random matrix theory. We investigate the prediction accuracy of marginal estimator with $R^2$ and generalize the results to consider cross-trait prediction and meta-analysis.
We also examine and compare the class of ridge-type estimators, and highlight the different or even reverse behaviors of in-sample and out-of-sample $R^2$.
Our theoretical results can be useful to evaluate the prediction accuracy in GWAS, and may also guide more future works in dense high-dimensional prediction.

A few interesting future problems can be studied in  high-dimensional dense signal settings. 
First, ridge-type estimators represent linear shrinkage estimation  
on $\bmSigma$ and its inverse $\bmSigma^{-1}$. It might be interesting to explore whether we can improve the prediction accuracy with nonlinear shrinkage estimators, such as \cite{ledoit2018optimal}. Second, if prior knowledge is known on the structure of $\bmSigma$, it is also possible to perform structured covariance estimation on $\bmSigma$, see  \cite{cai2016estimating} for a review of this area. For example, since  the $\bmSigma$ of SNP data is known to have a block-diagonal structure, it can be modeled as a bandable covariance matrix \citep{bickel2008regularized} with fast decay of feature correlation as their physical distance increases. 
Indeed, as mentioned before, marginal estimator can be viewed as a special banded covariance estimator of $\bmSigma$ with zero bandwidth, which may represent an extreme estimator that over-bands $\bmSigma$. 
Finally, other extensions such as binary outcomes, time-to-event data, and SNP annotations and selections are also of great interest following the presented framework.

%%%%%%%%%%%%%%%%%%%%%%%%%%%%%%%%%%%%%%%%%%%%%%%%%%%%%%%%%%%%%%%%%%%%
%%%%%%%%%%%%%%%%%%%%%%%%%%%%%%%%%%%%%%%%%%%%%%%%%%%%%%%%%%%%%%%%%%%%
%%%%%%%%%%%%%%%%%%%%%%%%%%%%%%%%%%%%%%%%%%%%%%%%%%%%%%%%%%%%%%%%%%%%
\section*{Acknowledgement}
We are grateful to Fei Zou for many helpful conversations, which motivate us to work on this problem in the first place.  
We would also like to thank Ziliang Zhu for helpful discussion on random matrix theory. 
This research was partially supported by U.S. NIH grants MH086633 and MH116527, and a grant from the Cancer Prevention Research Institute of Texas. 
This research has been conducted using the UK Biobank resource (application number $22783$), subject to a data transfer agreement.
We thank Tengfei Li and other members of the UNC BIG-S2 lab for processing the raw brain imaging data.
We thank the individuals represented in the UK Biobank and PING studies for their participation and the research teams for their work in collecting, processing and disseminating these datasets for analysis. More information of PING study can be found in supplementary file.

%%%%%%%%%%%%%%%%%%%%%%%%%%%%%%%%%%%%%%%%%%%%%%%%%%%%%%%%%%%%%%%%%%%%
%%%%%%%%%%%%%%%%%%%%%%%%%%%%%%%%%%%%%%%%%%%%%%%%%%%%%%%%%%%%%%%%%%%%
%%%%%%%%%%%%%%%%%%%%%%%%%%%%%%%%%%%%%%%%%%%%%%%%%%%%%%%%%%%%%%%%%%%%

%%%%%%%%%%%%%%%%%%%%%%%%%%%%%%%%%%%%%%%%%%%%%%%%%%%%%%%%
%%%%%%%%%%%%%%%%%%%%%%%%%%%%%%%%%%%%%%%%%%%%%%%%%%%%%%%%
%%%%%%%%%%%%%%%%%%%%%%%%%%%%%%%%%%%%%%%%%%%%%%%%%%%%%%%%
%%References
%\clearpage
\bibliographystyle{rss}
\bibliography{sample.bib}

\begin{thebibliography}{100}
\expandafter\ifx\csname natexlab\endcsname\relax\def\natexlab#1{#1}\fi
\expandafter\ifx\csname url\endcsname\relax
  \def\url#1{\texttt{#1}}\fi
\expandafter\ifx\csname urlprefix\endcsname\relax\def\urlprefix{URL: }\fi

\bibitem[{1000-Genomes-Project-Consortium.(2015)}]{10002015global}
1000-Genomes-Project-Consortium. (2015) A global reference for human genetic
  variation.
\newblock \textit{Nature}, \textbf{526}, 68--74.

\bibitem[{Avants et~al.(2011)Avants, Tustison, Song, Cook, Klein and
  Gee}]{avants2011reproducible}
Avants, B.~B., Tustison, N.~J., Song, G., Cook, P.~A., Klein, A. and Gee, J.~C.
  (2011) A reproducible evaluation of ants similarity metric performance in
  brain image registration.
\newblock \textit{Neuroimage}, \textbf{54}, 2033--2044.

\bibitem[{Bai and Silverstein(2010)}]{bai2010spectral}
Bai, Z. and Silverstein, J.~W. (2010) \textit{Spectral analysis of large
  dimensional random matrices}, vol.~20.
\newblock Springer.

\bibitem[{Barbeira et~al.(2018)Barbeira, Dickinson, Bonazzola, Zheng, Wheeler,
  Torres, Torstenson, Shah, Garcia, Edwards et~al.}]{barbeira2018exploring}
Barbeira, A.~N., Dickinson, S.~P., Bonazzola, R., Zheng, J., Wheeler, H.~E.,
  Torres, J.~M., Torstenson, E.~S., Shah, K.~P., Garcia, T., Edwards, T.~L.
  et~al. (2018) Exploring the phenotypic consequences of tissue specific gene
  expression variation inferred from gwas summary statistics.
\newblock \textit{Nature Communications}, \textbf{9}, 1825.

\bibitem[{Bickel and Levina(2008)}]{bickel2008regularized}
Bickel, P.~J. and Levina, E. (2008) Regularized estimation of large covariance
  matrices.
\newblock \textit{The Annals of Statistics}, \textbf{36}, 199--227.

\bibitem[{Bogdan et~al.(2018)Bogdan, Baranger and
  Agrawal}]{bogdan2018polygenic}
Bogdan, R., Baranger, D.~A. and Agrawal, A. (2018) Polygenic risk scores in
  clinical psychology: bridging genomic risk to individual differences.
\newblock \textit{Annual Review of Clinical Psychology}, \textbf{14}, 119--157.

\bibitem[{Boyle et~al.(2017)Boyle, Li and Pritchard}]{boyle2017expanded}
Boyle, E.~A., Li, Y.~I. and Pritchard, J.~K. (2017) An expanded view of complex
  traits: from polygenic to omnigenic.
\newblock \textit{Cell}, \textbf{169}, 1177--1186.

\bibitem[{Bulik-Sullivan et~al.(2015)Bulik-Sullivan, Finucane, Anttila, Gusev,
  Day, Loh, Duncan, Perry, Patterson, Robinson et~al.}]{bulik2015atlas}
Bulik-Sullivan, B., Finucane, H.~K., Anttila, V., Gusev, A., Day, F.~R., Loh,
  P.-R., Duncan, L., Perry, J.~R., Patterson, N., Robinson, E.~B. et~al. (2015)
  An atlas of genetic correlations across human diseases and traits.
\newblock \textit{Nature Genetics}, \textbf{47}, 1236--1241.

\bibitem[{Cai et~al.(2016)Cai, Ren and Zhou}]{cai2016estimating}
Cai, T.~T., Ren, Z. and Zhou, H.~H. (2016) Estimating structured
  high-dimensional covariance and precision matrices: Optimal rates and
  adaptive estimation.
\newblock \textit{Electronic Journal of Statistics}, \textbf{10}, 1--59.

\bibitem[{de~los Campos et~al.(2013)de~los Campos, Vazquez, Fernando,
  Klimentidis and Sorensen}]{de2013prediction}
de~los Campos, G., Vazquez, A.~I., Fernando, R., Klimentidis, Y.~C. and
  Sorensen, D. (2013) Prediction of complex human traits using the genomic best
  linear unbiased predictor.
\newblock \textit{PLoS Genetics}, \textbf{9}, e1003608.

\bibitem[{Chatterjee et~al.(2013)Chatterjee, Wheeler, Sampson, Hartge, Chanock
  and Park}]{chatterjee2013projecting}
Chatterjee, N., Wheeler, B., Sampson, J., Hartge, P., Chanock, S.~J. and Park,
  J.-H. (2013) Projecting the performance of risk prediction based on polygenic
  analyses of genome-wide association studies.
\newblock \textit{Nature Genetics}, \textbf{45}, 400--405.

\bibitem[{Choi et~al.(2018)Choi, Heng~Mak and
  O{\textquoteright}Reilly}]{Choi416545}
Choi, S.~W., Heng~Mak, T.~S. and O{\textquoteright}Reilly, P.~F. (2018) A guide
  to performing polygenic risk score analyses.
\newblock \textit{bioRxiv}.
\newblock
  \urlprefix\url{https://www.biorxiv.org/content/early/2018/09/14/416545}.

\bibitem[{Cortes and Vapnik(1995)}]{cortes1995support}
Cortes, C. and Vapnik, V. (1995) Support-vector networks.
\newblock \textit{Machine Learning}, \textbf{20}, 273--297.

\bibitem[{Daetwyler et~al.(2010)Daetwyler, Pong-Wong, Villanueva and
  Woolliams}]{daetwyler2010impact}
Daetwyler, H.~D., Pong-Wong, R., Villanueva, B. and Woolliams, J.~A. (2010) The
  impact of genetic architecture on genome-wide evaluation methods.
\newblock \textit{Genetics}, \textbf{185}, 1021--1031.

\bibitem[{Daetwyler et~al.(2008)Daetwyler, Villanueva and
  Woolliams}]{daetwyler2008accuracy}
Daetwyler, H.~D., Villanueva, B. and Woolliams, J.~A. (2008) Accuracy of
  predicting the genetic risk of disease using a genome-wide approach.
\newblock \textit{PLoS One}, \textbf{3}, e3395.

\bibitem[{Das et~al.(2016)Das, Forer, Sch{\"o}nherr, Sidore, Locke, Kwong,
  Vrieze, Chew, Levy, McGue et~al.}]{das2016next}
Das, S., Forer, L., Sch{\"o}nherr, S., Sidore, C., Locke, A.~E., Kwong, A.,
  Vrieze, S.~I., Chew, E.~Y., Levy, S., McGue, M. et~al. (2016) Next-generation
  genotype imputation service and methods.
\newblock \textit{Nature Genetics}, \textbf{48}, 1284--1287.

\bibitem[{Dicker(2013)}]{dicker2013optimal}
Dicker, L.~H. (2013) Optimal equivariant prediction for high-dimensional linear
  models with arbitrary predictor covariance.
\newblock \textit{Electronic Journal of Statistics}, \textbf{7}, 1806--1834.

\bibitem[{Dicker(2016)}]{dicker2016ridge}
--- (2016) Ridge regression and asymptotic minimax estimation over spheres of
  growing dimension.
\newblock \textit{Bernoulli}, \textbf{22}, 1--37.

\bibitem[{Dicker and Erdogdu(2017)}]{dicker2017flexible}
Dicker, L.~H. and Erdogdu, M.~A. (2017) Flexible results for quadratic forms
  with applications to variance components estimation.
\newblock \textit{The Annals of Statistics}, \textbf{45}, 386--414.

\bibitem[{Dobriban and Sheng(2018)}]{dobriban2018distributed}
Dobriban, E. and Sheng, Y. (2018) Distributed linear regression by averaging.
\newblock \textit{arXiv preprint arXiv:1810.00412}.

\bibitem[{Dobriban and Sheng(2019)}]{dobriban2019one}
--- (2019) One-shot distributed ridge regression in high dimensions.
\newblock \textit{arXiv preprint arXiv:1903.09321}.

\bibitem[{Dobriban and Wager(2018)}]{dobriban2018high}
Dobriban, E. and Wager, S. (2018) High-dimensional asymptotics of prediction:
  Ridge regression and classification.
\newblock \textit{The Annals of Statistics}, \textbf{46}, 247--279.

\bibitem[{Dudbridge(2013)}]{dudbridge2013power}
Dudbridge, F. (2013) Power and predictive accuracy of polygenic risk scores.
\newblock \textit{PLoS Genetics}, \textbf{9}, e1003348.

\bibitem[{El~Karoui(2013)}]{karoui2013asymptotic}
El~Karoui, N. (2013) Asymptotic behavior of unregularized and ridge-regularized
  high-dimensional robust regression estimators: rigorous results.
\newblock \textit{arXiv preprint arXiv:1311.2445}.

\bibitem[{El~Karoui(2018)}]{el2018impact}
--- (2018) On the impact of predictor geometry on the performance on
  high-dimensional ridge-regularized generalized robust regression estimators.
\newblock \textit{Probability Theory and Related Fields}, \textbf{170},
  95--175.

\bibitem[{Evans et~al.(2018)Evans, Tahmasbi, Vrieze, Abecasis, Das, Gazal,
  Bjelland, Goddard, Neale, Yang et~al.}]{evans2018comparison}
Evans, L., Tahmasbi, R., Vrieze, S., Abecasis, G., Das, S., Gazal, S.,
  Bjelland, D., Goddard, M., Neale, B., Yang, J. et~al. (2018) Comparison of
  methods that use whole genome data to estimate the heritability and genetic
  architecture of complex traits.
\newblock \textit{Nature Genetics}, \textbf{50}, 737--745.

\bibitem[{Fan et~al.(2012)Fan, Guo and Hao}]{fan2012variance}
Fan, J., Guo, S. and Hao, N. (2012) Variance estimation using refitted
  cross-validation in ultrahigh dimensional regression.
\newblock \textit{Journal of the Royal Statistical Society: Series B
  (Statistical Methodology)}, \textbf{74}, 37--65.

\bibitem[{Fan and Lv(2008)}]{fan2008sure}
Fan, J. and Lv, J. (2008) Sure independence screening for ultrahigh dimensional
  feature space.
\newblock \textit{Journal of the Royal Statistical Society: Series B
  (Statistical Methodology)}, \textbf{70}, 849--911.

\bibitem[{Fan et~al.(2018)Fan, Shao and Zhou}]{fan2018discoveries}
Fan, J., Shao, Q.-M. and Zhou, W.-X. (2018) Are discoveries spurious?
  distributions of maximum spurious correlations and their applications.
\newblock \textit{The Annals of Statistics}, \textbf{46}, 989--1017.

\bibitem[{Feng and Zhang(2017)}]{feng2017sorted}
Feng, L. and Zhang, C.-H. (2017) Sorted concave penalized regression.
\newblock \textit{arXiv preprint arXiv:1712.09941}.

\bibitem[{Gamazon et~al.(2015)Gamazon, Wheeler, Shah, Mozaffari,
  Aquino-Michaels, Carroll, Eyler, Denny, Nicolae, Cox
  et~al.}]{gamazon2015gene}
Gamazon, E.~R., Wheeler, H.~E., Shah, K.~P., Mozaffari, S.~V., Aquino-Michaels,
  K., Carroll, R.~J., Eyler, A.~E., Denny, J.~C., Nicolae, D.~L., Cox, N.~J.
  et~al. (2015) A gene-based association method for mapping traits using
  reference transcriptome data.
\newblock \textit{Nature Genetics}, \textbf{47}, 1091--1098.

\bibitem[{Goddard(2009)}]{goddard2009genomic}
Goddard, M. (2009) Genomic selection: prediction of accuracy and maximisation
  of long term response.
\newblock \textit{Genetica}, \textbf{136}, 245--257.

\bibitem[{Guo and Cheng(2018)}]{guo2018moderate}
Guo, X. and Cheng, G. (2018) Moderate-dimensional inferences on quadratic
  functionals in ordinary least squares.
\newblock \textit{arXiv preprint arXiv:1810.01323}.

\bibitem[{Guo et~al.(2019)Guo, Wang, Cai and Li}]{guo2019optimal}
Guo, Z., Wang, W., Cai, T.~T. and Li, H. (2019) Optimal estimation of genetic
  relatedness in high-dimensional linear models.
\newblock \textit{Journal of the American Statistical Association},
  \textbf{114}, 358--369.

\bibitem[{Gurdasani et~al.(2019)Gurdasani, Barroso, Zeggini and
  Sandhu}]{gurdasani2019genomics}
Gurdasani, D., Barroso, I., Zeggini, E. and Sandhu, M.~S. (2019) Genomics of
  disease risk in globally diverse populations.
\newblock \textit{Nature Reviews Genetics}, in press.

\bibitem[{Gusev et~al.(2016)Gusev, Ko, Shi, Bhatia, Chung, Penninx, Jansen,
  De~Geus, Boomsma, Wright et~al.}]{gusev2016integrative}
Gusev, A., Ko, A., Shi, H., Bhatia, G., Chung, W., Penninx, B.~W., Jansen, R.,
  De~Geus, E.~J., Boomsma, D.~I., Wright, F.~A. et~al. (2016) Integrative
  approaches for large-scale transcriptome-wide association studies.
\newblock \textit{Nature Genetics}, \textbf{48}, 245--252.

\bibitem[{Hastie et~al.(2019)Hastie, Montanari, Rosset and
  Tibshirani}]{hastie2019surprises}
Hastie, T., Montanari, A., Rosset, S. and Tibshirani, R.~J. (2019) Surprises in
  high-dimensional ridgeless least squares interpolation.
\newblock \textit{arXiv preprint arXiv:1903.08560}.

\bibitem[{Henderson(1950)}]{henderson1950best}
Henderson, C.~R. (1950) Estimation of genetic parameters (abstract).
\newblock \textit{Annals of Mathematical Statistics}, \textbf{21}, 309--310.

\bibitem[{Henderson(1975)}]{henderson1975best}
--- (1975) Best linear unbiased estimation and prediction under a selection
  model.
\newblock \textit{Biometrics}, \textbf{31}, 423--447.

\bibitem[{Hoerl and Kennard(1970)}]{hoerl1970ridge}
Hoerl, A.~E. and Kennard, R.~W. (1970) Ridge regression: Biased estimation for
  nonorthogonal problems.
\newblock \textit{Technometrics}, \textbf{12}, 55--67.

\bibitem[{Holmes et~al.(2019)Holmes, Speed and Balding}]{Holmes532069}
Holmes, J.~B., Speed, D. and Balding, D.~J. (2019) Summary statistic analyses
  can mistake confounding bias for heritability.
\newblock \textit{bioRxiv}.
\newblock
  \urlprefix\url{https://www.biorxiv.org/content/early/2019/06/04/532069}.

\bibitem[{Hsu et~al.(2011)Hsu, Kakade and Zhang}]{hsu2011random}
Hsu, D., Kakade, S.~M. and Zhang, T. (2011) Random design analysis of ridge
  regression.
\newblock \textit{arXiv preprint arXiv:1106.2363}.

\bibitem[{Hu et~al.(2019)Hu, Li, Lu, Weng, Wang, Zekavat, Yu, Li, Gu, Muchnik
  et~al.}]{hu2019statistical}
Hu, Y., Li, M., Lu, Q., Weng, H., Wang, J., Zekavat, S., Yu, Z., Li, B., Gu,
  J., Muchnik, S. et~al. (2019) A statistical framework for cross-tissue
  transcriptome-wide association analysis.
\newblock \textit{Nature Genetics}, \textbf{51}, 568--576.

\bibitem[{Jernigan et~al.(2016)Jernigan, Brown, Hagler~Jr, Akshoomoff, Bartsch,
  Newman, Thompson, Bloss, Murray, Schork et~al.}]{jernigan2016pediatric}
Jernigan, T.~L., Brown, T.~T., Hagler~Jr, D.~J., Akshoomoff, N., Bartsch, H.,
  Newman, E., Thompson, W.~K., Bloss, C.~S., Murray, S.~S., Schork, N. et~al.
  (2016) The pediatric imaging, neurocognition, and genetics (ping) data
  repository.
\newblock \textit{Neuroimage}, \textbf{124}, 1149--1154.

\bibitem[{Jiang et~al.(2016)Jiang, Li, Paul, Yang and Zhao}]{jiang2016high}
Jiang, J., Li, C., Paul, D., Yang, C. and Zhao, H. (2016) On high-dimensional
  misspecified mixed model analysis in genome-wide association study.
\newblock \textit{The Annals of Statistics}, \textbf{44}, 2127--2160.

\bibitem[{Khera et~al.(2018)Khera, Chaffin, Aragam, Haas, Roselli, Choi,
  Natarajan, Lander, Lubitz, Ellinor et~al.}]{khera2018genome}
Khera, A.~V., Chaffin, M., Aragam, K.~G., Haas, M.~E., Roselli, C., Choi,
  S.~H., Natarajan, P., Lander, E.~S., Lubitz, S.~A., Ellinor, P.~T. et~al.
  (2018) Genome-wide polygenic scores for common diseases identify individuals
  with risk equivalent to monogenic mutations.
\newblock \textit{Nature Genetics}, \textbf{50}, 1219--1224.

\bibitem[{Klein and Tourville(2012)}]{klein2012101}
Klein, A. and Tourville, J. (2012) 101 labeled brain images and a consistent
  human cortical labeling protocol.
\newblock \textit{Frontiers in Neuroscience}, \textbf{6}, 171.

\bibitem[{Ledoit and P{\'e}ch{\'e}(2011)}]{ledoit2011eigenvectors}
Ledoit, O. and P{\'e}ch{\'e}, S. (2011) Eigenvectors of some large sample
  covariance matrix ensembles.
\newblock \textit{Probability Theory and Related Fields}, \textbf{151},
  233--264.

\bibitem[{Ledoit and Wolf(2004)}]{ledoit2004well}
Ledoit, O. and Wolf, M. (2004) A well-conditioned estimator for
  large-dimensional covariance matrices.
\newblock \textit{Journal of Multivariate Analysis}, \textbf{88}, 365--411.

\bibitem[{Ledoit and Wolf(2018)}]{ledoit2018optimal}
--- (2018) Optimal estimation of a large-dimensional covariance matrix under
  stein’s loss.
\newblock \textit{Bernoulli}, \textbf{24}, 3791--3832.

\bibitem[{Lee et~al.(2018)Lee, Wedow, Okbay, Kong, Maghzian, Zacher,
  Nguyen-Viet, Bowers, Sidorenko, Linn{\'e}r et~al.}]{lee2018gene}
Lee, J.~J., Wedow, R., Okbay, A., Kong, E., Maghzian, O., Zacher, M.,
  Nguyen-Viet, T.~A., Bowers, P., Sidorenko, J., Linn{\'e}r, R.~K. et~al.
  (2018) Gene discovery and polygenic prediction from a genome-wide association
  study of educational attainment in 1.1 million individuals.
\newblock \textit{Nature Genetics}, \textbf{50}, 1112--1121.

\bibitem[{Li et~al.(2014)Li, Yang, Gelernter and Zhao}]{li2014improving}
Li, C., Yang, C., Gelernter, J. and Zhao, H. (2014) Improving genetic risk
  prediction by leveraging pleiotropy.
\newblock \textit{Human Genetics}, \textbf{5}, 639--650.

\bibitem[{Loh et~al.(2015)Loh, Bhatia, Gusev, Finucane, Bulik-Sullivan,
  Pollack, de~Candia, Lee, Wray, Kendler et~al.}]{loh2015contrasting}
Loh, P.-R., Bhatia, G., Gusev, A., Finucane, H.~K., Bulik-Sullivan, B.~K.,
  Pollack, S.~J., de~Candia, T.~R., Lee, S.~H., Wray, N.~R., Kendler, K.~S.
  et~al. (2015) Contrasting genetic architectures of schizophrenia and other
  complex diseases using fast variance-components analysis.
\newblock \textit{Nature Genetics}, \textbf{47}, 1385--1392.

\bibitem[{Ma and Dicker(2019)}]{ma2019mahalanobis}
Ma, R. and Dicker, L.~H. (2019) The mahalanobis kernel for heritability
  estimation in genome-wide association studies: fixed-effects and
  random-effects methods.
\newblock \textit{arXiv preprint arXiv:1901.02936}.

\bibitem[{Marchenko and Pastur(1967)}]{marchenko1967distribution}
Marchenko, V.~A. and Pastur, L.~A. (1967) Distribution of eigenvalues for some
  sets of random matrices.
\newblock \textit{Matematicheskii Sbornik}, \textbf{114}, 507--536.

\bibitem[{Martin et~al.(2018)Martin, Daly, Robinson, Hyman and
  Neale}]{martin2018predicting}
Martin, A.~R., Daly, M.~J., Robinson, E.~B., Hyman, S.~E. and Neale, B.~M.
  (2018) Predicting polygenic risk of psychiatric disorders.
\newblock \textit{Biological Psychiatry}, in press.

\bibitem[{Martin et~al.(2019)Martin, Kanai, Kamatani, Okada, Neale and
  Daly}]{martin2019clinical}
Martin, A.~R., Kanai, M., Kamatani, Y., Okada, Y., Neale, B.~M. and Daly, M.~J.
  (2019) Clinical use of current polygenic risk scores may exacerbate health
  disparities.
\newblock \textit{Nature Genetics}, \textbf{51}, 584--591.

\bibitem[{Mavaddat et~al.(2019)Mavaddat, Michailidou, Dennis, Lush, Fachal,
  Lee, Tyrer, Chen, Wang, Bolla et~al.}]{mavaddat2019polygenic}
Mavaddat, N., Michailidou, K., Dennis, J., Lush, M., Fachal, L., Lee, A.,
  Tyrer, J.~P., Chen, T.-H., Wang, Q., Bolla, M.~K. et~al. (2019) Polygenic
  risk scores for prediction of breast cancer and breast cancer subtypes.
\newblock \textit{The American Journal of Human Genetics}, \textbf{104},
  21--34.

\bibitem[{Miller et~al.(2016)Miller, Alfaro-Almagro, Bangerter, Thomas, Yacoub,
  Xu, Bartsch, Jbabdi, Sotiropoulos, Andersson et~al.}]{miller2016multimodal}
Miller, K.~L., Alfaro-Almagro, F., Bangerter, N.~K., Thomas, D.~L., Yacoub, E.,
  Xu, J., Bartsch, A.~J., Jbabdi, S., Sotiropoulos, S.~N., Andersson, J.~L.
  et~al. (2016) Multimodal population brain imaging in the uk biobank
  prospective epidemiological study.
\newblock \textit{Nature neuroscience}, \textbf{19}, 1523.

\bibitem[{O'Connor et~al.(2019)O'Connor, Schoech, Hormozdiari, Gazal, Patterson
  and Price}]{o2019extreme}
O'Connor, L.~J., Schoech, A.~P., Hormozdiari, F., Gazal, S., Patterson, N. and
  Price, A.~L. (2019) Extreme polygenicity of complex traits is explained by
  negative selection.
\newblock \textit{The American Journal of Human Genetics}, \textbf{105},
  456--476.

\bibitem[{Pasaniuc and Price(2017)}]{pasaniuc2017dissecting}
Pasaniuc, B. and Price, A.~L. (2017) Dissecting the genetics of complex traits
  using summary association statistics.
\newblock \textit{Nature Reviews Genetics}, \textbf{18}, 117--127.

\bibitem[{Paul and Aue(2014)}]{paul2014random}
Paul, D. and Aue, A. (2014) Random matrix theory in statistics: A review.
\newblock \textit{Journal of Statistical Planning and Inference}, \textbf{150},
  1--29.

\bibitem[{Pluta et~al.(2017)Pluta, Ombao, Chen, Xue, Moyzis and
  Yu}]{pluta2017adaptive}
Pluta, D., Ombao, H., Chen, C., Xue, G., Moyzis, R. and Yu, Z. (2017) Adaptive
  mantel test for associationtesting in imaging genetics data.
\newblock \textit{arXiv preprint arXiv:1712.07270}.

\bibitem[{Purcell et~al.(2007)Purcell, Neale, Todd-Brown, Thomas, Ferreira,
  Bender, Maller, Sklar, De~Bakker, Daly et~al.}]{purcell2007plink}
Purcell, S., Neale, B., Todd-Brown, K., Thomas, L., Ferreira, M.~A., Bender,
  D., Maller, J., Sklar, P., De~Bakker, P.~I., Daly, M.~J. et~al. (2007) Plink:
  a tool set for whole-genome association and population-based linkage
  analyses.
\newblock \textit{The American Journal of Human Genetics}, \textbf{81},
  559--575.

\bibitem[{Purcell et~al.(2009)Purcell, Wray, Stone, Visscher, O'Donovan,
  Sullivan, Sklar, Ruderfer, McQuillin, Morris et~al.}]{purcell2009common}
Purcell, S.~M., Wray, R., Stone, L., Visscher, M., O'Donovan, C., Sullivan, F.,
  Sklar, P., Ruderfer, M., McQuillin, A., Morris, W. et~al. (2009) Common
  polygenic variation contributes to risk of schizophrenia and bipolar
  disorder.
\newblock \textit{Nature}, \textbf{460}, 748--752.

\bibitem[{Quick et~al.(2018)Quick, Fuchsberger, Taliun, Abecasis, Boehnke and
  Kang}]{quick2018emerald}
Quick, C., Fuchsberger, C., Taliun, D., Abecasis, G., Boehnke, M. and Kang,
  H.~M. (2018) emerald: rapid linkage disequilibrium estimation with massive
  datasets.
\newblock \textit{Bioinformatics}, \textbf{35}, 164--166.

\bibitem[{van Rheenen et~al.(2019)van Rheenen, Peyrot, Schork, Lee and
  Wray}]{van2019genetic}
van Rheenen, W., Peyrot, W.~J., Schork, A.~J., Lee, S.~H. and Wray, N.~R.
  (2019) Genetic correlations of polygenic disease traits: from theory to
  practice.
\newblock \textit{Nature Reviews Genetics}, in press.

\bibitem[{Robinson(1991)}]{robinson1991blup}
Robinson, G.~K. (1991) That blup is a good thing: the estimation of random
  effects.
\newblock \textit{Statistical Science}, \textbf{6}, 15--32.

\bibitem[{Schaid et~al.(2018)Schaid, Chen and Larson}]{schaid2018genome}
Schaid, D., Chen, W. and Larson, N. (2018) From genome-wide associations to
  candidate causal variants by statistical fine-mapping.
\newblock \textit{Nature Reviews Genetics}, \textbf{19}, 491--504.

\bibitem[{Silverstein(1995)}]{silverstein1995strong}
Silverstein, J.~W. (1995) Strong convergence of the empirical distribution of
  eigenvalues of large dimensional random matrices.
\newblock \textit{Journal of Multivariate Analysis}, \textbf{55}, 331--339.

\bibitem[{Speed and Balding(2014)}]{speed2014multiblup}
Speed, D. and Balding, D. (2014) Multiblup: improved snp-based prediction for
  complex traits.
\newblock \textit{Genome Research}, \textbf{24}, 1550--1557.

\bibitem[{Speed and Balding(2019)}]{speed2019sumher}
--- (2019) Sumher better estimates the snp heritability of complex traits from
  summary statistics.
\newblock \textit{Nature Genetics}, \textbf{51}, 277--284.

\bibitem[{Steinsaltz et~al.(2018)Steinsaltz, Dahl and
  Wachter}]{steinsaltz2018statistical}
Steinsaltz, D., Dahl, A. and Wachter, K.~W. (2018) Statistical properties of
  simple random-effects models for genetic heritability.
\newblock \textit{Electronic Journal of Statistics}, \textbf{12}, 321--358.

\bibitem[{Sudlow et~al.(2015)Sudlow, Gallacher, Allen, Beral, Burton, Danesh,
  Downey, Elliott, Green, Landray et~al.}]{sudlow2015uk}
Sudlow, C., Gallacher, J., Allen, N., Beral, V., Burton, P., Danesh, J.,
  Downey, P., Elliott, P., Green, J., Landray, M. et~al. (2015) Uk biobank: an
  open access resource for identifying the causes of a wide range of complex
  diseases of middle and old age.
\newblock \textit{PLoS Medicine}, \textbf{12}, e1001779.

\bibitem[{Sugrue and Desikan(2019)}]{sugrue2019polygenic}
Sugrue, L.~P. and Desikan, R.~S. (2019) What are polygenic scores and why are
  they important?
\newblock \textit{JAMA}, \textbf{321}, 1820--1821.

\bibitem[{Sullivan and Geschwind(2019)}]{sullivan2019defining}
Sullivan, P.~F. and Geschwind, D.~H. (2019) Defining the genetic, genomic,
  cellular, and diagnostic architectures of psychiatric disorders.
\newblock \textit{Cell}, \textbf{177}, 162--183.

\bibitem[{Sun and Lin(2017)}]{sun2017set}
Sun, R. and Lin, X. (2017) Set-based tests for genetic association using the
  generalized berk-jones statistic.
\newblock \textit{arXiv preprint arXiv:1710.02469}.

\bibitem[{Tam et~al.(2019)Tam, Patel, Turcotte, Boss{\'e}, Par{\'e} and
  Meyre}]{tam2019benefits}
Tam, V., Patel, N., Turcotte, M., Boss{\'e}, Y., Par{\'e}, G. and Meyre, D.
  (2019) Benefits and limitations of genome-wide association studies.
\newblock \textit{Nature Reviews Genetics}, in press.

\bibitem[{Tibshirani(1996)}]{tibshirani1996regression}
Tibshirani, R. (1996) Regression shrinkage and selection via the lasso.
\newblock \textit{Journal of the Royal Statistical Society. Series B
  (Methodological)}, \textbf{58}, 267--288.

\bibitem[{Tikhonov(1963)}]{tikhonov1963solution}
Tikhonov, A.~N. (1963) On the solution of ill-posed problems and the method of
  regularization.
\newblock In \textit{Doklady Akademii Nauk}, vol. 151, 501--504. Russian
  Academy of Sciences.

\bibitem[{Timpson et~al.(2018)Timpson, Greenwood, Soranzo, Lawson and
  Richards}]{timpson2018genetic}
Timpson, N.~J., Greenwood, C.~M., Soranzo, N., Lawson, D.~J. and Richards,
  J.~B. (2018) Genetic architecture: the shape of the genetic contribution to
  human traits and disease.
\newblock \textit{Nature Reviews Genetics}, \textbf{19}, 110--125.

\bibitem[{Torkamani et~al.(2018)Torkamani, Wineinger and
  Topol}]{torkamani2018personal}
Torkamani, A., Wineinger, N.~E. and Topol, E.~J. (2018) The personal and
  clinical utility of polygenic risk scores.
\newblock \textit{Nature Reviews Genetics}, \textbf{19}, 581--590.

\bibitem[{Tustison et~al.(2014)Tustison, Cook, Klein, Song, Das, Duda, Kandel,
  van Strien, Stone, Gee et~al.}]{tustison2014large}
Tustison, N.~J., Cook, P.~A., Klein, A., Song, G., Das, S.~R., Duda, J.~T.,
  Kandel, B.~M., van Strien, N., Stone, J.~R., Gee, J.~C. et~al. (2014)
  Large-scale evaluation of ants and freesurfer cortical thickness
  measurements.
\newblock \textit{Neuroimage}, \textbf{99}, 166--179.

\bibitem[{Visscher et~al.(2017)Visscher, Wray, Zhang, Sklar, McCarthy, Brown
  and Yang}]{visscher201710}
Visscher, P.~M., Wray, N.~R., Zhang, Q., Sklar, P., McCarthy, M.~I., Brown,
  M.~A. and Yang, J. (2017) 10 years of gwas discovery: biology, function, and
  translation.
\newblock \textit{The American Journal of Human Genetics}, \textbf{101}, 5--22.

\bibitem[{Wang et~al.(2015)Wang, Pan, Tong and Zhu}]{wang2015shrinkage}
Wang, C., Pan, G., Tong, T. and Zhu, L. (2015) Shrinkage estimation of large
  dimensional precision matrix using random matrix theory.
\newblock \textit{Statistica Sinica}, \textbf{25}, 993--1008.

\bibitem[{Wang and Leng(2016)}]{wang2016high}
Wang, X. and Leng, C. (2016) High dimensional ordinary least squares projection
  for screening variables.
\newblock \textit{Journal of the Royal Statistical Society: Series B
  (Statistical Methodology)}, \textbf{78}, 589--611.

\bibitem[{Watanabe et~al.(2019)Watanabe, Stringer, Frei, Mirkov, de~Leeuw,
  Polderman, van~der Sluis, Andreassen, Neale and
  Posthuma}]{watanabe2019global}
Watanabe, K., Stringer, S., Frei, O., Mirkov, M.~U., de~Leeuw, C., Polderman,
  T.~J., van~der Sluis, S., Andreassen, O.~A., Neale, B.~M. and Posthuma, D.
  (2019) A global overview of pleiotropy and genetic architecture in complex
  traits.
\newblock \textit{Nature Genetics}, \textbf{51}, 1339--1348.

\bibitem[{Wheeler et~al.(2014)Wheeler, Aquino-Michaels, Gamazon, Trubetskoy,
  Dolan, Huang, Cox and Im}]{wheeler2014poly}
Wheeler, H.~E., Aquino-Michaels, K., Gamazon, E.~R., Trubetskoy, V.~V., Dolan,
  M.~E., Huang, R.~S., Cox, N.~J. and Im, H.~K. (2014) Poly-omic prediction of
  complex traits: Omickriging.
\newblock \textit{Genetic Epidemiology}, \textbf{38}, 402--415.

\bibitem[{Wray et~al.(2018)Wray, Wijmenga, Sullivan, Yang and
  Visscher}]{wray2018common}
Wray, N.~R., Wijmenga, C., Sullivan, P.~F., Yang, J. and Visscher, P.~M. (2018)
  Common disease is more complex than implied by the core gene omnigenic model.
\newblock \textit{Cell}, \textbf{173}, 1573--1580.

\bibitem[{Yang et~al.(2010)Yang, Benyamin, McEvoy, Gordon, Henders, Nyholt,
  Madden, Heath, Martin, Montgomery et~al.}]{yang2010common}
Yang, J., Benyamin, B., McEvoy, B.~P., Gordon, S., Henders, A.~K., Nyholt,
  D.~R., Madden, P.~A., Heath, A.~C., Martin, N.~G., Montgomery, G.~W. et~al.
  (2010) Common snps explain a large proportion of the heritability for human
  height.
\newblock \textit{Nature Genetics}, \textbf{42}, 565--569.

\bibitem[{Yang et~al.(2011)Yang, Lee, Goddard and Visscher}]{yang2011gcta}
Yang, J., Lee, S.~H., Goddard, M.~E. and Visscher, P.~M. (2011) Gcta: a tool
  for genome-wide complex trait analysis.
\newblock \textit{The American Journal of Human Genetics}, \textbf{88}, 76--82.

\bibitem[{Yang et~al.(2017)Yang, Zeng, Goddard, Wray and
  Visscher}]{yang2017concepts}
Yang, J., Zeng, J., Goddard, M.~E., Wray, N.~R. and Visscher, P.~M. (2017)
  Concepts, estimation and interpretation of snp-based heritability.
\newblock \textit{Nature Genetics}, \textbf{49}, 1304--1310.

\bibitem[{Yang and Cheng(2018)}]{yang2018quadratic}
Yang, Q. and Cheng, G. (2018) Quadratic discriminant analysis under moderate
  dimension.
\newblock \textit{arXiv preprint arXiv:1808.10065}.

\bibitem[{Yao et~al.(2015)Yao, Zheng and Bai}]{yao2015sample}
Yao, J., Zheng, S. and Bai, Z. (2015) \textit{Sample covariance matrices and
  high-dimensional data analysis}, vol.~2.
\newblock Cambridge University Press Cambridge.

\bibitem[{Zhao et~al.(2019)Zhao, Luo, Li, Li, Zhang, Shan, Wang, Yang, Zhou,
  Zhu et~al.}]{zhao2019genome}
Zhao, B., Luo, T., Li, T., Li, Y., Zhang, J., Shan, Y., Wang, X., Yang, L.,
  Zhou, F., Zhu, Z. et~al. (2019) Genome-wide association analysis of 19,629
  individuals identifies variants influencing regional brain volumes and
  refines their genetic co-architecture with cognitive and mental health
  traits.
\newblock \textit{Nature Genetics}, \textbf{51}, 1637--1644.

\bibitem[{Zhao and Zou(2019)}]{Zhao447797}
Zhao, B. and Zou, F. (2019) On prs for complex polygenic trait prediction.
\newblock \textit{bioRxiv}.
\newblock
  \urlprefix\url{https://www.biorxiv.org/content/early/2019/06/04/447797}.

\bibitem[{Zhao and Yu(2006)}]{zhao2006model}
Zhao, P. and Yu, B. (2006) On model selection consistency of lasso.
\newblock \textit{Journal of Machine Learning Research}, \textbf{7},
  2541--2563.

\bibitem[{Zhao et~al.(2018)Zhao, Wang, Hemani, Bowden and
  Small}]{zhao2018statistical}
Zhao, Q., Wang, J., Hemani, G., Bowden, J. and Small, D.~S. (2018) Statistical
  inference in two-sample summary-data mendelian randomization using robust
  adjusted profile score.
\newblock \textit{arXiv preprint arXiv:1801.09652}.

\bibitem[{Zhou et~al.(2013)Zhou, Carbonetto and Stephens}]{zhou2013polygenic}
Zhou, X., Carbonetto, P. and Stephens, M. (2013) Polygenic modeling with
  bayesian sparse linear mixed models.
\newblock \textit{PLoS Genetics}, \textbf{9}, e1003264.

\bibitem[{Zou and Hastie(2005)}]{zou2005regularization}
Zou, H. and Hastie, T. (2005) Regularization and variable selection via the
  elastic net.
\newblock \textit{Journal of the Royal Statistical Society: Series B
  (Statistical Methodology)}, \textbf{67}, 301--320.

\end{thebibliography}
\clearpage
%%%%%%%%%%%%%%%%%%%%%%%%%%%%%%%%%%%%%%%%%%%%%%%%%%%%%%%%
%%%%%%%%%%%%%%%%%%%%%%%%%%%%%%%%%%%%%%%%%%%%%%%%%%%%%%%%
%%%%%%%%%%%%%%%%%%%%%%%%%%%%%%%%%%%%%%%%%%%%%%%%%%%%%%%%
\section{\LARGE Supplementary material}\label{sec9}
%%%%%%%%%%%%%%%%%%%%%%%%%%%%%%%%%%%%%%%%%%%%%%%%%%%%%%%%%%%%%%%%%%%%%%%%%%%%%%%%
%%%%%%%%%%%%%%%%%%%%%%%%%%%%%%%%%%%%%%%%%%%%%%%%%%%%%%%%%%%%%%%%%%%%%%%%%%%%%%%%
\subsection{RMT lemmas}
\label{sec9.1}
We introduce some known results from classic random matrix theory (RMT, e.g., \cite{bai2010spectral,paul2014random,yao2015sample}) and some recent advances of trace functionals (e.g., \cite{ledoit2011eigenvectors,wang2015shrinkage,dobriban2018high,hastie2019surprises}),
which are foundations for our theoretical analysis of the large-scale GWAS data $(\X,\Z)$. Below we mainly use the training data $\X$ as an example, but all the lemmas are applicable for the testing SNP data $\Z$ as well.

The ESD of $\widehat{\bmSigma}=n^{-1}\X^T\X$
is given by $
F^{\widehat{\bmSigma}}_p(x)=p^{-1}\sum^{p}_{i=1}\bmI\big\{\lambda_i(\widehat{\bmSigma})\le x \big\}, x\in \bbR$.
We are interested in the limit behavior of $F^{\widehat{\bmSigma}}_p(x)$, which has one-to-one correspondence with the limit behavior of its Stieltjes transform. 
For a general distribution $G(x)$ with support $I \subset \bbR$, the Stieltjes transform (e.g., page 514 of \cite{bai2010spectral})  and its first order derivative (evaluated at $z$) are given by 
$s_{G}(z)=\int_{x \in I} (x-z)^{-1}d G(x)$ and $ \quad s_{G}^{'}(z)=\int_{x \in I} (x-z)^{-2}d G(x)$,
respectively, for $z \in\bbC \backslash I$.
Therefore, let $I=[0,\infty)$, as $\mbox{min}(n,p) \to \infty$, the Stieltjes transform of $F^{\widehat{\bmSigma}}_p(x)$ and its first order derivative are given by 
$s_{F_p}(z)=p^{-1}\tr \big\{\big(\widehat{\bmSigma}-z\I_p \big)^{-1}\big\}$ and 
$s_{F_p}^{'}(z)=p^{-1}\tr \big\{\big(\widehat{\bmSigma}-z\I_p \big)^{-2}\big\}$,
respectively, for $z \in\bbC \backslash I$.
The asymptotic behavior of $F^{\widehat{\bmSigma}}_p(x)$ can be characterized in the following lemma \citep{marchenko1967distribution,silverstein1995strong} by its Stieltjes transform. See, for example, Theorem~2.4 of \cite{yao2015sample}. 
\begin{lem}
\label{lemma1}
Under Condition~1, as $\mbox{min}(n,p) \to \infty$,  $F^{\widehat{\bmSigma}}_p(x)$ converges weakly to a limit probability distribution $M(x)$ with probability one, $x \in \bbR$.
The Stieltjes transform of $M(x)$, donated as $s_{M}(z)$, is implicitly defined by the Marchenko-Pastur (M-P) equation 
\begin{flalign*}
s_{M}(z)=\int \frac{1}{t\{1-\omega-\omega z s_{M}(z)\}-z} dH(t).
\end{flalign*}
In general,
$s_{M}(z)$ has no closed-form expression, but all information about the LSD $M(x)$ is contained in this equation. 
For the special case $\bmSigma=\I_{p}$, we have $s_{M}(z)=\{1-\omega-\omega z s_{M}(z)-z\}^{-1}$, it follows that (e.g., page 52 of \cite{bai2010spectral}) 
\begin{flalign}
\label{equ2.3.1}
& s_{M}(z)= \frac{\sqrt{(1-\omega-z)^2-4\omega z}-(1-\omega-z)}{-2\omega z},
\quad \mbox{and}\\
&
\label{equ2.3.2}
 s_{M}^{'}(z)=\frac{(\omega-1)\sqrt{(-z-\omega+1)^2-4\omega z}-(\omega+1)z+(\omega-1)^2}{2\omega z^2\sqrt{(-z-\omega+1)^2-4\omega z}},
\end{flalign}
respectively, for $z \in\bbC \backslash I$.
Moreover, the LDS $M(x)$ in this special case is named the M-P law, and its probability density function is given by 
\begin{flalign*}
p(x)_{\omega}=\frac{1}{2\pi\omega x}\cdot \sqrt{\big\{b_{+}(\omega)-x\big\}\big\{x-b_{-}(\omega)\big\}}, 
\end{flalign*}
if $x \in[b_{-}(\omega),b_{+}(\omega)]$; and 
$p(x)_{\omega}=0$ if $x \notin[b_{-}(\omega),b_{+}(\omega)]$; and 
$p(x)_{\omega}$ has a point mass $1-1/\omega$ at the origin if $\omega>1$, where $b_{\pm}(\omega)=(1\pm \sqrt{\omega})^2$.
\end{lem}
Stieltjes transforms can also be used to study the limit of trace functionals of $F^{\widehat{\bmSigma}}_p(x)$.  
Below we donate $g(z)\equiv s_{M}(z)$ and $g^{'}(z) \equiv s_{M}^{'}(z) $, respectively. 
Sometimes, it is more convenient to use the notation defined on the $n\times n$ \textit{companion} matrix  $\widehat{\bmPhi}=n^{-1}\X\X^T$ \citep{dobriban2018high}. 
Let $F^{\widehat{\bmPhi}}_n(x)=n^{-1}\sum^{n}_{i=1}\bmI\big\{\lambda_i(\widehat{\bmPhi})\le x\big\}$ be the ESD of 
$\widehat{\bmPhi}$, $x\in \bbR$, and let $P(x)$ be the limiting distribution of $F^{\widehat{\bmPhi}}_n(x)$. 
Define the Stieltjes transform of $F^{\widehat{\bmPhi}}_n(x)$ and its first order derivative as $
s_{F_n}(z)=n^{-1}\tr \big\{\big(\widehat{\bmPhi}-z\I_n \big)^{-1}\big\}$ and $s_{F_n}^{'}(z)=n^{-1}\tr \big\{\big(\widehat{\bmPhi}-z\I_n \big)^{-2}\big\}$, respectively, and define the Stieltjes transform of $P(x)$ and its first order derivative as $ v(z) \equiv s_{P}(z)$, and $v^{'}(z) \equiv s_{P}^{'}(z) $, respectively. 
We summarize the connections among
$s_{F_p}(z)$, $s^{'}_{F_p}(z)$, 
$g(z)$, $g^{'}(z)$, 
$s_{F_n}(z)$, $s_{F_n}^{'}(z)$,
$v(z)$ and $v^{'}(z)$ in the following lemma \citep{ledoit2011eigenvectors,dobriban2018high}. 
\begin{lem}
\label{lemma2}
Under Condition~1, as $\mbox{min}(n,p) \to \infty$, for any $z \in\bbC \backslash I$, we have 
\begin{gather*}
s_{F_p}(z)=p^{-1} \tr \big\{\big(\widehat{\bmSigma}-z\I_p \big)^{-1}\big\} \to_{a.s.} g(z), \quad
s^{'}_{F_p}(z)=p^{-1} \tr \big\{\big(\widehat{\bmSigma}-z\I_p \big)^{-2}\big\} \to_{a.s.} g^{'}(z),\\
s_{F_n}(z)=n^{-1} \tr \big\{\big(\widehat{\bmPhi}-z\I_n \big)^{-1}\big\} \to_{a.s.} v( z),\quad 
s^{'}_{F_n}(z)=n^{-1} \tr \big\{\big(\widehat{\bmPhi}-z\I_n \big)^{-2}\big\} \to_{a.s.} v^{'}( z),\\
p^{-1} \tr \big\{\bmSigma\big(\widehat{\bmSigma}-z\I_p \big)^{-1}\big\} \to_{a.s.} \frac{1}{\omega}\Big\{ \frac{1}{-zv(z)}-1 \Big\}, \quad
p^{-1} \tr \big\{\bmSigma\big(\widehat{\bmSigma}-z\I_p \big)^{-2}\big\} \to_{a.s.} \frac{v(z)+zv^{'}(z)}{\omega\big\{-zv(z) \big\}^2},\\
\omega \big\{g(z)+z^{-1}\big\}=v(z)+z^{-1},
\quad \mbox{and} \quad
\omega \big\{g^{'}(z)-z^{-2}\big\}=v^{'}(z)-z^{-2}.
\end{gather*}
\end{lem}
In general, little is known about the connection between population LSD $H(t)$ and empirical LDS $M(t)$. However, there is one-to-one correspondence between the moments of $H(t)$ and those of $M(t)$. 
For any positive integer $k$, define the $k$th moment of $H(t)$ as $b_k(\bmSigma)=\int_{\bbR}t^kdH(t)=p^{-1}\tr(\bmSigma^k)$, and the $k$th moment of $M(t)$ as $b_k(\widehat{\bmSigma})=\int_{\bbR}t^kdM(t)=p^{-1}\tr(\widehat{\bmSigma}^k)$. 
Then by Lemma~\ref{lemma1}, we have the following Lemma on the two sets of moments (Lemma~2.16 of \cite{yao2015sample}).
\begin{lem}
\label{lemma3}
Under Condition~1, as $\mbox{min}(n,p) \to \infty$, for any positive integer $k$, $b_k(\widehat{\bmSigma})$ is a function of $b_l(\bmSigma)$, for $0<l \le k$, and $\omega$. Specifically, the first three moments of $H(t)$ and the first three moments of  $M(t)$ are linked as
$b_1(\widehat{\bmSigma})=b_1(\bmSigma)$, 
$b_2(\widehat{\bmSigma})=b_2(\bmSigma)+\omega b_1(\bmSigma)^2$, and
$b_3(\widehat{\bmSigma})=b_3(\bmSigma)+3\omega b_1(\bmSigma)b_2(\bmSigma)+\omega^2 b_1(\bmSigma)^3$. 
Moreover, when $\bmSigma=\I_{p}$, we have $b_k(\bmSigma)\equiv1$ and
$
b_k(\widehat{\bmSigma})=\sum_{r=0}^{k-1}(r+1)^{-1}
\begin{psmallmatrix} 
k  \\
r  
\end{psmallmatrix}
\begin{psmallmatrix} 
k-1  \\
r  
\end{psmallmatrix}
\omega^r.
$
\end{lem}
%More general results can be found in .
%We note that $b_1(\bmSigma)=b_1(\widehat{\bmSigma})=1$ when $\bmSigma_{ii}=1$, $i=1,\cdots,p$. 
For any positive integer $k$,
since $\lambda_i(\bmSigma)$ is uniformly bounded, $i=1,\cdots,p$, 
$b_k(\bmSigma)$ and $b_k(\widehat{\bmSigma})$ are also bounded for any $\omega \in (0,\infty)$. Thus, we have 
the following lemma on the concentration of quadratic forms. 
\begin{lem}
\label{lemma4}
Under Condition~1, as $\mbox{min}(n,p) \to \infty$, for any positive integer $k$, we have  $0<c\le b_k(\bmSigma)\le b_k(\widehat{\bmSigma}) \le C$. In addition, let $\widehat{\bmSigma}_{\X}=n^{-1}\X^T\X$,
$\widehat{\bmSigma}_{\Z}=n_z^{-1}\Z^T\Z$, $\B_{k_1,k_2}=\widehat{\bmSigma}_{\X}^{k_1}\widehat{\bmSigma}_{\Z}^{k_2}$,
and define $\A^0=\I$ for any matrix $\A$. Then 
for any non-negative integers $k_1,k_2$, 
we have 
\begin{flalign*}
\frac{\tr\big( \B_{k_1,k_2}\B_{k_1,k_2}^T\big)}
{\big\{\tr(\B_{k_1,k_2})\big\}^2}=O(\frac{1}{p})=o(1).
\end{flalign*}
Moreover, let $\bmalpha$ be a $p$-dimensional random vector of i.i.d. elements with mean zero, variance $\sigma_{\alpha}^2$, and finite fourth order moment, we have 
\begin{flalign*}
% \var\Big\{\frac{\bmalpha^T\B_{k_1,k_2}\bmalpha }{\tr(\B_{k_1,k_2})}\Big\}=
% \tE\Big[\big\{\frac{\bmalpha^T\B_{k_1,k_2}\bmalpha }{\sigma_{\alpha}^2\tr(\B_{k_1,k_2} )}-1\big\}^2 \Big] =o(1)\quad \mbox{and} \quad
\bmalpha^T\B_{k_1,k_2}\bmalpha=\sigma_{\alpha}^2\cdot\tr(\B_{k_1,k_2} )\cdot\{1+o_p(1)\}.
\end{flalign*}

\end{lem}
The proof of Lemma~\ref{lemma4} is based on the Lemma~B.26 of \cite{bai2010spectral} and the Markov's inequality. 
Lemma~\ref{lemma4} shows that the quadratic forms of $\B_{k_1,k_2}$ concentrate around their means. 
We note that  $\omega \in(0,\infty)$ is a key condition. When $\omega=\infty$, the concentration still holds when either $k_1$ or $k_2$ is zero, but may not hold when both $k_1$ and $k_2$ are  nonzero.

\subsection{Heritability and genetic correlation}
\label{sec9.2}
In GWAS, genetics signal strength and their genetics overlaps are often quantified as heritability and genetic correlation, and are standard measures to report. 
Reliable estimators for $\h_{\beta}^2$, $\h_{\eta}^2$ and  $\varphi_{\beta\eta}$ are available from various models in genetics, using either individual-level data \citep{yang2011gcta,loh2015contrasting} or summary-level data \citep{bulik2015atlas,speed2019sumher}. 
Particularly, 
\cite{jiang2016high} shows that the REML estimator of $\h^2$ is consistent in high-dimensional LMM regardless of $m$, and this estimator $\widehat{h}^2$ (named GREML) has been implemented in the popular genetic tool GCTA (\url{http://cnsgenomics.com/software/gcta/#GREML}). 
The theoretical results on $\h^2$ in \cite{jiang2016high} are built on the special case $\bmSigma=\I_p$, and the $\widehat{h}^2$ might be biased given general $\bmSigma$ \citep{ma2019mahalanobis} or hidden confounding effects \citep{Holmes532069},
though such bias is often believed to be small and acceptable in practice. 
See \cite{yang2017concepts} and \cite{van2019genetic} for overviews of these genetic concepts, and a detailed numerical comparison of population methods in \cite{evans2018comparison}.

\subsection{Relative prediction accuracy}\label{sec9.3}
In this section, we study the relative prediction accuracy of marginal estimator compared to the optimal ridge estimator. 
We focus on the case $\bmSigma=\I_p$, in which we can quantify the efficiency loss in closed-form expressions. 
%We have the following corollary. 
\begin{cor.s}\label{cor2}
Let $R_R(\h_{\beta}^2,\omega)=A^2_{R}(\lambda^{*})/A^2_S$. 
Under polygenic model~(\ref{equ1.2.1}) and Conditions~\ref{con1} and~\ref{con2}, as $\mbox{min}(n, n_z,m_{\beta\eta},p)\rightarrow\infty$, for any $\h_{\beta}^2, \h_{\eta}^2\in (0,1]$, $\varphi_{\beta\eta} \in [-1,1]$, and $\bmSigma=\I_p$,
we have 
\begin{flalign*}
R_R(\h_{\beta}^2,\omega)&=
\frac{(\omega+\h_{\beta}^2)\big\{(\omega+\h^2)-\sqrt{(\omega+\h_{\beta}^2)^2-4\omega \h_{\beta}^4}\big\}}{2\h_{\beta}^4\omega}+o_p(1)
> 1+o_p(1).
\end{flalign*}
Moreover, for any fixed $\omega \in (0,\infty)$, $\diff{R}{\h_{\beta}^2}R_B(\h_{\beta}^2,\omega)>0$ on $\h_{\beta}^2 \in (0,1)$; and for any given $\h_{\beta}^2\in (0,1]$, we have 
\begin{flalign*}
\diff{R}{\omega}R_R(\h_{\beta}^2,\omega)  \left\{ 
\begin{array}{lll}
>0, & \mbox{if \quad $0<\omega<\h_{\beta}^2$;}\\ 
=0, & \mbox{if \quad$\omega=\h_{\beta}^2$;} \\
<0, & \mbox{if \quad $\omega>\h_{\beta}^2$.} \\
\end{array} \right .
\end{flalign*}
\end{cor.s}
%\begin{remark}\label{rmk10}
Corollary~S\ref{cor2} shows that 
$\widehat{\bmbeta}_{R}(\lambda^*)$ always has better asymptotic out-of-sample $R^2$ than $\widehat{\bmbeta}_{S}$.
$R_R(\h_{\beta}^2,\omega)$ is higher for larger $\h_{\beta}^2$ and is not a monotone function of $\omega$. 
For given $\h_{\beta}^2$, $R_R(\h_{\beta}^2,\omega)$ is maximized at $\omega=\h_{\beta}^2$, with the maximum value 
\begin{flalign*}
R_R^*(\h_{\beta}^2,\omega)=\frac{2-2\sqrt{1-\h_{\beta}^2}}{\h_{\beta}^2}+o_p(1).
\end{flalign*}
$R_R^*(\h_{\beta}^2,\omega)$ is an increasing function of $\h_{\beta}^2$ on $\h_{\beta}^2 \in (0,1]$ and the maximun value is $2$ at $\h_{\beta}^2=1$. 
That is, for a fully heritable trait and $p=n$ in the training GWAS, we have $R_B(\h_{\beta}^2,\omega)=A^2_{R}(\lambda^{*})/A^2_S=2$. This represents the difference between $n/p$ and $n/(n+p)$. 
As $\omega$ becomes large, $R_B(\h_{\beta}^2,\omega)$ decreases, which can be viewed as a blessing of dimensionality due to the fact that the difference between $n/p$ and $n/(n+p)$ decreases. 
%\end{remark}
Another interesting question is the relative prediction accuracy between $\widehat{\bmbeta}_{S}$ and  $\widehat{\bmbeta}_{R}(0^{+})$, which is quantified in the following corollary. 
\begin{cor.s}\label{cor3}
Let $R_{R^0}(\h_{\beta}^2,\omega)=A^2_{R}(0^{+})/A^2_S$. 
Under polygenic model~(\ref{equ1.2.1}) and Conditions~\ref{con1} and~\ref{con2}, as $\mbox{min}(n, n_z,m_{\beta\eta},p)\rightarrow\infty$, for any $\h_{\beta}^2, \h_{\eta}^2\in (0,1]$, $\varphi_{\beta\eta} \in [-1,1]$, and $\bmSigma=\I_p$,
we have 
\begin{flalign*}
R_{R^0}(\h_{\beta}^2,\omega)&=
\left\{ 
\begin{array}{lll}
\frac{\h_{\beta}^2+\frac{\omega}{1-\omega}\cdot(1-\h_{\beta}^2)}{\h_{\beta}^2+\omega}+o_p(1), & \mbox{if \quad $\omega<1$;}\\ 
\frac{\h_{\beta}^2\omega+(1-\h_{\beta}^2)\frac{\omega^2}{\omega-1}}{\h_{\beta}^2+\omega}+o_p(1), & \mbox{if \quad $\omega>1$.} \\
\end{array} \right .
\end{flalign*}
It follows that for $\h_{\beta}^2\in (0,1/2]$, we have 
\begin{flalign*}
R_{R^0}(\h_{\beta}^2,\omega)&  \left\{ 
\begin{array}{lll}
>1+o_p(1), & \mbox{if \quad $\omega<\h_{\beta}^2$;}\\ 
=1+o_p(1), & \mbox{if \quad$\omega=\h_{\beta}^2$;} \\
<1+o_p(1), & \mbox{if \quad $\omega>\h_{\beta}^2$;} \\
\end{array} \right .
\end{flalign*}
and for $\h_{\beta}^2\in (1/2,1)$, we have 
\begin{flalign*}
R_{R^0}(\h_{\beta}^2,\omega)&  \left\{ 
\begin{array}{lll}
>1+o_p(1), & \mbox{if \quad $\omega<\h_{\beta}^2$\quad or \quad$\omega>\h_{\beta}^2/(2\h_{\beta}^2-1)$;}\\ 
=1+o_p(1), & \mbox{if \quad$\omega=\h_{\beta}^2$\quad or \quad$\omega=\h_{\beta}^2/(2\h_{\beta}^2-1)$;} \\
<1+o_p(1), & \mbox{if \quad $\h_{\beta}^2<\omega<\h_{\beta}^2/(2\h_{\beta}^2-1)$.} \\
\end{array} \right .
\end{flalign*}
And if $\h_{\beta}^2=1$, then $R_{R^0}(\h_{\beta}^2,\omega)>1+o_p(1)$ for any $\omega$.
\end{cor.s}
As $\widehat{\bmbeta}_{R}(0^{+})$ reduces to $\widehat{\bmbeta}_{O}$ when $\omega<1$, our results indicate that $\widehat{\bmbeta}_{S}$ can have better out-of-sample $R^2$ than $\widehat{\bmbeta}_{O}$ when $1>\omega>\h_{\beta}^2$. 
Thus, $\widehat{\bmbeta}_{S}$ can easily outperform $\widehat{\bmbeta}_{O}$ when $\h_{\beta}^2$ is low. 
Moreover, if $\h_{\beta}^2\le 0.5$,  $\widehat{\bmbeta}_{R}(0^{+})$ is worse than $\widehat{\bmbeta}_{S}$ for $1<\omega$. 
If $\h_{\beta}^2>0.5$, however, $\widehat{\bmbeta}_{R}(0^{+})$ is better when $\omega$ is large.

%%%%%%%%%%%%%%%%%%%%%%%%%%%%%%%%%%%%%%%%%%%%%%%%%%%%%%
%%%%%%%%%%%%%%%%%%%%%%%%%%%%%%%%%%%%%%%%%%%%%%%%%%%%%%
%%%%%%%%%%%%%%%%%%%%%%%%%%%%%%%%%%%%%%%%%%%%%%%%%%%%%%
\subsection{Mean squared prediction errors}
\label{sec9.4}
In this section, we study the MSE of marginal estimator and illustrate the bias-variance trade-off of each estimator. 
We focus on the same trait prediction case in which $\bmbeta=\bmeta$.
The MSE and bias-variance decomposition (e.g., \cite{hastie2019surprises}) of a generic $p\times 1$ estimator $\widehat{\bmbeta}$ trained on GWAS dataset $(\X,\y)$ can be defined as 
\begin{flalign*}
M^2&=\tE\big\{\Vert\widehat{\bmbeta}-\bmbeta\Vert_{\bmSigma}^2|\X,\y \big\}=
\tE\big\{\big(\widehat{\bmbeta}-\bmbeta\big)^T\bmSigma\big(\widehat{\bmbeta}-\bmbeta\big)|\X,\y \big\} \\
&=\big\{\tE\big(\widehat{\bmbeta}|\X,\y\big)-\bmbeta\big\}^T\bmSigma
\big\{\tE\big(\widehat{\bmbeta}|\X,\y\big)-\bmbeta\big\}
+\tr\big[\cov\big(\widehat{\bmbeta}|\X,\y\big)\bmSigma \big]
\equiv
B^2+V^2,
\end{flalign*}
where $B^2$ represents the squared bias of $\widehat{\bmbeta}$, and $V^2$ measures the total variance of the $\widehat{\bmbeta}$ due to the random error term. 
We define
$M^2_{S}=B^2_{S}+V^2_{S}$,
$M^2_{R}(\lambda)=B^2_{R}(\lambda)+V^2_{R}(\lambda)$,
$M^2_{R}(0^{+})=B^2_{R}(0^{+})+V^2_{R}(0^{+})$,
$M^2_{O}=B^2_{O}+V^2_{O}$, 
$M^2_{B}(\tau)= B^2_{B}(\tau)+V^2_{B}(\tau)$, and
$M^2_{B}(0^{+})=B^2_{B}(0^{+})+V^2_{B}(0^{+})$,
for marginal, ridge, ridge-less, OLS, BLUP and BLUP-less estimators, respectively. 

\begin{proposition.s}
\label{prop1}
Under polygenic model~(\ref{equ1.2.1}) and Conditions~\ref{con1} and~\ref{con2},
as $\mbox{min}(n, n_z,m_{\beta},p)$ $\rightarrow\infty$, for any $\h_{\beta}^2 \in (0,1]$ and $\bm\bmSigma$, we have 
\begin{flalign*}
&M^2_{S}=
\Big[ 
m\sigma_{\beta}^2/p \cdot \big\{ \omega b_2(\bmSigma)+b_3(\bmSigma)-2b_2(\bmSigma)+1 \big\}+
\sigma^2_{\epsilon}\cdot\omega b_2(\bmSigma) \Big]\cdot\{1+o_p(1)\},\\
&M^2_{R}(\lambda)=M^2_{B}(\lambda/\omega)=\Big[m\sigma_{\beta}^2/p \cdot \frac{v(-\lambda)-\lambda v^{'}(-\lambda)}{v(-\lambda)^2\omega} +\sigma^2_{\epsilon}\cdot\big\{\frac{v^{'}(-\lambda)}{v(-\lambda)^2}-1 \big\}\Big]\cdot\{1+o_p(1)\},\\
&M^2_{R}(0^{+})=M^2_{B}(0^{+})=\Big[ m\sigma_{\beta}^2/p \cdot \frac{1}{v(0^{+})\omega}+
\sigma^2_{\epsilon}\cdot\big\{\frac{v^{'}(0^{+})}{v(0^{+})^2}-1 \big\}\Big]\cdot\{1+o_p(1)\}, \quad \mbox{and}\\
&M^2_{O}=\sigma^2_{\epsilon}\cdot \frac{\omega}{1-\omega} \cdot\{1+o_p(1)\} \quad (\omega<1).
\end{flalign*}
Moreover, if $\h_{\beta}^2\in (0,1)$, $M^2_{R}(\lambda)$ is minimized at 
$\lambda^*$ with the minimize value  
\begin{flalign*}
M^2_{R}(\lambda^*)=M^2_{B}(\lambda^*/\omega)
=m\sigma^2_{\beta}/p\cdot \Big\{ \frac{1}{v(-\lambda^{*})\omega}-\frac{\lambda^{*}}{\omega} \Big\} \cdot\{1+o_p(1)\}.
\end{flalign*}
And if $\h_{\beta}^2=1$, $M^2_{R}(\lambda)$ is minimized as $\lambda \to 0^{+}$, and thus the minimize value is $M^2_{R}(0^{+})$.
\end{proposition.s}

Supplementary Figure~\ref{sfig4} illustrates the different patterns of MSE for optimal ridge, ridge-less/OLS, and marginal estimators when $\bmSigma=\I_p$. 
For OLS estimator $\widehat{\bmbeta}_{O}$, we have $B^2_{O}=0$, but $V^2_{O}=\sigma^2_{\epsilon}\cdot\omega/(1-\omega)$ can become unbounded as $\omega \to 1^{-}$. 
Similar issue occurs for ridge-less and BLUP-less estimators as $\omega \to 1^{+}$. 
On the other hand, the MSE of marginal estimator $\widehat{\bmbeta}_{S}$ is linear in $\omega$, and thus can be much larger than those of other ridge-type estimators when $\omega$ is large. 
Specifically, $V^2_{S}$ is linear in $\omega$ and is always smaller than $V^2_{O}$. 
However, $B^2_{S}$ is also linear in $\omega$, and thus the MSE of $\widehat{\bmbeta}_{S}$ linearly grows up with $\omega$.
When $\bmSigma=\I_p$, we have the following closed-form expressions on MSE of ridge-type estimators. 

%%%%%%%%%%%%%%%%%%%%%%%%%%%%%%%%%%%%%%%%%%%%%%%%%%%%%%%%%%%%%%%%%%%%
%%%%%%%%%%%%%%%%%%%%%%%%%%%%%%%%%%%%%%%%%%%%%%%%%%%%%%%%%%%%%%%%%%%%
%%%%%%%%%%%%%%%%%%%%%%%%%%%%%%%%%%%%%%%%%%%%%%%%%%%%%%%%%%%%%%%%%%%%

\begin{proposition.s}
\label{prop.s1}
Under polygenic model~(\ref{equ1.2.1}) and Conditions~\ref{con1} and~\ref{con2},
as $\mbox{min}(n, n_z,m_{\beta},p)\rightarrow\infty$, when $\bmSigma=\I_p$, for any $\h_{\beta}^2\in (0,1)$, we have 
\begin{flalign*}
&M^2_{S}=\big(m\sigma_{\beta}^2/p \cdot \omega+\sigma^2_{\epsilon}\cdot\omega\big)\cdot\{1+o_p(1)\},\\
&M^2_{R}(\lambda)=M^2_{B}(\lambda/\omega)=\Big[m\sigma_{\beta}^2/p \cdot \lambda^2g^{'}(-\lambda)+\sigma^2_{\epsilon}\cdot\omega\big\{g(-\lambda)-\lambda g^{'}(-\lambda)\big\}\Big]\cdot\{1+o_p(1)\},
\end{flalign*}
and
\begin{flalign*}
M^2_{R}(0^{+})=M^2_{B}(0^{+})&=\Big[ m\sigma_{\beta}^2/p \cdot \frac{(\omega-1)+|\omega-1|}{2\omega}+
\sigma^2_{\epsilon}\cdot\frac{\omega+1-|\omega-1|}{2|\omega-1|} \Big]\cdot\{1+o_p(1)\}\\
&=\left\{ 
\begin{array}{lll}
M^2_{O}, & \mbox{if \quad $\omega<1$;}\\ 
\Big\{m\sigma_{\beta}^2/p \cdot \frac{\omega-1}{\omega}+
\sigma^2_{\epsilon}\cdot\frac{1}{\omega-1} \Big\}\cdot\{1+o_p(1)\}, & \mbox{if \quad $\omega>1$.} \\
\end{array} \right .
\end{flalign*}
Moreover, for $\h_{\beta}^2\in (0,1)$, 
$M^2_{R}(\lambda)$ is minimized at 
$\lambda^*$ with the minimize value  
\begin{flalign*}
M^2_{R}(\lambda^*)&=M^2_{B}(\lambda^*/\omega)= m\sigma^2_{\beta}\cdot \lambda^{*}g(-\lambda^{*})\\
&=
m\sigma^2_{\beta}/p\cdot  \frac{2\omega\h_{\beta}^2+\sqrt{(\omega-\h_{\beta}^2)^2+4\omega\h_{\beta}^2(1-\h_{\beta}^2)}-\omega-\h_{\beta}^2}{2\omega\h_{\beta}^2}\cdot\{1+o_p(1)\}.
\end{flalign*}
For $\h_{\beta}^2=1$, $M^2_{R}(\lambda)$ is minimized at $\lambda^* \to 0^{+}$, and the minimize value is 
\begin{flalign*}
M^2_{R}(0^{+})=
\left\{ 
\begin{array}{lll}
o_p(1), & \mbox{if \quad $\omega<1$;}\\ 
m\sigma_{\beta}^2/p \cdot \frac{\omega-1}{\omega} \cdot\{1+o_p(1)\}, & \mbox{if \quad $\omega>1$.} \\
\end{array} \right .
\end{flalign*}
\end{proposition.s}

\begin{proposition.s}
\label{prop.s2}
Under the same conditions as in Proposition~S\ref{prop.s1}, we have for any $\omega$
\begin{flalign*}
\frac{M^2_{S}}{M^2_{R}(\lambda^*)}=\frac{1}{(1-\h_{\beta}^2)g(-\lambda^*)}=\frac{2\omega^2}{\sqrt{(\omega+\h_{\beta}^2)^2-4\h_{\beta}^4\omega}+\h_{\beta}^2(2\omega-1)-\omega}>1+o_p(1).
\end{flalign*}
Moreover, we have 
\begin{flalign*}
\frac{M^2_{S}}{M^2_{O}}=\frac{1-\omega}{1-\h_{\beta}^2}
 \left\{ 
\begin{array}{lll}
>1+o_p(1), & \mbox{if \quad $0<\omega<\h_{\beta}^2$;}\\ 
=1+o_p(1), & \mbox{if \quad$\omega=\h_{\beta}^2$;} \\
<1+o_p(1), & \mbox{if \quad $\omega>\h_{\beta}^2$.} \\
\end{array} \right .
\end{flalign*}
When $\h_{\beta}^2=1$, we have $M^2_{S}/M^2_{O}>1+o_p(1)$ and $M^2_{S}/{M^2_{B}(0^{+})}=M^2_{S}/{M^2_{R}(0^{+})}>1+o_p(1)$ for any $\omega$.
\end{proposition.s}
%%%%%%%%%%%%%%%%%%%%%%%%%%%%%%%%%%%%%%%%%%%%%%%%%%%%%%
%%%%%%%%%%%%%%%%%%%%%%%%%%%%%%%%%%%%%%%%%%%%%%%%%%%%%%
%%%%%%%%%%%%%%%%%%%%%%%%%%%%%%%%%%%%%%%%%%%%%%%%%%%%%%
\subsection{Relative goodness-of-fit}\label{sec9.5}
The following corollary provides the comparison between $E^2_{R}(\lambda^{*})$ and $E^2_S$, and some interesting properties of $E^2_S$.
\begin{cor.s}\label{cor4}
Let $Q_R(\h^2,\omega)=E^2_{R}(\lambda^{*})/E^2_S$. 
Under polygenic model~(\ref{equ1.2.1}) and Conditions~\ref{con1} and~\ref{con2}, as $\mbox{min}(n, n_z,m_{\beta},p)\rightarrow\infty$, for any $\h^2_{\beta}\in (0,1]$, $\omega \in (0,\infty)$, and $\bmSigma=\I_p$, 
we have 
\begin{flalign*}
Q_R(\h^2,\omega)
> 1+o_p(1).
\end{flalign*}
Moreover, if $\h^2\in (0,0.5]$,
we have 
\begin{flalign*}
\diff{E}{\omega}E^2_S(\h^2_{\beta},\omega)> 1+o_p(1).
\end{flalign*}
which indicates that $E^2_S(\h^2_{\beta},\omega)$ is a monotone function of $\omega$. 
If $\h^2_{\beta}\in (0.5,1]$, however, we have 
\begin{flalign*}
\diff{E}{\omega}E^2_S(\h^2_{\beta},\omega)  \left\{ 
\begin{array}{lll}
<0, & \mbox{if \quad $0<\omega<\h^2_{\beta}\cdot(2\h^2_{\beta}-1)$;}\\ 
=0, & \mbox{if \quad$\omega=\h^2_{\beta}\cdot(2\h^2_{\beta}-1)$;} \\
>0, & \mbox{if \quad $\omega>\h^2_{\beta}\cdot(2\h^2_{\beta}-1)$,} \\
\end{array} \right .
\end{flalign*}
and $E^2_S(\h^2_{\beta},\omega)=4\h^4_{\beta}/(4\h^4_{\beta}+1)+o_p(1)$ at $\omega=\h^2_{\beta}\cdot(2\h^2_{\beta}-1)$.
%we have 
% \begin{flalign*}
% E^2_S=\frac{}{4\h^4_{\beta}+1}. 
% \end{flalign*}
\end{cor.s}
%%%%%%%%%%%%%%%%%%%%%%%%%%%%%%%%%%%%%%%%%%%%%%%%%%%%%%
%%%%%%%%%%%%%%%%%%%%%%%%%%%%%%%%%%%%%%%%%%%%%%%%%%%%%%
%%%%%%%%%%%%%%%%%%%%%%%%%%%%%%%%%%%%%%%%%%%%%%%%%%%%%%
\subsection{Lemmas}\label{sec9.6}
\subsubsection{Marginal estimator }
\paragraph{Proof of Lemma~\ref{lemma4} (concentration of quadratic forms)}\mbox{}\\
Let $\bmalpha$ be a $p$-dimensional random vector of i.i.d. elements with mean zero and finite fourth order moment, and $
\A$ be a fixed $p\times p$ matrix.
Without loss of generality, we let $\tE(\alpha_1^2)=1$.
Then, by Lemma~B.26 of \cite{bai2010spectral}, for any $q\ge 1$, we have 
 \begin{flalign*}
\tE\Big[\big\{\bmalpha^T\A\bmalpha- \tr(\A)\big\}^q \Big] \le 
c_q \cdot \Big[ \big\{\tE(\alpha_1^4)\tr(\A\A^T)\big\}^{q/2} +\tE(\alpha_1^{2q})\tr\{(\A\A^T)^{q/2}\}
\Big],
\end{flalign*}
where $c_q$ is some constant that only depends on $q$. Let $q=2$, it follows that 
\begin{flalign*}
\var\Big\{\frac{\bmalpha^T\A\bmalpha }{\tr(\A)}\Big\}=
\tE\Big[\big\{\frac{\bmalpha^T\A\bmalpha }{\tr(\A)}-1\big\}^2 \Big] 
\le 
c_q \cdot \Big[ \frac{\tE(\alpha_1^4)\tr(\A\A^T)}{\tr(\A)^2}
+\frac{\tE(\alpha_1^4)\tr(\A\A^T)}{\tr(\A)^2}
\Big].
\end{flalign*}
Let $\B_{k_1,k_2}=\widehat{\bmSigma}_{\X}^{k_1}\widehat{\bmSigma}_{\Z}^{k_2}$, $\widehat{\bmSigma}_{\X}=n^{-1}\X^T\X$, and
$\widehat{\bmSigma}_{\Z}=n_z^{-1}\Z^T\Z$. Then, for bounded $\omega$ and any $\bmSigma$ with uniformaly bounded eigenvalues, we have  
\begin{flalign*}
\tr(\B_{k_1,k_2})=p\cdot b_1(\widehat{\bmSigma}_{\X}^{k_1}\widehat{\bmSigma}_{\Z}^{k_2}) \le 
p\cdot b_{{k_1+k_2}}(\widehat{\bmSigma}_{\X})=O(p).
\end{flalign*}
Then, we have 
\begin{flalign*}
\frac{\tr\big(\B_{k_1,k_2}\B_{k_1,k_2}^T\big)}{\big\{\tr(\B_{k_1,k_2})\big\}^2}=\frac{\tr(\B_{2k_1,2k_2})}{\big\{\tr(\B_{k_1,k_2})\big\}^2}=O(\frac{p}{p^2})=O(\frac{1}{p})=o(1).
\end{flalign*}
It follows that 
\begin{flalign*}
\var\Big\{\frac{\bmalpha^T\B_{k_1,k_2}\bmalpha }{\tr(\B_{k_1,k_2})}\Big\}=
\tE\Big[\big\{\frac{\bmalpha^T\B_{k_1,k_2}\bmalpha }{\sigma_{\alpha}^2\tr(\B_{k_1,k_2} )}-1\big\}^2 \Big] =o(1).
\end{flalign*}
Thus, by Markov's inequality, we have 
\begin{flalign*}
\bmalpha^T\B_{k_1,k_2}\bmalpha=\sigma_{\alpha}^2\cdot \tr(\B_{k_1,k_2} )\cdot\{1+o_p(1)\}.
\end{flalign*}

\subsubsection{Useful trace results for ridge-type estimators}
Here we summarize some results that are used frequently in our analysis of ridge-type estimators, which are based on Lemma~\ref{lemma2}.
\begin{lem.s}
\label{lemma.s1}
Under Condition~\ref{con1} of the main paper, as $\mbox{min}(n,p) \to \infty$, for any $\lambda>0$, we have 
\begin{gather*}
\tr\big\{\bmSigma (\X^T\X +\lambda n \I_p)^{-1} \big\} \to_{a.s.} \frac{1}{\lambda v(-\lambda)}-1,\quad \mbox{and}\\
\tr\big\{\bmSigma (\X^T\X +\lambda n \I_p)^{-2} \big\} \to_{a.s.} \frac{1}{n}\cdot \frac{v(-\lambda)-\lambda v^{'}(-\lambda)}{(\lambda v(-\lambda))^2}  
\end{gather*}
When $\bmSigma=\I_p$, we have closed-form limits 
\begin{gather*}
\tr\big\{ (\X^T\X +\lambda n \I_p)^{-1} \big\} \to_{a.s.} \omega g(-\lambda)=
\frac{\sqrt{(1-\omega+\lambda)^2+4\omega\lambda}-(1-\omega+\lambda)}{2\lambda}
, \quad \mbox{and} \\
\tr\big\{ (\X^T\X +\lambda n \I_p)^{-2} \big\} \to_{a.s.} \frac{p}{n^2}\cdot g^{'}(-\lambda)=
\frac{p}{n^2}\cdot \frac{(\omega-1)+\frac{(\omega+1)\lambda+(\omega-1)^2}{\sqrt{(1-\omega+\lambda)^2+4\omega\lambda}}}{2\omega\lambda^2}.  \\
\end{gather*}
Moreover, at the optimal $\lambda^*=\omega\cdot(1-\h^2_{\beta})/\h^2_{\beta}$, we have 
\begin{flalign*}
A^2_R(\lambda^*)&=\h^2_{\eta}\varphi^2_{\beta\eta} \cdot \{1-\lambda^* g(-\lambda^*)\} +o_p(1)\\
&=\h^2_{\eta}\varphi^2_{\beta\eta} \cdot
\frac{\omega+\h^2_{\beta}-\sqrt{(\h^2_{\beta}-2\omega \h^2_{\beta}+\omega)^2+4\omega^2\h^2_{\beta}(1-\h^2_{\beta})}}{2\omega\h^2_{\beta}} +o_p(1)\\
&=\h^2_{\eta}\varphi^2_{\beta\eta} \cdot\frac{\omega+\h^2_{\beta}-\sqrt{(\omega+\h^2_{\beta})^2-4\omega\h^4_{\beta}}}{2\omega\h^2_{\beta}} +o_p(1)\\
&=\h^2_{\eta}\varphi^2_{\beta\eta} \cdot\frac{\omega+\h^2_{\beta}-\sqrt{(\omega-\h^2_{\beta})^2+4\omega\h^2_{\beta}(1-\h^2_{\beta})}}{2\omega\h^2_{\beta}} +o_p(1).
\end{flalign*}
\end{lem.s}

\begin{lem.s}
\label{lemma.s2}
Under Condition~\ref{con1} of the main paper, when $\bmSigma=\I_p$, as $\mbox{min}(n,p) \to \infty$, 
$\lambda \to 0^{+}$, we have the following closed-form limits 
% \begin{gather*}
% \tr\big\{\bmSigma (\X^T\X +\lambda n \I_p)^{-1} \big\} \to_{a.s.} \frac{1}{\lambda v(0^{+})}-1,\quad \mbox{and}\\
% \tr\big\{\bmSigma (\X^T\X +\lambda n \I_p)^{-2} \big\} \to_{a.s.} \frac{1}{n}\cdot \frac{v(0^{+})-\lambda v^{'}(0^{+})}{(\lambda v(0^{+}))^2}  ,
% \end{gather*}
%where $v(0^{+})=\lim_{\lambda\to 0^+} v(-\lambda)$, and $v^{'}(0^{+})=\lim_{\lambda\to 0^+} v^{'}(-\lambda)$. 
\begin{gather*}
g(-\lambda) \to_{a.s.} \frac{\frac{1+\omega}{|1-\omega|}-1}{2\omega}, \quad \lambda g(-\lambda) \to_{a.s.} \frac{|1-\omega|-(1-\omega)}{2\omega} \\
\lambda g^{'}(-\lambda) \to_{a.s.} \frac{2\lambda}{\{(1-\omega+\lambda)^2+4\omega\lambda\}^{3/2}} =0, \quad \mbox{and} \quad
\lambda^2 g^{'}(-\lambda) \to_{a.s.} \frac{(\omega-1)+|1-\omega|}{2\omega} 
\end{gather*}
%Particular, when $\omega <1$, we have $\lim_{\lambda\to 0^+} \tr\big\{\bmSigma (\X^T\X +\lambda n \I_p)^{-1} \big\}=\tr\big\{\bmSigma (\X^T\X)^{-1} \big\} \to_{a.s.} \omega/(1-\omega)$ \citep{hastie2019surprises}.
\end{lem.s}

%%%%%%%%%%%%%%%%%%%%%%%%%%%%%%%%%%%%%%%%%%%%%%%%%%%%
The following lemma is used to show the equivalence of prediction accuracy of ridge estimator and BLUP, which can be easily proved by applying singular value decomposition on $\X$. 
\begin{lem.s}
\label{lemma.s3}
Under Condition~\ref{con1} of the main paper, as $\mbox{min}(n,p) \to \infty$, for any
$\lambda>0$ and arbitrary $\bmSigma$, we have 
\begin{gather*}
\tr\big\{ (\X\X^T +\lambda \I_n)^{-1} \X\bmSigma\X^T \big\}=
\tr\big\{ (\X^T\X +\lambda \I_p)^{-1} \X^T\X\bmSigma \big\} \quad \mbox{and} \\
\tr\big\{ (\X\X^T +\lambda \I_n)^{-2} \X\bmSigma\X^T \big\}=
\tr\big\{ (\X^T\X +\lambda \I_p)^{-2} \X^T\X\bmSigma \big\}.
\end{gather*}
\end{lem.s}

%%%%%%%%%%%%%%%%%%%%%%%%%%%%%%%%%%%%%%%%%%%%%%%%%%%%
%%%%%%%%%%%%%%%%%%%%%%%%%%%%%%%%%%%%%%%%%%%%%%%%%%%%
%%%%%%%%%%%%%%%%%%%%%%%%%%%%%%%%%%%%%%%%%%%%%%%%%%%%
\subsection{Proofs of marginal estimator}\label{sec9.7}
\paragraph{Out-of-sample}
\begin{proposition.s}\label{pop.s1}
Under polygenic model~(\ref{equ1.2.1}) and Conditions~\ref{con1} and~\ref{con2}, as $\mbox{min}(n$, $n_z$, $m_{\beta\eta}$, $p)\rightarrow\infty$,
for any $\omega \in (0,\infty)$, $\h_{\beta}^2, \h_{\eta}^2\in (0,1]$, $\varphi_{\beta\eta} \in [-1,1]$, and $\bmSigma$, we have 
\begin{flalign*}
&\frac{\big(\Z_{(1)}\bmeta_{(1)}+\bmeps_{z}\big)^T\big(\Z_{(1)}\bmeta_{(1)}+\bmeps_{z}\big)}{n_zm_{\eta}\cdot{\sigma^2_{\eta}}/p+n_z\cdot{\sigma^2_{\epsilon_{z}}}} =1 + o_p(1),  \\
&\frac{\big(\X_{(1)}\bmbeta_{(1)}+\bmeps\big)^T\X\Z^T\Z\X^T\big(\X_{(1)}\bmbeta_{(1)}+\bmeps\big)}{nn_zm_{\beta}\cdot \{(n+1)b_3(\bmSigma)+pb_2(\bmSigma)\}\cdot{\sigma^2_{\beta}}/p+nn_zpb_2(\bmSigma)\cdot{\sigma^2_{\epsilon}}} = 1 + o_p(1),
\end{flalign*}
and
\begin{flalign*}
\frac{\big(\Z_{(1)}\bmeta_{(1)}+\bmeps_{z}\big)^T\Z\X^T\big(\X_{(1)}\bmbeta_{(1)}+\bmeps\big)}{nn_zm_{\beta\eta}b_2(\bmSigma)\cdot{\sigma_{\beta\eta}}/p} =1 + o_p(1)\mathbf{.}
\end{flalign*}
\end{proposition.s}
By continuous mapping theorem, we have
\begin{flalign*}
A^2_{S}=\h_{\eta}^2\varphi_{\beta\eta}^2\cdot
\Big\{\frac{b_3(\bmSigma)}{b_2(\bmSigma)^2} + \frac{\omega}{\h_{\beta}^2} \cdot \frac{1}{b_2(\bmSigma)} \Big\}^{-1}
+ o_p(1).
\end{flalign*}
% When $\X$ and $\Z$ have different population-level correlation matrices $\bmSigma_{X}$ and $\bmSigma_{Z}$, then we replace $b_3(\bmSigma)$ by $b_1(\bmSigma_X\bmSigma_Z\bmSigma_X)$, and 
% $b_2(\bmSigma)$ by $b_1(\bmSigma_X\bmSigma_Z)$ in the above proposition. 
%%%%%%%%%%%%%%%%%%%%%%%%%%%%%%%%%%%%%%%%%%%%%%%%%%%%
\paragraph{Meta-analysis}\mbox{}\\
Under polygenic model~(\ref{equ1.2.1}) and Conditions~\ref{con1} and~\ref{con2},
suppose we have independent GWAS $(\X_i,\y_i)$, with sample sizes $(n_i,\cdots,n_k)$ and $p$ SNPs, $i=1,\cdots,k$, $k \in (0,\infty)$, let  $\widehat{\B}=[\widehat{\bmbeta}_1^T,\cdots,\widehat{\bmbeta}_k^T]$ be the $p\times k$ matrix of marginal estimators from the $k$ GWAS. Let $\bmd=(d_i,\cdots,d_k)^T$ be an $k\times 1$ vector of weights, and let $\widehat{\B}(\bmd)=\widehat{\B}\bmd$
be the aggregated summary statistics.
As $\mbox{min}(n_1$,$\cdots,n_k$, $n_z$, $m_{\beta\eta}$, $p)\rightarrow\infty$,
for any $\omega \in (0,\infty)$, $\h_{\beta}^2,\h_{\eta}^2 \in (0,1]$, $\varphi_{\beta\eta}\in [-1,1]$, $\bmd$, and $\bmSigma$, we have 
\begin{flalign*}
A^2_{S}(\bmd)&=\Big\{ \frac{\big(\sum_{i=1}^{k} d_i \widehat{\bmbeta}_i \big)^T\Z^T\big(\Z\eta+\bmeps_{z} \big) }{\Vert\Z\eta+\bmeps_{z} \Vert \Vert \Z\widehat{B}\bmd \Vert}\Big\}^2
=\h_{\eta}^2\varphi_{\beta\eta}^2\cdot
\frac{\big(\sum_{i=1}^{k} d_i \big)^2 \cdot b_2(\bmSigma)^2}{\big(\sum_{i=1}^{k} d_i \widehat{\bmbeta}_i \big)^T\Z^T\Z \big(\sum_{i=1}^{k} d_i \widehat{\bmbeta}_i \big)}+o_p(1).
\end{flalign*}
Note that for $i\neq j, (i, j) \in (1,\cdots,k)$, we have 
\begin{flalign*}
&\big(\Z\widehat{\bmbeta}_i\big)^T\big(\Z\widehat{\bmbeta}_j\big)=\frac{1}{n_in_j}\tE\big(\bmbeta^T\X_i^T\X_i\Z^T\Z\X_j^T\X_j\bmbeta\big)\cdot\{1+o_p(1)\}
=m_{\beta}n_z\cdot b_3(\bmSigma)\cdot \sigma_{\beta}^2/p\cdot\{1+o_p(1)\} 
\end{flalign*}
and
\begin{flalign*}
\big(\Z\widehat{\bmbeta}_i\big)^T\big(\Z\widehat{\bmbeta}_i\big)&=\frac{1}{n_i^2}\tE\big(\bmbeta^T\X_i^T\X_i\Z^T\Z\X_i^T\X_i\bmbeta+\bmeps_i^T\X_i^T\Z^T\Z\X_i^T\bmeps_i \big)\cdot\{1+o_p(1)\}\\
&=m_{\beta}n_z\cdot \big\{b_3(\bmSigma) + \frac{p}{n_i\h^2_{\beta}}b_2(\bmSigma)  \big\}\cdot \sigma_{\beta}^2/p\cdot\{1+o_p(1)\}. 
\end{flalign*}
It follows that 
\begin{flalign*}
A^2_{S}(\bmd)&=
\h_{\eta}^2\varphi_{\beta\eta}^2\cdot
\frac{\big(\sum_{i=1}^{k} d_i \big)^2 \cdot b_2(\bmSigma)^2}
{\sum_{i=1}^{k} d_i^2\cdot\{b_3(\bmSigma)+\frac{p}{n_i\h^2_{\beta}}b_2(\bmSigma) \} +2 \sum_{i\neq j}^{(i, j) \in (1,\cdots,k)} d_id_j \cdot b_3(\bmSigma)}+o_p(1)\\
&=\h_{\eta}^2\varphi_{\beta\eta}^2\cdot
\frac{\big(\sum_{i=1}^{k} d_i \big)^2 \cdot  b_2(\bmSigma)^2}
{\big(\sum_{i=1}^{k} d_i \big)^2\cdot b_3(\bmSigma) + \big(\sum_{i=1}^{k} d_i^2\frac{p}{n_i\h_{\beta}^2} \big)\cdot b_2(\bmSigma)}+o_p(1)\\
&=\h_{\eta}^2\varphi_{\beta\eta}^2\cdot
\Big\{ 
\frac{b_3(\bmSigma)}{b_2(\bmSigma)^2}+
\frac{\sum_{i=1}^k d_i^2/n_i}{(\sum_{i=1}^k d_i)^2}\cdot \frac{p}{b_2(\bmSigma)\h^2_{\beta}}
\Big\}^{-1}+o_p(1).
\end{flalign*}
Therefore, when $d_i=n_i$, we have 
\begin{flalign*}
A^2_{S}(\bma^*)=\h_{\eta}^2\varphi_{\beta\eta}^2\cdot \Big\{\frac{b_3(\bmSigma)}{b_2(\bmSigma)^2} +
\frac{p}{ \sum_{i=1}^k n_i} \cdot\frac{1}{b_2(\bmSigma)h^2_{\beta}}
\Big\}^{-1}+o_p(1).
\end{flalign*}
%%%%%%%%%%%%%%%%%%%%%%%%%%%%%%%%%%%%%%%%%%%%%%%%%%%%
\paragraph{In-sample}
\begin{proposition.s}\label{pop.s2}
Under polygenic model~(\ref{equ1.2.1}) and Conditions~\ref{con1} and~\ref{con2}, as $\mbox{min}(n$, $n_z$, $m_{\beta}$, $p)\rightarrow\infty$,
for any $\omega \in (0,\infty)$, $\h^2_{\beta} \in (0,1]$, and $\bmSigma$, we have 
\begin{flalign*}
&\frac{\big(\X_{(1)}\bmbeta_{(1)}+\bmeps\big)^T\big(\X_{(1)}\bmbeta_{(1)}+\bmeps\big)}{nm_{\beta}\cdot{\sigma^2_{\beta}}/p+n\cdot{\sigma^2_{\epsilon}}} =1 + o_p(1),  \\
&\frac{\big(\X_{(1)}\bmbeta_{(1)}+\bmeps\big)^T\X\X^T\X\X^T\big(\X_{(1)}\bmbeta_{(1)}+\bmeps\big)}{n^3m_{\beta}\cdot \{b_3(\bmSigma)+3\omega b_2(\bmSigma)+\omega^2\}\cdot{\sigma^2_{\beta}}/p+n^2p\cdot \{b_2(\bmSigma)+\omega\}\cdot{\sigma^2_{\epsilon}}} = 1 + o_p(1),
\end{flalign*}
and
\begin{flalign*}
\frac{\big(\X_{(1)}\bmbeta_{(1)}+\bmeps\big)^T\X\X^T\big(\X_{(1)}\bmbeta_{(1)}+\bmeps\big)}{n^2m_{\beta}\{b_2(\bmSigma)+\omega\}\cdot{\sigma^2_{\beta}}/p+np\cdot{\sigma^2_{\epsilon}}} =1 + o_p(1)\mathbf{.}
\end{flalign*}
\end{proposition.s}
By continuous mapping theorem, we have
\begin{flalign*}
E^2_{S}=
\frac{\{b_2(\bmSigma)\h^2_{\beta}+\omega\}^2}{\{b_2(\bmSigma)\h^2_{\beta}+\omega\}^2+b_2(\bmSigma)\omega+\{b_3(\bmSigma)-b_2(\bmSigma)^2\h^2_{\beta}\}\h^2_{\beta}}+o_p(1).
\end{flalign*}
For the special case $\bmSigma=\I_p$, we have $b_3(\bmSigma)=b_2(\bmSigma)=b_1(\bmSigma)=1$ in the above 
propositions.
%%%%%%%%%%%%%%%%%%%%%%%%%%%%%%%%%%%%%%%%%%%%%%%%%%%%
%%%%%%%%%%%%%%%%%%%%%%%%%%%%%%%%%%%%%%%%%%%%%%%%%%%%
%%%%%%%%%%%%%%%%%%%%%%%%%%%%%%%%%%%%%%%%%%%%%%%%%%%%
\subsection{Proofs of ridge-type estimators}\label{sec9.8}
\paragraph{OLS estimator, out-of-sample}
\begin{proposition.s}\label{pop.s3}
Under polygenic model~(\ref{equ1.2.1}) and Conditions~\ref{con1} and~\ref{con2}, as $\mbox{min}(n$, $n_z$, $m_{\beta\eta}$, $p)\rightarrow\infty$,
for any $\omega \in (0,\infty)$, $\h_{\beta}^2, \h_{\eta}^2\in (0,1]$, $\varphi_{\beta\eta} \in [-1,1]$, and $\bmSigma$, we have 
\begin{flalign*}
&\frac{\big(\X_{(1)}\bmbeta_{(1)}+\bmeps\big)^T\X(\X^T\X)^{-1}\Z^T\Z(\X^T\X)^{-1}\X^T\big(\X_{(1)}\bmbeta_{(1)}+\bmeps\big)}{n_zm_{\beta}\cdot{\sigma^2_{\beta}}/p+n_z\omega/(1-\omega)\cdot{\sigma^2_{\epsilon}}} = 1 + o_p(1),
\end{flalign*}
and
\begin{flalign*}
\frac{\big(\Z_{(1)}\bmeta_{(1)}+\bmeps_{z}\big)^T\Z(\X^T\X)^{-1}\X^T\big(\X_{(1)}\bmbeta_{(1)}+\bmeps\big)}{n_zm_{\beta\eta}\cdot{\sigma_{\beta\eta}}/p} =1 + o_p(1)\mathbf{.}
\end{flalign*}
\end{proposition.s}
By continuous mapping theorem, we have
\begin{flalign*}
&A^2_{O}=\h_{\eta}^2\varphi_{\beta\eta}^2\cdot 
\Big\{ 1+\frac{1-\h_{\beta}^2}{\h_{\beta}^2}\cdot \frac{\omega}{1-\omega} \Big\}^{-1}
+o_p(1)
\mathbf{.} \quad (\omega<1)
\end{flalign*}
%%%%%%%%%%%%%%%%%%%%%%%%%%%%%%%%%%%%%%%%%%%%%%%%%%%%
\paragraph{OLS estimator, in-sample}
\begin{proposition.s}\label{pop.s4}
Under polygenic model~(\ref{equ1.2.1}) and Conditions~\ref{con1} and~\ref{con2}, as $\mbox{min}(n$, $n_z$, $m_{\beta}$, $p)\rightarrow\infty$,
for any $\omega \in (0,\infty)$, $\h^2_{\beta} \in (0,1]$, and $\bmSigma$, we have 
\begin{flalign*}
&\frac{\big(\X_{(1)}\bmbeta_{(1)}+\bmeps\big)^T\X(\X^T\X)^{-1}\X^T\X(\X^T\X)^{-1}\X^T\big(\X_{(1)}\bmbeta_{(1)}+\bmeps\big)}{nm_{\beta}\cdot{\sigma^2_{\beta}}/p+p\cdot{\sigma^2_{\epsilon}}} = 1 + o_p(1),
\end{flalign*}
and
\begin{flalign*}
\frac{\big(\X_{(1)}\bmbeta_{(1)}+\bmeps\big)^T\X(\X^T\X)^{-1}\X^T\big(\X_{(1)}\bmbeta_{(1)}+\bmeps\big)}{nm_{\beta}\cdot{\sigma^2_{\beta}/p}+p\cdot{\sigma^2_{\epsilon}}} =1 + o_p(1)\mathbf{.}
\end{flalign*}
\end{proposition.s}
By continuous mapping theorem, we have
\begin{flalign*}
E^2_{O}=\big\{\h^2_{\beta}+\omega(1-h^2_{\beta})\big\} \cdot\{1+o_p(1)\}
\mathbf{.} \quad (\omega<1)
\end{flalign*}
%%%%%%%%%%%%%%%%%%%%%%%%%%%%%%%%%%%%%%%%%%%%%%%%%%%%
%%%%%%%%%%%%%%%%%%%%%%%%%%%%%%%%%%%%%%%%%%%%%%%%%%%%
\paragraph{Ridge estimator, out-of-sample}
\begin{proposition.s}\label{pop.s5}
Under polygenic model~(\ref{equ1.2.1}) and Conditions~\ref{con1} and~\ref{con2}, as $\mbox{min}(n$, $n_z$, $m_{\beta\eta}$, $p)\rightarrow\infty$,
for any $\omega \in (0,\infty)$, $\h_{\beta}^2, \h_{\eta}^2\in (0,1]$, $\varphi_{\beta\eta} \in [-1,1]$, and $\bmSigma$, we have 
\begin{flalign*}
&\frac{\big(\X_{(1)}\bmbeta_{(1)}+\bmeps\big)^T\X(\X^T\X+\lambda n \I_p)^{-1}\Z^T\Z(\X^T\X+\lambda n \I_p)^{-1}\X^T\big(\X_{(1)}\bmbeta_{(1)}+\bmeps\big)}{V_{R1}+V_{R2}} = 1 + o_p(1),
\end{flalign*}
where 
\begin{flalign*}
V_{R1}&=n_z m_{\beta}\cdot \Big[ 1-\frac{2\lambda}{\omega}\tr\{\bmSigma (\X^T\X+\lambda n \I_p)^{-1} \}+ \frac{\lambda^2 n}{\omega}\cdot \tr\{\bmSigma (\X^T\X+\lambda n \I_p)^{-2} \}\Big]\cdot \sigma^2_{\beta}/p\\
&=n_zm_{\beta}\cdot \Big[1- \frac{2\lambda}{\omega} \big\{\frac{1}{\lambda v(-\lambda)}-1 \big\}+\frac{\lambda^2}{\omega}\cdot \frac{v(-\lambda)-\lambda v^{'}(-\lambda)}{\{\lambda v(-\lambda)\}^2}
\Big] \cdot \sigma^2_{\beta}/p,
\end{flalign*}
and
\begin{flalign*}
V_{R2}&=n_z\cdot \Big[\tr\{\bmSigma (\X^T\X+\lambda n \I_p)^{-1} \} + \lambda n \cdot \tr\{\bmSigma (\X^T\X+\lambda n \I_p)^{-2} \}] \cdot \sigma^2_{\epsilon}\\
&=n_z\cdot \Big[\big\{ \frac{1}{\lambda v(-\lambda)} -1\big\}- \lambda \cdot \frac{v(-\lambda)-\lambda v^{'}(-\lambda)}{\{\lambda v(-\lambda)\}^2}
\Big] \cdot \sigma^2_{\epsilon},
\end{flalign*}
In addition, we have 
\begin{flalign*}
\frac{\big(\Z_{(1)}\bmeta_{(1)}+\bmeps_{z}\big)^T\Z(\X^T\X+\lambda n \I_p)^{-1}\X^T\big(\X_{(1)}\bmbeta_{(1)}+\bmeps\big)}{C_{R1}} =1 + o_p(1),
\end{flalign*}
where 
\begin{flalign*}
C_{R1}&=n_zm_{\beta\eta}\cdot [1-\frac{\lambda}{\omega}\cdot \tr\{\bmSigma (\X^T\X+\lambda n \I_p)^{-1} \} ]\cdot{\sigma_{\beta\eta}}/p\\
&=n_zm_{\beta\eta}\cdot [1-\frac{\lambda}{\omega}\cdot \{ \frac{1}{\lambda v(-\lambda)}-1 \} ]\cdot{\sigma_{\beta\eta}}/p.
\end{flalign*}
\end{proposition.s}
By continuous mapping theorem, we have
\begin{flalign*}
&A^2_{R}(\lambda)= \h_{\eta}^2\varphi_{\beta\eta}^2\cdot
\frac{\big[1+\frac{\lambda}{\omega}\{ 1-\frac{1}{\lambda v(-\lambda)}\} \big]^2\cdot \h_{\beta}^2}
{\h_{\beta}^2\cdot\Big[1+\frac{\lambda}{\omega}\big\{2-\frac{1}{\lambda v(-\lambda)}- \frac{ v^{'}(-\lambda)}{v(-\lambda)^2}\big\}\Big]
+(1-\h_{\beta}^2)   \cdot \big\{\frac{v^{'}(-\lambda)}{v(-\lambda)^2}-1 \big\}}
+o_p(1).
\end{flalign*}

Similar to Theorem~2.1 of \cite{dobriban2018high}, $A^2_{R}(\lambda)$ is optimized at 
$\lambda^*=\omega\cdot(1-\h_{\beta}^2)/\h_{\beta}^2$, where the second order term $v^{'}(-\lambda)$ disappears.  In Theorem~2.1 of \cite{dobriban2018high}, they set $\gamma=p/n$, and the signal to noise ratio to $\alpha^2$, thus, their optimal $\lambda$ is $\lambda^*=\gamma\alpha^{-2}$.
The $A^2_{R}(0^{+})$ can be obtained by taking $\lambda \to 0^{+}$, with careful exchanging limits as $n,p \to \infty$ and $\lambda \to 0^{+}$, detailed in Threorem~4 of \cite{hastie2019surprises}.
When $\bmSigma=\I_p$, using the results in Lemmas~S\ref{lemma.s1}~and~S\ref{lemma.s2}, we have closed-form expressions for $A^2_{R}(\lambda)$, $A^2_{R}(\lambda^*)$, and $A^2_{R}(0^{+})$. 

%%%%%%%%%%%%%%%%%%%%%%%%%%%%%%%%%%%%%%%%%%%%%%%%%%%%
\paragraph{Ridge estimator, in-sample}
\begin{proposition.s}\label{pop.s6}
Under polygenic model~(\ref{equ1.2.1}) and Conditions~\ref{con1} and~\ref{con2}, as $\mbox{min}(n$, $n_z$, $m_{\beta}$, $p)\rightarrow\infty$,
for any $\omega \in (0,\infty)$, $\h^2_{\beta} \in (0,1]$ and $\bmSigma$, we have 
\begin{flalign*}
&\frac{\big(\X_{(1)}\bmbeta_{(1)}+\bmeps\big)^T\X(\X^T\X+\lambda n \I_p)^{-1}\X^T\X(\X^T\X+\lambda n \I_p)^{-1}\X^T\big(\X_{(1)}\bmbeta_{(1)}+\bmeps\big)}{V_{R3}+V_{R4}} = 1 + o_p(1),
\end{flalign*}
where 
\begin{flalign*}
V_{R3}&=nm_{\beta}\cdot \Big[1- \frac{2\lambda}{p}\cdot\tr\big\{\X^T\X (\X^T\X+\lambda n \I_p)^{-1} \big\} +
\frac{\lambda^2}{\omega} \cdot\tr\big\{\X^T\X (\X^T\X+\lambda n \I_p)^{-2} \big\} \Big]    \cdot \sigma^2_{\beta}/p\\
&=nm_{\beta}\cdot \big\{1- 2\lambda+3\lambda^2g(-\lambda)-\lambda^3g^{'}(-\lambda)
\big\} \cdot \sigma^2_{\beta}/p
\end{flalign*}
and
\begin{flalign*}
V_{R4}&=p \cdot \Big[1-\frac{2\lambda}{\omega}\cdot\tr\big\{(\X^T\X+\lambda n \I_p)^{-1}  \big\} +
\frac{\lambda^2 n}{\omega}\cdot\tr\big\{(\X^T\X+\lambda n \I_p)^{-2} \big\} \Big] \cdot \sigma^2_{\epsilon}\\
&=p \cdot \{ 1-2\lambda g(-\lambda)+\lambda^2 g^{'}(-\lambda)\} \cdot \sigma^2_{\epsilon}.
\end{flalign*}
In addition, we have 
\begin{flalign*}
\frac{\big(\X_{(1)}\bmbeta_{(1)}+\bmeps\big)^T\X(\X^T\X+\lambda n \I_p)^{-1}\X^T\big(\X_{(1)}\bmbeta_{(1)}+\bmeps\big)}
{C_{R2}} =1 + o_p(1),
\end{flalign*}
where 
\begin{flalign*}
C_{R2}&= nm_{\beta}\cdot \Big[1- \frac{\lambda}{p}\cdot\tr\big\{\X^T\X (\X^T\X+\lambda n \I_p)^{-1} \big\} \Big]\cdot{\sigma_{\beta}^2}/p+ \tr\big\{\X^T\X (\X^T\X+\lambda n \I_p)^{-1} \big\} \cdot \sigma^2_{\epsilon} \\
&=nm_{\beta}\cdot \{1- \lambda+\lambda^2 g(-\lambda)\}\cdot{\sigma_{\beta}^2}/p+ p\cdot \{1-\lambda g(-\lambda) \} \cdot \sigma^2_{\epsilon}. 
\end{flalign*}
\end{proposition.s}
By continuous mapping theorem, we have
\begin{flalign*}
&E^2_{R}(\lambda)=
\frac{\Big[\h^2_{\beta}\cdot \big\{1-\lambda+\lambda^2g(-\lambda)\big\} +(1-\h^2_{\beta})\cdot \omega  \big\{ 1-\lambda g(-\lambda)\big\} \Big]^2}
{ \h^2_{\beta}\cdot\big\{1-2\lambda+3\lambda^2g(-\lambda)-\lambda^3 g^{'}(-\lambda) \big\}
+(1-\h^2_{\beta})\cdot \omega \big\{1-2\lambda+\lambda^2g^{'}(-\lambda)\big\}}
+o_p(1).
\end{flalign*}
Different from $A^2_{R}(\lambda)$, $E^2_{R}(\lambda)$ is minimized as $\lambda \to 0$, which means the over-fitting of the model in high-dimensions.
%%%%%%%%%%%%%%%%%%%%%%%%%%%%%%%%%%%%%%%%%%%%%%%%%%%%
%%%%%%%%%%%%%%%%%%%%%%%%%%%%%%%%%%%%%%%%%%%%%%%%%%%%
%%%%%%%%%%%%%%%%%%%%%%%%%%%%%%%%%%%%%%%%%%%%%%%%%%%%
\subsection{Proofs of MSE}\label{sec9.9}
The main work of studying the out-of-sample MSE of ridge and ridge-less estimators has been done in \cite{dobriban2018high} and \cite{hastie2019surprises}, which use the key results of \cite{ledoit2011eigenvectors}.
\subsubsection*{Marginal estimator}
\begin{flalign*}
B^2_S&=\big\{\tE\big(\widehat{\bmbeta}_S|\X,\y\big)-\bmbeta\big\}^T\bmSigma
\big\{\tE\big(\widehat{\bmbeta}_S|\X,\y\big)-\bmbeta\big\}&\\
&=\big(n^{-1}\X^T\X\bmbeta-\bmbeta\big)^T\bmSigma\big(n^{-1}\X^T\X\bmbeta-\bmbeta\big)&\\
&=n^{-2}\bmbeta^T\X^T\X\bmSigma\X^T\X\bmbeta
+\bmbeta^T\bmSigma\bmbeta
-2n^{-1}\bmbeta^T\bmSigma\X^T\X\bmbeta&\\
%&=n^{-2}m_{\beta}\cdot[n\{(n+1)b_3(\bmSigma)+p b_2(\bmSigma) \}]\cdot \sigma^2_{\beta}+m_{\beta}\sigma^2_{\beta}-2n^{-1}m_{\beta}\sigma^2_{\beta}n b_2(\bmSigma)&\\
&=\Big[m_{\beta}\cdot\big\{ b_3(\bmSigma)+\omega b_2(\bmSigma)-2b_2(\bmSigma)+1\big\}\cdot \sigma^2_{\beta}/p\Big]\cdot\{1+o_p(1)\}.
\end{flalign*}

\begin{flalign*}
V^2_S&=\tr\big\{\cov\big(\widehat{\bmbeta}_S|\X,\y\big)\bmSigma \big\}&\\
&=\tr\big[\{\tE\big(\widehat{\bmbeta}_S\widehat{\bmbeta}_S^T|\X,\y\big)-
\tE\big(\widehat{\bmbeta}_S|\X,\y\big)\tE\big(\widehat{\bmbeta}_S|\X,\y\big)^T\big \}\bmSigma]\\
&=\tr\Big[ 
\big[\tE\big\{n^{-1}\X^T(\X\bmbeta+\bmeps)\cdot n^{-1}(\X\bmbeta+\bmeps)^T\X\}-
(n^{-1}\X^T\X\bmbeta)(n^{-1}\X^T\X\bmbeta)^T
\big]\bmSigma  
\Big]\\
&=n^{-2} \tr\big(\X^T\bmeps\bmeps^T\X\bmSigma\big)& \\
&=n^{-2} \tr\big( \bmeps^T\X\bmSigma\X^T\bmeps\big)&\\
&=\big\{\omega b_2(\bmSigma)\cdot  \sigma^2_{\epsilon} \big\}\cdot\{1+o_p(1)\}.
\end{flalign*}
\subsubsection*{OLS estimator}

\begin{flalign*}
B^2_O&=\big\{\tE\big(\widehat{\bmbeta}_O|\X,\y\big)-\bmbeta\big\}^T\bmSigma
\big\{\tE\big(\widehat{\bmbeta}_O|\X,\y\big)-\bmbeta\big\}&\\
&=\big(\bmbeta-\bmbeta\big)^T\bmSigma\big(\bmbeta-\bmbeta\big)=0.&
\end{flalign*}

\begin{flalign*}
V^2_O&=\tr\big\{\cov\big(\widehat{\bmbeta}_O|\X,\y\big)\bmSigma \big\}&\\
&=\tr\big[\{\tE\big(\widehat{\bmbeta}_O\widehat{\bmbeta}_O^T|\X,\y\big)-
\tE\big(\widehat{\bmbeta}_O|\X,\y\big)\tE\big(\widehat{\bmbeta}_O|\X,\y\big)^T\big \}\bmSigma]\\
&=
\tr\big\{(\X^T\X)^{-1}\X^T\bmeps 
\bmeps^T\X(\X^T\X)^{-1}\bmSigma\big\}& \\
&=\tr\{(\X^T\X)^{-1}\bmSigma \} \cdot \sigma^2_{\epsilon}&\\
&=\frac{\omega}{1-\omega} \cdot \sigma^2_{\epsilon} \cdot\{1+o_p(1)\}. \quad (\omega <1)
\end{flalign*}
\subsubsection*{Ridge estimator}

\begin{flalign*}
B^2_R(\lambda)&=\big[\tE\big\{\widehat{\bmbeta}_R(\lambda)|\X,\y\big\}-\bmbeta\big]^T\bmSigma
\big[\tE\big\{\widehat{\bmbeta}_R(\lambda)|\X,\y\big\}-\bmbeta\big]&\\
&=\big\{(\X^T\X+\lambda n \I_p)^{-1}\X^T\X\bmbeta-\bmbeta
\big\}^T\bmSigma
\big\{(\X^T\X+\lambda n \I_p)^{-1}\X^T\X\bmbeta-\bmbeta
\big\}&\\
&=\bmbeta^T\big\{(\X^T\X+\lambda n \I_p)^{-1}\X^T\X-\I_p
\big\}^T\bmSigma\big\{(\X^T\X+\lambda n \I_p)^{-1}\X^T\X-\I_p
\big\}\bmbeta&\\
&=\Big[m_{\beta}\cdot \big\{ 1+2\lambda\omega^{-1}-\frac{1}{v(-\lambda)\omega}-\frac{\lambda v^{'}(-\lambda)}{v(-\lambda)^2\omega} \big\}\cdot \sigma^2_{\beta}/p+
m_{\beta}\cdot\sigma^2_{\beta}/p&\\
&\qquad-2m_{\beta}\cdot\big\{1-\frac{1}{v(-\lambda)\omega}+\frac{\lambda}{\omega} \big\}\cdot\sigma^2_{\beta}/p\Big]\cdot\{1+o_p(1)\}. 
&\\
&=m_{\beta}\cdot \frac{v(-\lambda)-\lambda v^{'}(-\lambda)}{v(-\lambda)^2 \omega}\cdot \sigma^2_{\beta}/p\cdot\{1+o_p(1)\}. 
\end{flalign*}

\begin{flalign*}
V^2_R(\lambda)&=\tr\big[\cov\big\{\widehat{\bmbeta}_R(\lambda)|\X,\y\big\}\bmSigma \big]&\\
&=\tr\big\{
(\X^T\X+\lambda n \I_p)^{-1}\X^T\bmeps \bmeps^T\X (\X^T\X+\lambda n \I_p)^{-1}
\bmSigma  
\big\}\\
&=\tr\big\{
(\X^T\X+\lambda n \I_p)^{-1}\X^T\X (\X^T\X+\lambda n \I_p)^{-1}
\bmSigma  
\big\} \cdot \sigma^2_{\epsilon}\cdot\{1+o_p(1)\}&\\
&= \big\{ \frac{1}{\lambda v(-\lambda)}-1-\frac{v(-\lambda)-\lambda v(-\lambda)^{'}}{\lambda v(-\lambda)^2} \big\} \cdot \sigma^2_{\epsilon}\cdot\{1+o_p(1)\}&\\
&= \big\{ \frac{v(-\lambda)^{'}}{v(-\lambda)^2} -1\big\} \cdot \sigma^2_{\epsilon}\cdot\{1+o_p(1)\}.  &
\end{flalign*}
Again, similar to Theorem~2.1 of \cite{dobriban2018high} and Theorem~4 of \cite{hastie2019surprises}, $M^2_R(\lambda)$ is optimized at 
$\lambda^*=\omega\cdot(1-\h_{\beta}^2)/\h_{\beta}^2$. 
The $B^2_R(0^{+})$ and $V^2_R(0^{+})$ can be obtained by taking $\lambda \to 0^{+}$. 
When $\bmSigma=\I_p$, using the results in Lemmas~S\ref{lemma.s1}~and~S\ref{lemma.s2}, we have closed-form expressions for $M^2_{R}(\lambda)$, $M^2_{R}(\lambda^*)$, and $M^2_{R}(0^{+})$. 
%%%%%%%%%%%%%%%%%%%%%%%%%%%%%%%%%%%%%%%%%%%%%%%%%%%%%%%%
%%%%%%%%%%%%%%%%%%%%%%%%%%%%%%%%%%%%%%%%%%%%%%%%%%%%%%%%
%%%%%%%%%%%%%%%%%%%%%%%%%%%%%%%%%%%%%%%%%%%%%%%%%%%%%%%%
\subsection{Intermediate results}\label{sec9.10}
%%%%%%%%%%%%%%%%%%%%%%%%%%%%%%%%%%%%%%%%%%%%%%%%%%%%%%%%
%%%%%%%%%%%%%%%%%%%%%%%%%%%%%%%%%%%%%%%%%%%%%%%%%%%%%%%%
%%%%%%%%%%%%%%%%%%%%%%%%%%%%%%%%%%%%%%%%%%%%%%%%%%%%%%%%
\subsubsection*{Marginal estimator}
\begin{proposition.s}\label{pop.i1}
Under polygenic model~(\ref{equ1.2.1}) and Conditions~\ref{con1} and~\ref{con2}, as $\mbox{min}(n$, $n_z$, $m_{\beta\eta}$, $p)\rightarrow\infty$,
for any $\omega \in (0,\infty)$, $\h_{\beta}^2, \h_{\eta}^2\in (0,1]$, $\varphi_{\beta\eta} \in [-1,1]$, and $\bmSigma$, we have 
\begin{flalign*}
&\tE \Big\{\big(\Z_{(1)}\bmeta_{(1)}+\bmeps_{z}\big)^T\Z\X^T\big(\X_{(1)}\bmbeta_{(1)}+\bmeps\big) \Big\} = nn_zm_{\beta\eta}b_2(\bmSigma)\cdot{\sigma_{\beta\eta}}/p, &\\
&\tE\Big\{\big(\Z_{(1)}\bmeta_{(1)}+\bmeps_{z}\big)^T\big(\Z_{(1)}\bmeta_{(1)}+\bmeps_{z}\big)\Big\}=n_zm_{\eta}\cdot{\sigma^2_{\eta}}/p+n_z\cdot{\sigma^2_{\epsilon_{z}}}, &\\
&\tE\Big\{\big(\X_{(1)}\bmbeta_{(1)}+\bmeps\big)^T\X\Z^T\Z\X^T\big(\X_{(1)}\bmbeta_{(1)}+\bmeps\big)\Big\}&\\
&=nn_zm_{\beta}\cdot \{(n+1)b_3(\bmSigma)+pb_2(\bmSigma)\}\cdot{\sigma^2_{\beta}}/p+nn_zpb_2(\bmSigma)\cdot{\sigma^2_{\epsilon}},&\\~\\
%%%%%
&\tE \Big\{\big(\X_{(1)}\bmbeta_{(1)}+\bmeps\big)^T\X\X^T\big(\X_{(1)}\bmbeta_{(1)}+\bmeps\big) \Big\} = n^2m_{\beta}\{b_2(\bmSigma)+\omega\}\cdot{\sigma^2_{\beta}}/p+np\cdot{\sigma^2_{\epsilon}}, &\\
&\tE\Big\{\big(\X_{(1)}\bmbeta_{(1)}+\bmeps\big)^T\big(\X_{(1)}\bmbeta_{(1)}+\bmeps\big)\Big\}=
nm_{\beta}\cdot{\sigma^2_{\beta}}/p+n\cdot{\sigma^2_{\epsilon}}, \quad \mbox{and} &\\
&\tE\Big\{\big(\X_{(1)}\bmbeta_{(1)}+\bmeps\big)^T\X\X^T\X\X^T\big(\X_{(1)}\bmbeta_{(1)}+\bmeps\big)\Big\}&\\
&=n^3m_{\beta}\cdot \{b_3(\bmSigma)+3\omega b_2(\bmSigma)+\omega^2\}\cdot{\sigma^2_{\beta}}/p+n^2p\cdot \{b_2(\bmSigma)+\omega\}\cdot{\sigma^2_{\epsilon}},&
\end{flalign*}
Then Propositions~S\ref{pop.s1}~and~S\ref{pop.s2} follow from Lemma~\ref{lemma4}, the concentration of marginal estimator quadratic forms. 
\end{proposition.s}
% We note that the uniform boundness of $\lambda_i(\bmSigma)$ is not a necessary condition to have the asymptotic limits of $A^2_S$ and $E^2_S$. 
% Given boundness condition on high order moments of $H(t)$, i.e., $b_6(\bmSigma)<\infty$, both $A^2_S$ and $E^2_S$ can have asymptotic limits by Markov's inequality. However, the uniform boundness of $\lambda_i(\bmSigma)$ in Condition~\ref{con1} is required for ridge-less and Blup-less estimators in which $\lambda \to 0^+$, see \cite{hastie2019surprises} for more details. 

\subsubsection*{Ridge-type estimators}
\begin{proposition.s}\label{pop.i2}
Under polygenic model~(\ref{equ1.2.1}) and Conditions~\ref{con1} and~\ref{con2}, as $\mbox{min}(n$, $n_z$, $m_{\beta\eta}$, $p)\rightarrow\infty$,
for any $\omega \in (0,\infty)$, $\h_{\beta}^2, \h_{\eta}^2\in (0,1]$, $\varphi_{\beta\eta} \in [-1,1]$, and $\bmSigma$, we have almost surely that 
\begin{flalign*}
&\tE \Big\{\big(\Z_{(1)}\bmeta_{(1)}+\bmeps_{z}\big)^T\Z(\X^T\X)^{-1}\X^T\big(\X_{(1)}\bmbeta_{(1)}+\bmeps\big) \Big\} = n_zm_{\beta\eta}\cdot{\sigma_{\beta\eta}}/p, &\\
&\tE\Big\{\big(\X_{(1)}\bmbeta_{(1)}+\bmeps\big)^T\X(\X^T\X)^{-1}\Z^T\Z(\X^T\X)^{-1}\X^T\big(\X_{(1)}\bmbeta_{(1)}+\bmeps\big)
\Big\}&\\
&=n_zm_{\beta}\cdot{\sigma^2_{\beta}}/p+n_z\omega/(1-\omega)\cdot{\sigma^2_{\epsilon}},&\\~\\
%%%%%
&\tE \Big\{\big(\X_{(1)}\bmbeta_{(1)}+\bmeps\big)^T\X(\X^T\X)^{-1}\X^T\big(\X_{(1)}\bmbeta_{(1)}+\bmeps\big) \Big\} = nm_{\beta}\cdot{\sigma^2_{\beta}}/p+p\cdot{\sigma^2_{\epsilon}}, &\\
&\tE\Big\{\big(\X_{(1)}\bmbeta_{(1)}+\bmeps\big)^T\X(\X^T\X)^{-1}\X^T\X(\X^T\X)^{-1}\X^T\big(\X_{(1)}\bmbeta_{(1)}+\bmeps\big)
\Big\}=nm_{\beta}\cdot{\sigma^2_{\beta}}/p+p\cdot{\sigma^2_{\epsilon}},&
%%%%%
\end{flalign*}
\begin{flalign*}
&\tE \Big\{\big(\Z_{(1)}\bmeta_{(1)}+\bmeps_{z}\big)^T\Z(\X^T\X+\lambda n \I_p)^{-1}\X^T\big(\X_{(1)}\bmbeta_{(1)}+\bmeps\big)
\Big\}&\\
&= n_zm_{\beta\eta}\cdot [1-\frac{\lambda}{\omega}\cdot \{ \frac{1}{\lambda v(-\lambda)}-1 \} ]\cdot{\sigma_{\beta\eta}}/p, &\\
&\tE\Big\{\big(\X_{(1)}\bmbeta_{(1)}+\bmeps\big)^T\X(\X^T\X+\lambda n \I_p)^{-1}\Z^T\Z(\X^T\X+\lambda n \I_p)^{-1}\X^T\big(\X_{(1)}\bmbeta_{(1)}+\bmeps\big)
\Big\}&\\
&=n_zm_{\beta}\cdot \Big[1- \frac{2\lambda}{\omega} \big\{\frac{1}{\lambda v(-\lambda)}-1 \big\}+\frac{\lambda^2}{\omega}\cdot \frac{v(-\lambda)-\lambda v^{'}(-\lambda)}{\{\lambda v(-\lambda)\}^2}
\Big] \cdot \sigma^2_{\beta}/p+\\
&\qquad n_z\cdot \Big[\big\{ \frac{1}{\lambda v(-\lambda)} -1\big\}- \lambda \cdot \frac{v(-\lambda)-\lambda v^{'}(-\lambda)}{\{\lambda v(-\lambda)\}^2}
\Big] \cdot \sigma^2_{\epsilon},&
\end{flalign*}

\begin{flalign*}
&\tE \Big\{\big(\X_{(1)}\bmbeta_{(1)}+\bmeps\big)^T\X(\X^T\X+\lambda n \I_p)^{-1}\X^T\big(\X_{(1)}\bmbeta_{(1)}+\bmeps\big)
\Big\}&\\
&=nm_{\beta}\cdot \{1- \lambda+\lambda^2 g(-\lambda)\}\cdot{\sigma_{\beta}^2}/p+ p\cdot \{1-\lambda g(-\lambda) \} \cdot \sigma^2_{\epsilon}, \quad \mbox{and}  &\\
&\tE\Big\{\big(\X_{(1)}\bmbeta_{(1)}+\bmeps\big)^T\X(\X^T\X+\lambda n \I_p)^{-1}\X^T\X(\X^T\X+\lambda n \I_p)^{-1}\X^T\big(\X_{(1)}\bmbeta_{(1)}+\bmeps\big)
\Big\}&\\
&=nm_{\beta}\cdot \big\{1- 2\lambda+3\lambda^2g(-\lambda)-\lambda^3g^{'}(-\lambda)
\big\} \cdot \sigma^2_{\beta}/p+p \cdot \{ 1-2\lambda g(-\lambda)+\lambda^2 g^{'}(-\lambda)\} \cdot \sigma^2_{\epsilon},&
\end{flalign*}
These results are based on Lemmas~S\ref{lemma.s1}~and~S\ref{lemma.s2}.
Then Propositions~S\ref{pop.s3}~-~S\ref{pop.s6} follow from the concentration of ridge-type quadratic forms.
\end{proposition.s}

%%%%%%%%%%%%%%%%%%%%%%%%%%%%%%%%%%%%%%%%%%%%%%%%%%%%%%%%
%%%%%%%%%%%%%%%%%%%%%%%%%%%%%%%%%%%%%%%%%%%%%%%%%%%%%%%%
%%%%%%%%%%%%%%%%%%%%%%%%%%%%%%%%%%%%%%%%%%%%%%%%%%%%%%%%
\subsection{Additional technical details for marginal estimator}\label{sec9.11}
The following technical details are useful in proving our theoretical results of marginal estimator. 

\begin{flalign*}
&\tE \Big\{\big(\Z_{(1)}\bmeta_{(1)}+\bmeps_{z}\big)^T\Z\X^T\big(\X_{(1)}\bmbeta_{(1)}+\bmeps\big) \Big\} &\\
&= (m_{\beta\eta}/p) n_z n \tr(\widehat{\bmSigma}_X \widehat{\bmSigma}_Z)\cdot{\sigma_{\beta\eta}}/p= 
(m_{\beta\eta}/p) n_z n \tr(\bmSigma^2)\cdot{\sigma_{\beta\eta}}/p=
m_{\beta\eta} n_z n b_2(\bmSigma)\cdot{\sigma_{\beta\eta}}/p, &\\
&\tE\Big\{\big(\Z_{(1)}\bmeta_{(1)}+\bmeps_{z}\big)^T\big(\Z_{(1)}\bmeta_{(1)}+\bmeps_{z}\big)\Big\}&\\
&=(m_{\eta}/p)n_z\tr(\widehat{\bmSigma}_Z)\cdot{\sigma^2_{\eta}}/p+n_z\cdot{\sigma^2_{\epsilon_{z}}}=n_zm_{\eta}\cdot{\sigma^2_{\eta}}/p+n_z\cdot{\sigma^2_{\epsilon_{z}}}, &\\
&\tE\Big\{\big(\X_{(1)}\bmbeta_{(1)}+\bmeps\big)^T\X\Z^T\Z\X^T\big(\X_{(1)}\bmbeta_{(1)}+\bmeps\big)\Big\}&\\
&=(m_{\beta}/p)n^2n_z\tr(\widehat{\bmSigma}_X\widehat{\bmSigma}_Z\widehat{\bmSigma}_X)\cdot{\sigma^2_{\beta}}/p+nn_z\tr(\widehat{\bmSigma}_X\widehat{\bmSigma}_Z)\cdot{\sigma^2_{\epsilon}}&\\
&=(m_{\beta}/p)n^2n_z\big\{\frac{n+1}{n}\tr(\bmSigma^3)+\frac{1}{n}\tr(\bmSigma)\tr(\bmSigma^2)\big\} \cdot{\sigma^2_{\beta}}/p+nn_z\tr(\bmSigma^2)\cdot{\sigma^2_{\epsilon}}&\\
&=nn_zm_{\beta}\cdot \{(n+1)b_3(\bmSigma)+pb_2(\bmSigma)\}\cdot{\sigma^2_{\beta}}/p+nn_zpb_2(\bmSigma)\cdot{\sigma^2_{\epsilon}}
\end{flalign*}
%%%%%
\begin{flalign*}
&\tE \Big\{\big(\X_{(1)}\bmbeta_{(1)}+\bmeps\big)^T\X\X^T\big(\X_{(1)}\bmbeta_{(1)}+\bmeps\big) \Big\}&\\
&= (m_{\beta}/p) n^2 \tr(\widehat{\bmSigma}_X^2)\cdot{\sigma_{\beta}^2}/p+n \tr(\widehat{\bmSigma})\cdot {\sigma^2_{\epsilon}}&\\
&= 
(m_{\beta}/p) n^2p b_2(\widehat{\bmSigma}_X)\cdot{\sigma_{\beta}^2}/p+npb_1(\widehat{\bmSigma})\cdot {\sigma^2_{\epsilon}}&\\
&=
m_{\beta} n^2 \{b_2(\bmSigma)+\omega\}\cdot{\sigma_{\beta}^2}/p+np\cdot {\sigma^2_{\epsilon}}&\\
&\tE\Big\{\big(\X_{(1)}\bmbeta_{(1)}+\bmeps\big)^T\big(\X_{(1)}\bmbeta_{(1)}+\bmeps\big)\Big\}&\\
&=(m_{\beta}/p)n\tr(\widehat{\bmSigma}_X)\cdot{\sigma^2_{\beta}}/p+n_z\cdot{\sigma^2_{\epsilon}}
=nm_{\beta}\cdot{\sigma^2_{\beta}}/p+n\cdot{\sigma^2_{\epsilon}},\quad \mbox{and}&\\
&\tE\Big\{\big(\X_{(1)}\bmbeta_{(1)}+\bmeps\big)^T\X\X^T\X\X^T\big(\X_{(1)}\bmbeta_{(1)}+\bmeps\big)\Big\} &\\ 
&= (m_{\beta}/p) n^3 \tr(\widehat{\bmSigma}_X^3)\cdot{\sigma_{\beta}^2}/p+n^2 \tr(\widehat{\bmSigma}^2)\cdot {\sigma^2_{\epsilon}}&\\
&=n^3m_{\beta}\cdot \{b_3(\bmSigma)+3\omega b_2(\bmSigma)+\omega^2\}\cdot{\sigma^2_{\beta}}/p+n^2p\cdot \{b_2(\bmSigma)+\omega\}\cdot{\sigma^2_{\epsilon}}.&
\end{flalign*}

%%%%%%%%%%%%%%%%%%%%%%%%%%%%%%%%%%%%%%%%%%%%%%%%%%%%%%%%
%%%%%%%%%%%%%%%%%%%%%%%%%%%%%%%%%%%%%%%%%%%%%%%%%%%%%%%%
%%%%%%%%%%%%%%%%%%%%%%%%%%%%%%%%%%%%%%%%%%%%%%%%%%%%%%%%

\subsection{Real data analysis}\label{sec9.12}

\subsubsection*{Data processing}
The raw MRI, covariates and genetic data are downloaded from data resources. We process the MRI data locally using consistent procedures via advanced normalization tools (ANTs, \citep{avants2011reproducible}) to generate ROI volumes for each dataset. 
Normalization steps using the ANTs software are detailed in \cite{tustison2014large} and \cite{avants2011reproducible}. We use the standard OASIS-30 Atropos template for registration and Mindboggle-101 atlases for labeling. 
Details of these templates and processing steps can be found in \url{https://mindboggle.info/data.html}, \cite{klein2012101} and \cite{tustison2014large}. 
For each phenotype and continuous covariate variable, we further remove values greater than five times the median absolute deviation from the median value. 
We use imputed SNP data in real data analysis. We perform the following genetic variants data quality controls on each dataset: 1) exclude subjects with more than $10\%$ missing genotypes; 2) exclude variants with minor allele frequency less than $0.01$; 3) exclude variants with larger than $10\%$ missing genotyping rate; 4) exclude variants that fail the Hardy-Weinberg test at $1\times 10^{-7}$ level; and 5) remove variants with imputation INFO score less than $0.8$. 
All individuals are aged between $3$ and $92$ years. More cohort information of these studies can be found in \cite{zhao2019genome}.

\subsubsection*{Data acknowledgements}
Part of the data collection and sharing for this project was funded by the Pediatric Imaging, Neurocognition and Genetics Study (PING) (U.S. National Institutes of Health Grant RC2DA029475). PING is funded by the National Institute on Drug Abuse and the Eunice Kennedy Shriver National Institute of Child Health \& Human Development. PING data are disseminated by the PING Coordinating Center at the Center for Human Development, University of California, San Diego. 
\subsubsection*{PING methods}
Part of the data used in the preparation of this article were obtained from the Pediatric Imaging, Neurocognition and Genetics (PING) Study database (\url{http://ping.chd.ucsd.edu/}). PING was launched in 2009 by the National Institute on Drug Abuse (NIDA) and the Eunice Kennedy Shriver National Institute Of Child Health \& Human Development (NICHD) as a 2-year project of the American Recovery and Reinvestment Act. The primary goal of PING has been to create a data resource of highly standardized and carefully curated magnetic resonance imaging (MRI) data, comprehensive genotyping data, and developmental and neuropsychological assessments for a large cohort of developing children aged 3 to 20 years. The scientific aim of the project is, by openly sharing these data, to amplify the power and productivity of investigations of healthy and disordered development in children, and to increase understanding of the origins of variation in neurobehavioral phenotypes. For up-to-date information, see \url{http://ping.chd.ucsd.edu/}.

\subsection*{Pediatric Imaging, Neurocognition and Genetics (PING) authors}
{Connor McCabe$^1$, Linda Chang$^2$, Natacha Akshoomoff$^{3}$, Erik Newman$^{1}$, Thomas Ernst$^{2}$, Peter Van Zijl$^{4}$, Joshua Kuperman$^{5}$, Sarah Murray$^{6}$, Cinnamon Bloss$^{6}$, Mark Appelbaum$^{1}$, Anthony Gamst$^{1}$, Wesley Thompson$^{3}$, Hauke Bartsch$^{5}$.}\\

{$^{1}$UC San Diego, La Jolla, CA 92093, USA.}
{$^{2}$U Hawaii, Honolulu, HI 96822, USA.}
{$^{3}$Department of Psychiatry, University of California, San Diego, La Jolla, California 92093, USA.}
{$^{4}$Kennedy Krieger Institute, Baltimore, MD 21205, USA.}
{$^{5}$Multimodal Imaging Laboratory, Department of Radiology, University of California San Diego, La Jolla, California 92037, USA.}
{$^{6}$Scripps Translational Science Institute, La Jolla, CA 92037, USA.}

%%%%%%%%%%%%%%%%%%%%%%%%%%%%%%%%%%%%%%%%%%%%%%%%%%%%%%%%
%%%%%%%%%%%%%%%%%%%%%%%%%%%%%%%%%%%%%%%%%%%%%%%%%%%%%%%%
%%%%%%%%%%%%%%%%%%%%%%%%%%%%%%%%%%%%%%%%%%%%%%%%%%%%%%%%
\subsection{Supplementary figures}
\clearpage
%%%%%%%%%%%%%%%%%
\begin{suppfigure}
\includegraphics[page=1,width=1\linewidth]{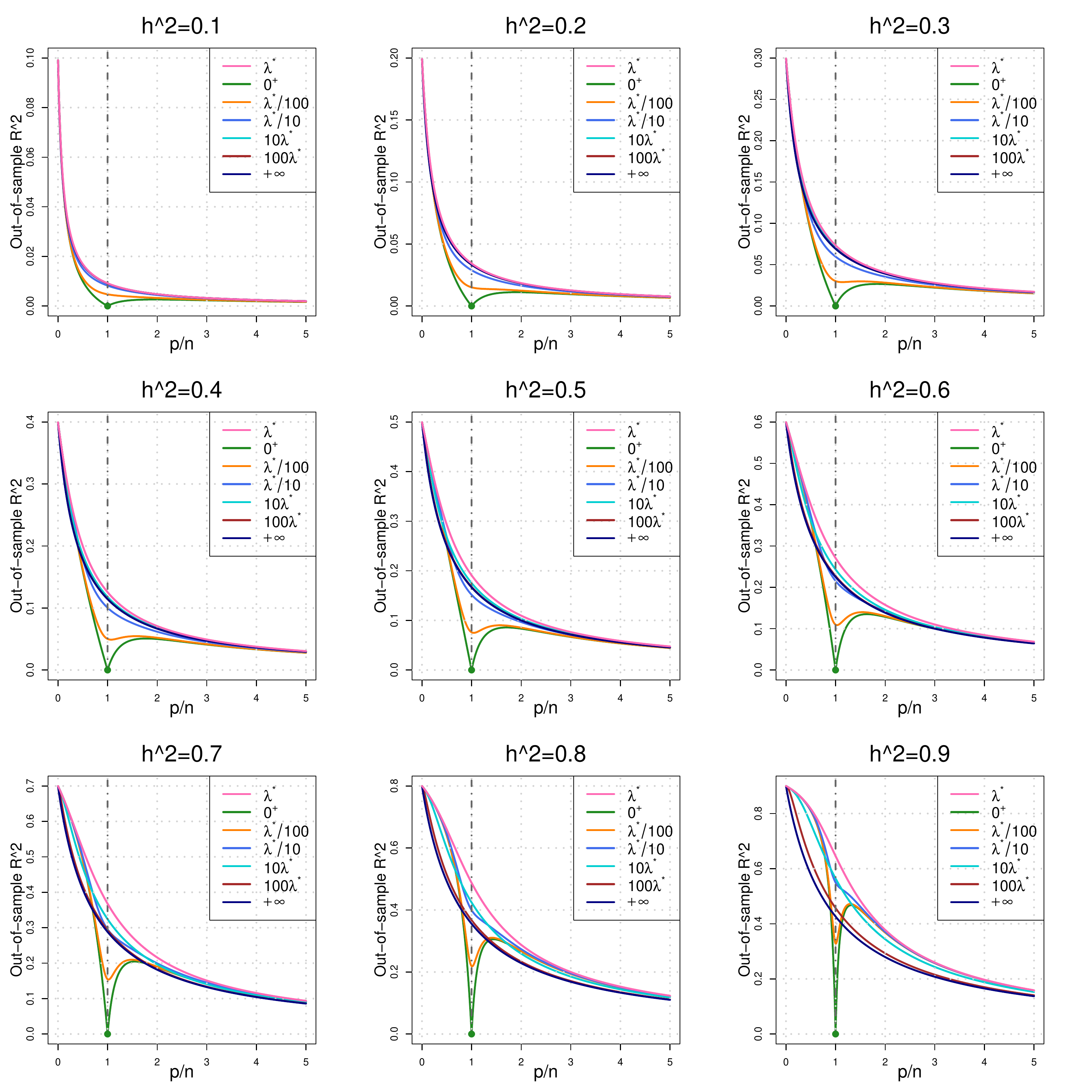}
  \caption{Out-of-sample $R$-squared $A^2_{R}(\lambda)$ of ridge-type estimators given different $\lambda$ and heritability  when $\bmSigma=\I_p$. $\lambda^*$ is the optimal $\lambda$ value, $0^{+}$ corresponds to the case $\lambda \to 0^+$, and $+\infty$ represents $\lambda \to +\infty$. 
  We set $\varphi_{\beta\eta}=1$, and vary $\h_{\beta}^2=\h_{\eta}^2$ from $0.1$ to $0.9$. 
}
\label{sfig1}
\end{suppfigure}
%%%%%%%%%%%%%%%%%%
%%%%%%%%%%%%%%%%%
\begin{suppfigure}
\includegraphics[page=1,width=1\linewidth]{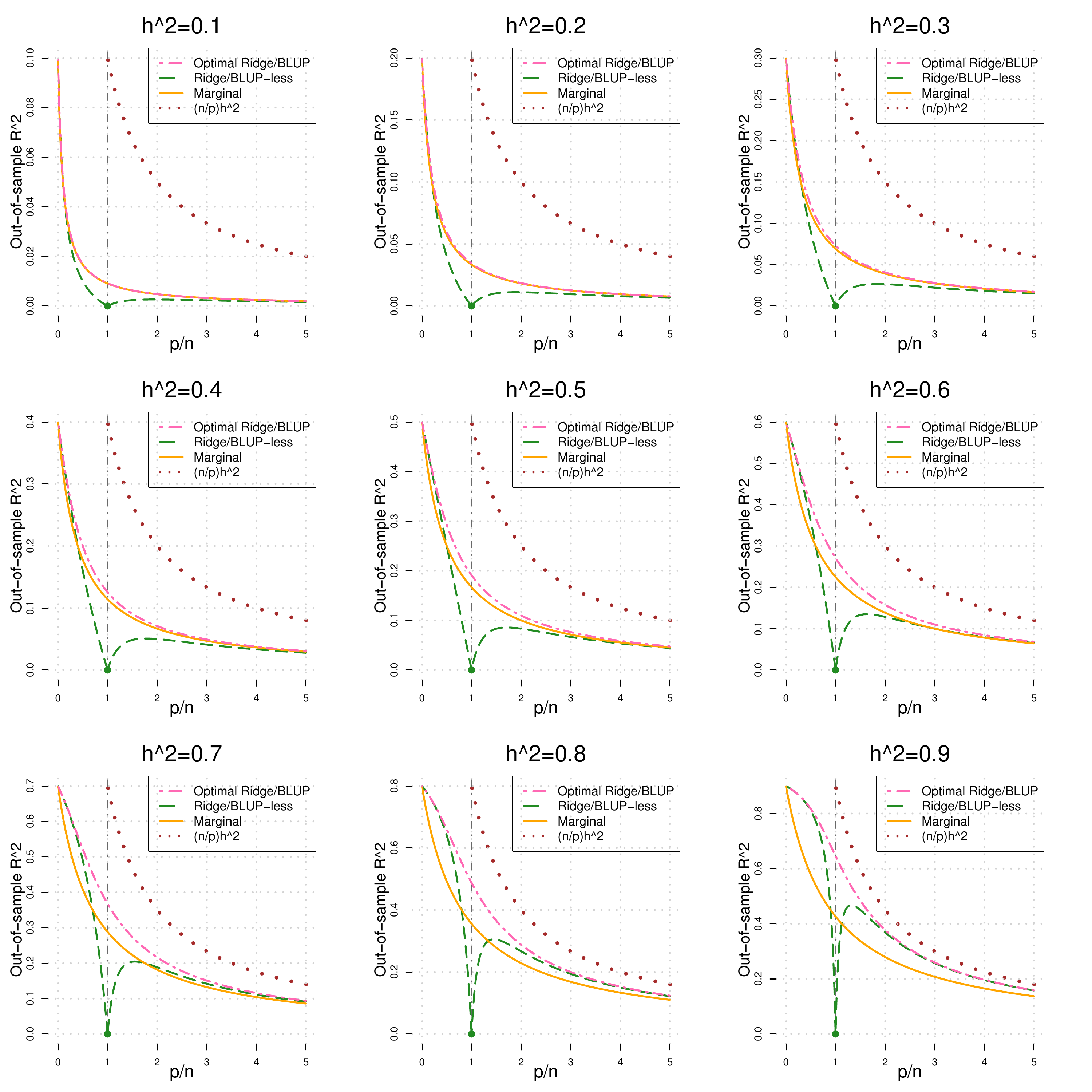}
  \caption{Out-of-sample  $R$-squared of optimal ridge/BLUP estimators ($A^2_{R}(\lambda^*)$=$A^2_{B}(\lambda^*/\omega)$), ridge/BLUP-less estimators ($A^2_{R}(0^{+})$=$A^2_{B}(0^{+})$), and marginal estimator ($A^2_{S}$) when $\bmSigma=\I_p$.   
  We set $\varphi_{\beta\eta}=1$, and vary $\h_{\beta}^2=\h_{\eta}^2$ from $0.1$ to $0.9$. 
}
\label{sfig2}
\end{suppfigure}
%%%%%%%%%%%%%%%%%%
%%%%%%%%%%%%%%%%%
\begin{suppfigure}
\includegraphics[page=1,width=1\linewidth]{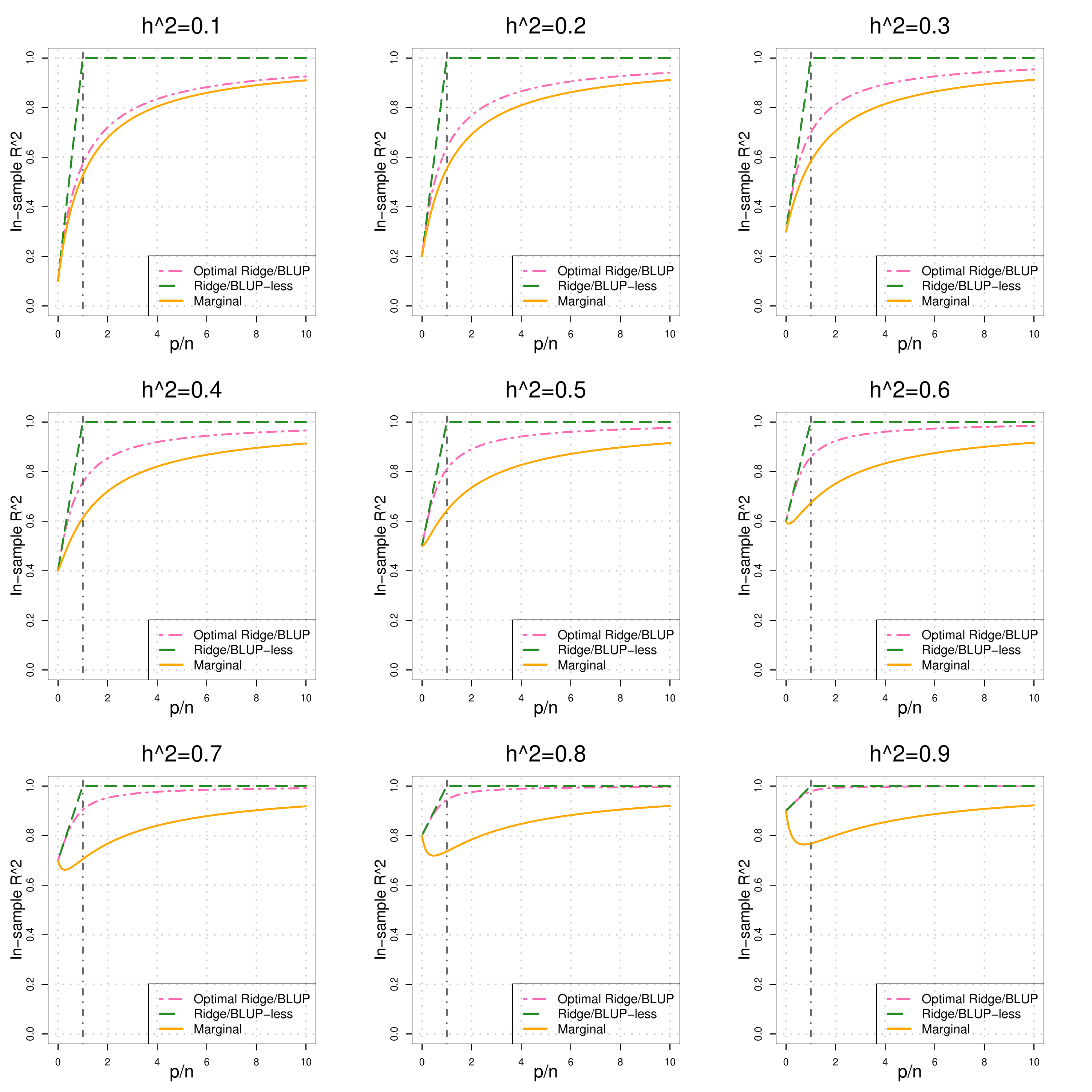}
  \caption{In-sample $R$-squared  of ridge/BLUP-less estimators ($E^2_{R}(0^{+})=E^2_{B}(0^{+})$), optimal (out-of-sample) ridge/BLUP estimators ($E^2_{R}(\lambda^*)=E^2_{B}(\lambda^*/\omega)$), and marginal estimator ($E^2_{S}$) when $\bmSigma=\I_p$. 
   We vary $\h_{\beta}^2$ from $0.1$ to $0.9$. 
}
\label{sfig3}
\end{suppfigure}
%%%%%%%%%%%%%%%%%%

%%%%%%%%%%%%%%%%%
\begin{suppfigure}
\includegraphics[page=1,width=1\linewidth]{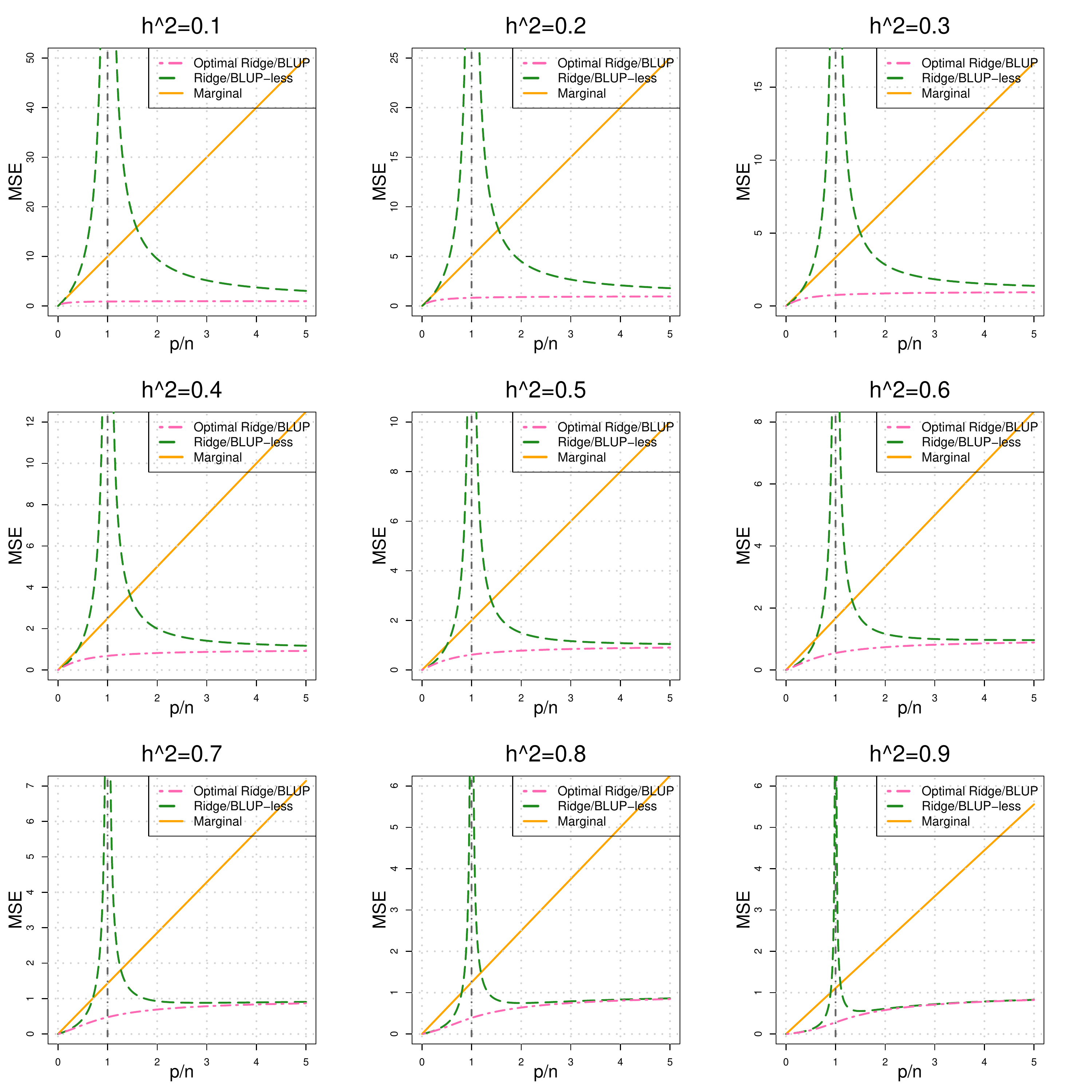}
  \caption{Mean squared prediction error (MSE) of marginal, ridge/BLUP, and ridge-less/BLUP-less estimators when $\bmSigma=\I_p$.
  We set $m_{\beta}\sigma^2_{\beta}/p=1$, and vary $\h_{\beta}^2$ from $0.1$ to $0.9$. 
}
\label{sfig4}
\end{suppfigure}
%%%%%%%%%%%%%%%%%%
%%%%%%%%%%%%%%%%%%
%%%%%%%%%%%%%%%%%%
\begin{suppfigure}
\includegraphics[page=1,width=1\linewidth]{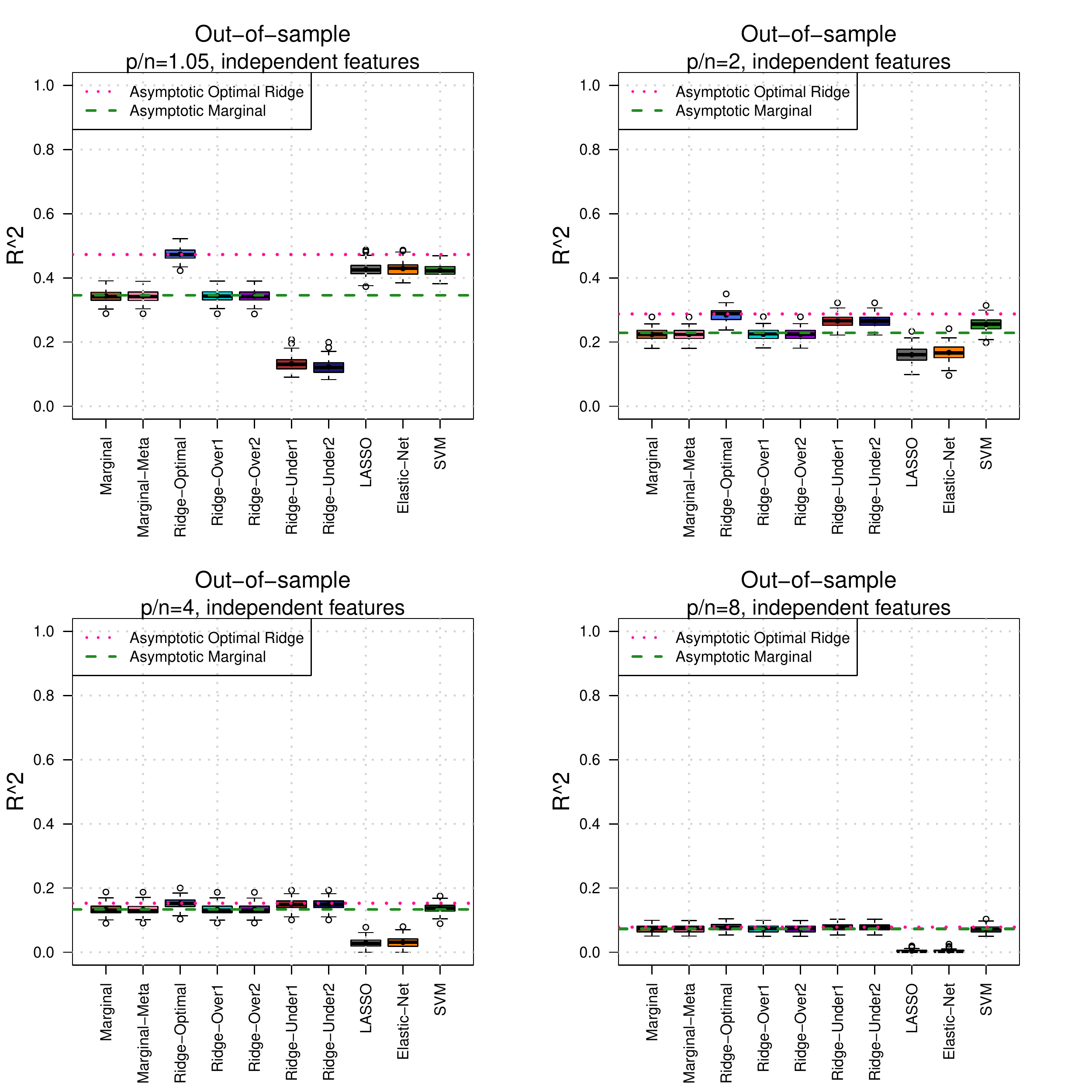}
  \caption{Out-of-sample $R$-squared of different estimators for independent features. See Figure~\ref{fig5} for figure notations.
  We set $n=2000$, and vary $\omega=$ from $1.05$ to $8$.
  The dash lines represent the asymptotic limits of ridge (red) and  marginal (green) estimators.}
\label{sfig5}
\end{suppfigure}
%%%%%%%%%%%%%%%%%%%
%%%%%%%%%%%%%%%%%%
\begin{suppfigure}
\includegraphics[page=2,width=1\linewidth]{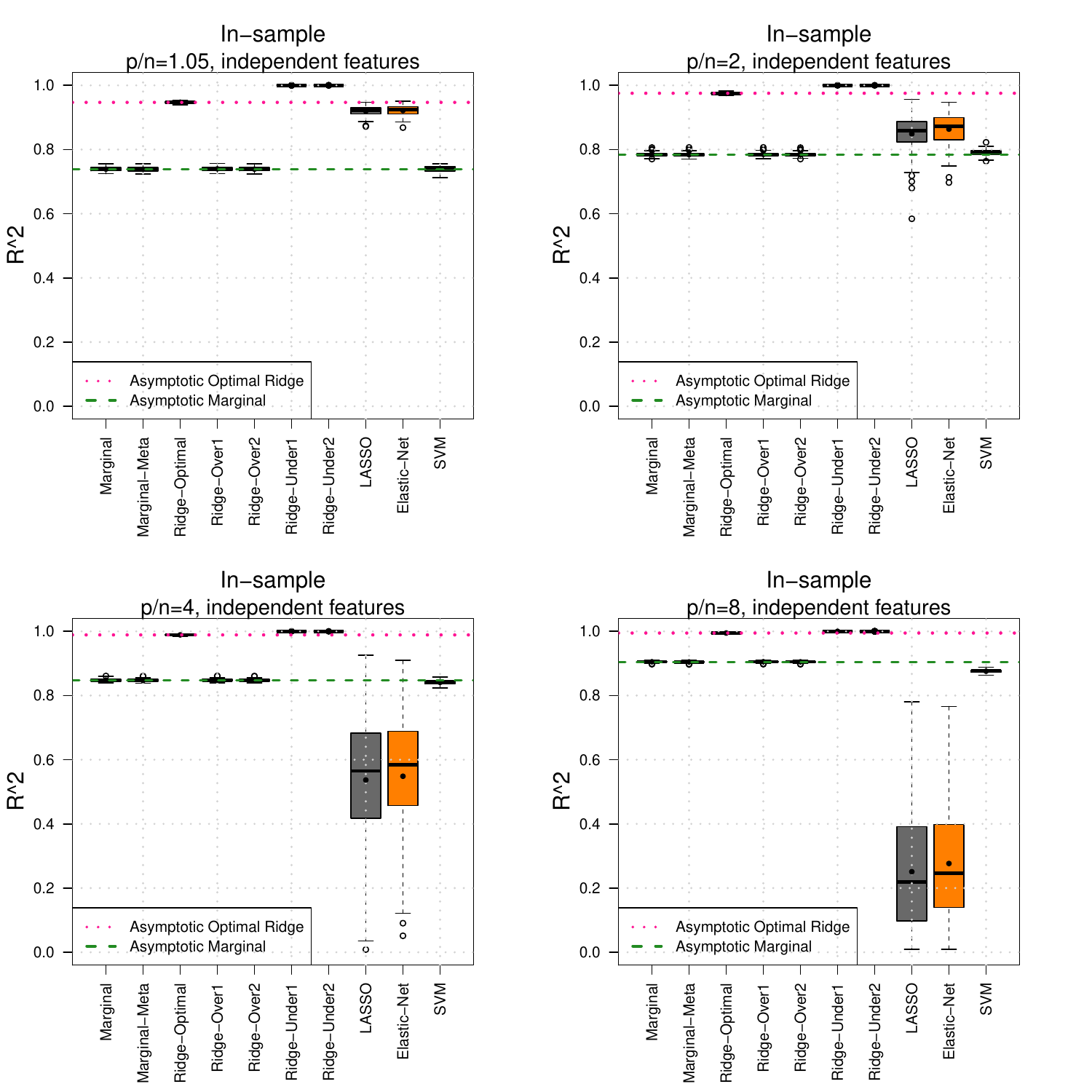}
  \caption{In-sample $R$-squared of different estimators for independent features.
   See Figure~\ref{fig5} for figure notations.
  We set $n=2000$, and vary $\omega=$ from $1.05$ to $8$.
  The dash lines represent the asymptotic limits of ridge (red) and  marginal (green) estimators.}
\label{sfig6}
\end{suppfigure}
%%%%%%%%%%%%%%%%%%
%%%%%%%%%%%%%%%%%%
\begin{suppfigure}
\includegraphics[page=1,width=1\linewidth]{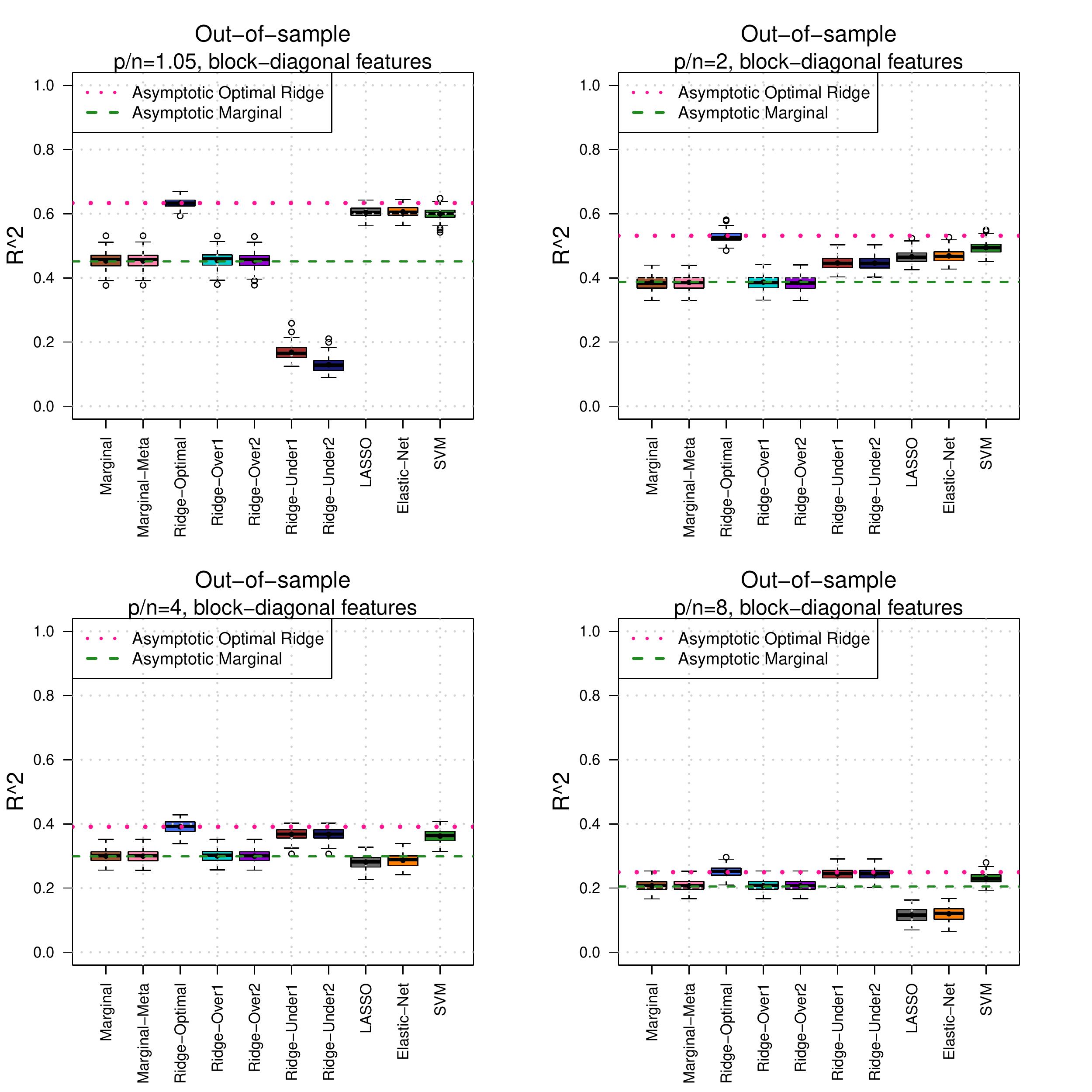}
  \caption{Out-of-sample $R$-squared of different estimators for features with block-diagonal correlation structure. See Figure~\ref{fig5} for figure notations.
  We set $n=2000$, and vary $\omega=$ from $1.05$ to $8$.
  The dash lines represent the asymptotic limits of ridge (red) and  marginal (green) estimators.}
\label{sfig7}
\end{suppfigure}
%%%%%%%%%%%%%%%%%%%
%%%%%%%%%%%%%%%%%%
\begin{suppfigure}
\includegraphics[page=2,width=1\linewidth]{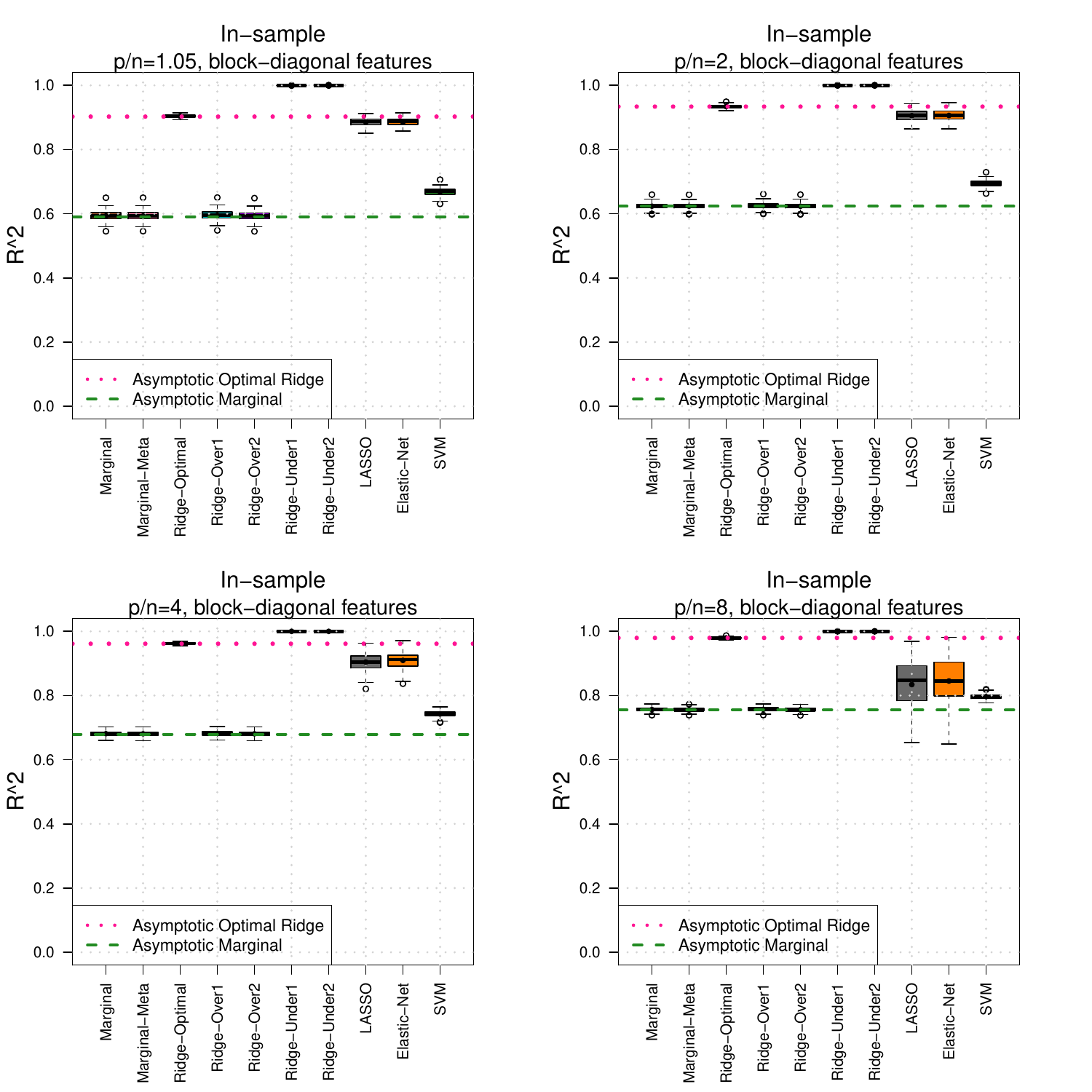}
  \caption{In-sample $R$-squared of different estimators for features with block-diagonal correlation structure. See Figure~\ref{fig5} for figure notations.
  We set $n=2000$, and vary $\omega$ from $1.05$ to $8$.
  The dash lines represent the asymptotic limits of ridge (red) and  marginal (green) estimators.}
\label{sfig8}
\end{suppfigure}

%%%%%%%%%%%%%%%%%%%
%%%%%%%%%%%%%%%%%%
\clearpage
\section*{Supplementary table}

% latex table generated in R 3.5.0 by xtable 1.8-2 package
% Sun Nov 17 22:43:06 2019
\begin{supptable}[ht]
\centering
\begin{tabular}{rrrrr}
  \hline
ROI ID & BLUP-IBAM & Marginal-IBAM & BLUP-LST & Marginal-LST \\
  \hline
left.thalamus.proper & 1.739E-02 & 3.501E-02 & 5.185E-02 & 3.895E-02 \\ 
left.caudate & 4.712E-02 & 1.377E-01 & 5.536E-02 & 4.042E-05 \\ 
left.putamen & 1.864E-01 & 2.224E-01 & 1.396E-02 & 3.509E-01 \\ 
left.pallidum & 1.839E-02 & 4.233E-02 & 9.342E-02 & 3.600E-02 \\ 
left.hippocampus & 8.727E-01 & 7.419E-02 & 4.957E-01 & 3.823E-02 \\ 
left.amygdala & 6.878E-01 & 9.384E-01 & 7.425E-01 & 7.848E-01 \\ 
left.accumbens.area & 6.475E-04 & 6.923E-02 & 1.647E-02 & 9.124E-03 \\ 
right.thalamus.proper & 4.950E-04 & 2.509E-04 & 1.205E-01 & 6.544E-03 \\ 
right.caudate & 1.549E-02 & 2.900E-01 & 1.513E-02 & 2.731E-02 \\ 
right.putamen & 2.607E-01 & 2.474E-01 & 3.173E-02 & 3.328E-01 \\ 
right.pallidum & 8.900E-02 & 9.409E-02 & 1.796E-01 & 6.048E-03 \\ 
right.hippocampus & 2.344E-01 & 3.116E-03 & 1.044E-01 & 1.666E-02 \\ 
right.amygdala & 2.369E-02 & 1.942E-01 & 1.181E-02 & 1.040E-01 \\ 
right.accumbens.area & 1.591E-02 & 2.564E-03 & 1.092E-02 & 4.209E-05 \\ 
   \hline
\end{tabular}
\caption{Partial $R$-squared ($\times 100\%$) of $14$ subcortical ROI volumes to predict IBAM and LST cognitive test scores in the PING cohort. 
The partial $R$-squared is estimated from linear regression while adjusting for the effects of age and gender.
BLUP: best linear unbiased prediction; Marginal: marginal estimator. 
IBAM: IBAM score; LST: list sort total score.}
\label{tables1}
\end{supptable}
%%%%%%%%%%%%%%%%%%%%%%%%%%%%%%%%%%%%%%%%%%%%%%%%%%%%%%%%
%%%%%%%%%%%%%%%%%%%%%%%%%%%%%%%%%%%%%%%%%%%%%%%%%%%%%%%%
%%%%%%%%%%%%%%%%%%%%%%%%%%%%%%%%%%%%%%%%%%%%%%%%%%%%%%%%
\end{document}